\documentclass[11pt,a4paper]{article}

\PassOptionsToPackage{linktocpage,pagebackref}{hyperref}
\usepackage{jheppub}
\hypersetup{
    pdftitle={Bounds on scattering amplitudes of non-identical scalar particles in 4d},
    pdfauthor={Gabriele Ferretti, Denis Karateev, Alessandro Piazza, Marco Serone}
}

\usepackage[utf8]{inputenc}
\usepackage[T1]{fontenc}
\usepackage{subcaption}
\usepackage{braket}
\usepackage{booktabs}
\usepackage{placeins}

\usepackage{tikz}
\usepackage{pgfplots}
\usepackage{relsize}
\usetikzlibrary{arrows}
\usetikzlibrary{arrows.meta}
\usetikzlibrary{calc}
\usetikzlibrary{decorations.pathmorphing}
\usetikzlibrary{decorations.markings}
\usetikzlibrary{pgfplots.fillbetween}
\usetikzlibrary{tikzmark}
\usetikzlibrary{trees}
\usetikzlibrary{shapes.misc}
\usetikzlibrary{shapes.multipart}
\pgfplotsset{compat=1.18}

\renewcommand*{\backref}[1]{}
\renewcommand*{\backrefalt}[4]{\ifcase #1 \or {\scriptsize\,[p.~#2]} \else {\scriptsize\,[pp.~#2]} \fi}

\makeatletter
\patchcmd\NAT@citexnum{\let\NAT@last@num\NAT@num}{\MakeLinkTarget[cite]{}\Hy@backout{\@citeb\@extra@b@citeb}\let\NAT@last@num\NAT@num}{}{\fail}
\makeatother

\newcommand{\nn}{\nonumber}
\newcommand{\ped}[1]{\ifmmode{_{\mathrm{#1}}}\else\textsubscript{#1}\fi}
\newcommand{\ap}[1]{\ifmmode{^{\mathrm{#1}}}\else\textsuperscript{#1}\fi}

\let\originalleft\left
    \let\originalright\right
\renewcommand{\left}{\mathopen{}\mathclose\bgroup\originalleft}
    \renewcommand{\right}{\aftergroup\egroup\originalright}

\DeclareMathOperator*{\res}{Res}
\DeclareMathOperator{\disc}{Disc}
\DeclareMathOperator{\ddisc}{dDisc}

\newcommand{\Z}{\mathbb{Z}}
\newcommand{\id}{\text{\usefont{U}{bbold}{m}{n}1}}
\newcommand{\mom}{\boldsymbol{p}}
\newcommand{\momPrime}{\boldsymbol{q}}

\renewcommand{\Re}{\operatorname{Re}}
\renewcommand{\Im}{\operatorname{Im}}

\usepackage{xifthen}
\newcommand{\dif}[1][]{
    \ifthenelse{\NOT\isempty{#1}}{
        {\mathrm{d}}^{#1}
    }{
        \mathrm{d}
    }
}

\title{\boldmath Bounds on scattering amplitudes of \\ non-identical scalar particles in \( 4d \)}

\author[a]{Gabriele Ferretti,}
\author[b]{Denis Karateev,}
\author[c,d]{Alessandro Piazza,}
\author[c,d]{and Marco Serone}

\affiliation[a]{Department of Physics and Astronomy, Chalmers University of Technology,\\
	Fysikg{\char229}rden 1, 41296 G{\"o}teborg, Sweden}

\affiliation[b]{D\'epartment de Physique Th\'eorique, Universit\'e de Gen\`eve,\\
	24 quai Ernest-Ansermet, 1211 Gen\`eve, Switzerland}

\affiliation[c]{SISSA, Via Bonomea 265, I-34136, Trieste, Italy}

\affiliation[d]{INFN, Sezione di Trieste, Via Valerio 2, I-34127, Trieste, Italy}

\emailAdd{gabriele.ferretti@chalmers.se}
\emailAdd{denis.karateev@unige.ch}
\emailAdd{apiazza@sissa.it}
\emailAdd{serone@sissa.it}

\abstract{%
    We initiate the study of two-to-two scattering amplitudes involving two distinct scalar particles in $d=4$ spacetime dimensions using the primal $S$-matrix bootstrap. We impose a $\Z_2\times \Z_2$ global symmetry under which the two particles are the lightest ones carrying charges $(-,+)$ and $(+,-)$, guaranteeing their absolute stability and absence of triangular anomalous thresholds.
    For unequal masses, the presence of a pseudo-physical cut whose discontinuity is not directly constrained by physical unitarity qualitatively changes the bootstrap problem. We find several observables to be unbounded, while others obey non-trivial one-sided bounds or become bounded once another observable is fixed. In the equal-mass limit, where the pseudo-physical region disappears, two-sided bounds are recovered. As a proof of concept, we show that supplying information about the pseudo-physical discontinuity through an Omnès parametrisation also restores boundedness. Our results provide a starting point for future studies of physical processes such as pion-kaon and pion-nucleon scattering.}

\dedicated{Dedicated to the memory of Daniele Amati}

\begin{document}
\maketitle

\clearpage

\section{Introduction}
\label{sec:introduction}

The \( S \)-matrix bootstrap is a collection of methods for studying scattering amplitudes in quantum field theory (QFT) in a model-independent way, without committing to a particular action. Its basic input consists of Lorentz invariance, crossing symmetry, causality/analyticity and unitarity, and most applications have focused on two-to-two scattering amplitudes. See~\cite{Eden:1966dnq,Martin:102663} for classic references.
This program has recently been revived with the seminal papers \cite{Paulos:2016fap,Paulos:2016but}.
For a modern overview, see the white paper~\cite{Kruczenski:2022lot}.
One may think of amplitudes as forming an infinite-dimensional space parametrised by dimensionless observables.
The general principles of QFT carve out the allowed region in this space.
The bootstrap aims both to bound this region and to construct amplitudes that lie on its boundary.
The first non-perturbative bounds of this kind were obtained by Lopez and Mennessier~\cite{Lopez:1975ca,Lopez:1976zs}, while a simple modern implementation was introduced in~\cite{Paulos:2017fhb}.

For present purposes, it is useful to organise \( S \)-matrix bootstrap methods into two broad classes, called \emph{primal} and \emph{dual}. Dual methods rule out inconsistent amplitudes, whereas primal methods ``rule in'' amplitudes by constructing explicit solutions. Dual approaches include~\cite{Lopez:1975ca,Lopez:1976zs,Guerrieri:2020kcs,Guerrieri:2021tak,He:2021eqn}, while primal approaches include~\cite{Paulos:2017fhb,Paulos:2016but,He:2021eqn,deRham:2025vaq}. The two viewpoints are complementary: dual methods offer more direct control over the rigour of exclusion bounds, while primal methods are convenient for constructing amplitudes in practice. In this paper we use the primal method of~\cite{Paulos:2017fhb}. Bootstrap methods based on neural networks were proposed in~\cite{Gumus:2024lmj,Gumus:2026mhb}, while other constructive approaches were developed in~\cite{Tourkine:2021fqh,Tourkine:2023xtu}.

The \( S \)-matrix bootstrap is particularly powerful in $d=2$, where many bounds are saturated by integrable models~\cite{Paulos:2016but}; further developments include~\cite{Cordova:2018uop,Cordova:2019lot,Chen:2021pgx,Cordova:2023wjp,Copetti:2024dcz,Cordova:2025bah,Homrich:2019cbt,Bercini:2019vme}. In higher dimensions, bounds on identical neutral massive particles were obtained in~\cite{Lopez:1975ca,Lopez:1976zs,Paulos:2017fhb}, with further results in $d=4$~\cite{EliasMiro:2022xaa,Chen:2022nym,EliasMiro:2026utl} and in general dimension~\cite{Chen:2022nym,Gumus:2025hwq}. Most of these works concern massive scalars. Spinning particles in $d=4$ were treated in~\cite{Hebbar:2020ukp}, see also~\cite{deRham:2017zjm,Buric:2023ykg}. In particular, bounds on identical massive Majorana fermions were obtained in~\cite{Hebbar:2020ukp}. For massless particles, universal soft expansions, which follow from the shift symmetry at low energy, must also be imposed. A practical implementation was introduced in~\cite{Guerrieri:2020bto}, with further developments in~\cite{Acanfora:2023axz,Bocchia:2026xxx}; bounds on photons were obtained in~\cite{Haring:2022sdp}. Other applications include QCD flux tubes~\cite{EliasMiro:2019kyf,EliasMiro:2021nul,Guerrieri:2024ckc}, the QCD world-sheet axion~\cite{Gaikwad:2023hof}, glueballs~\cite{Guerrieri:2023qbg}, the $c$-anomaly in $d=2$ and the $a$-anomaly in $d=4$~\cite{Karateev:2019ymz,Karateev:2022jdb}, Froissart growth~\cite{Correia:2025uvc} and scattering amplitudes in string theory and M-theory~\cite{Guerrieri:2021ivu,Guerrieri:2022sod}.

Under the standard assumptions of analyticity and Froissart boundedness~\cite{Froissart:1961ux}, scattering amplitudes obey dispersion relations. Combining these relations with positivity, which captures a restricted subset of unitarity, gives non-perturbative bounds on the space of observables. These dispersive methods should be viewed as dual $S$-matrix bootstrap methods: they generally give weaker bounds because they do not use the full unitarity constraints, but are much easier to implement in practice. Modern formulations were developed in~\cite{Bellazzini:2020cot,Caron-Huot:2020cmc,Tolley:2020gtv}, with further developments in~\cite{Sinha:2020win,Bellazzini:2021oaj,Beadle:2024hqg,Bellazzini:2025bay,EliasMiro:2025rqo,Peng:2025klv}. Many applications concern massless amplitudes, where one is often forced to assume a tree-level-like analytic structure. Examples include string-like amplitudes~\cite{Figueroa:2022onw,Haring:2023zwu,Eckner:2024ggx,Eckner:2024pqt,Huang:2025icl,Alday:2025pmg,Eckner:2025kve,deRham:2026lvc} and pion amplitudes in the $N_c\to\infty$ limit of QCD~\cite{Albert:2022oes,Fernandez:2022kzi,Albert:2023jtd,Ma:2023vgc,Albert:2023seb,Albert:2026xyz,Bocchia:2026kew}.\footnote{Dispersive methods are also powerful for bounding effective field theories. For early works see~\cite{Pham:1985cr,Adams:2006sv}; further developments include~\cite{deRham:2017avq,deRham:2017imi,deRham:2018qqo}, and a review is given in~\cite{Baumgart:2022yty}. More recent works include~\cite{Arkani-Hamed:2020blm,Caron-Huot:2021rmr,Henriksson:2021ymi,Caron-Huot:2022ugt,Henriksson:2022oeu,Bellazzini:2023nqj,CarrilloGonzalez:2023cbf,McPeak:2023wmq,Albert:2024yap,Dong:2024omo,Bellazzini:2025shd,Beadle:2025cdx,Dong:2025dpy}.} For massive amplitudes, dispersion relations were studied systematically for identical scalars in~\cite{Chen:2022nym} and have recently been applied to spin-one particles~\cite{Bertucci:2024qzt} and particles of unequal masses~\cite{deRham:2025htd}.

\paragraph{What is this paper about?}
In this paper we are particularly interested in applications of the non-perturbative \( S \)-matrix bootstrap to low-energy QCD.
Theoretical bounds on pion-pion scattering were obtained in the pioneering work of~\cite{Guerrieri:2018uew}.
By injecting experimental data into the \( S \)-matrix bootstrap setup one can isolate the pion-pion amplitude and study the resonances it contains, see~\cite{Guerrieri:2024jkn}.\footnote{%
    An interesting new direction is to combine the \( S \)-matrix and form-factor bootstrap in the study of pion-pion scattering; see~\cite{He:2023lyy,He:2024nwd,He:2025gws}.
}

A natural next step is to study processes involving different hadrons, in particular pion-kaon scattering, $\pi K\to\pi K$, and pion-nucleon scattering, $\pi N\to\pi N$. Both processes involve two non-identical particles. Apart from the $d=2$ studies of~\cite{Homrich:2019cbt,Guerrieri:2020kcs}, the \( S \)-matrix bootstrap literature has so far focused almost entirely on the scattering of identical particles.\footnote{%
    See however~\cite{Karateev:2022jdb}, which already used the primal \( S \)-matrix bootstrap in a setup of non-identical scalars in \( 4d \).
    In that work, very special assumptions were made on the particles involved which forbid some of the features that we discuss later.
}
The reason is that scattering of non-identical particles brings new analytic structures and a larger system of amplitudes and unitarity constraints.

The main new issue concerns analyticity. For elastic scattering of the lightest particle in the spectrum with itself, one can adopt the standard assumption of maximal analyticity: the amplitude on the principal sheet is analytic for generic complex values of $s$ and $t$, apart from physical-threshold cuts and possible bound-state poles. For scattering involving heavier or non-identical particles, this analytic structure is no longer sufficient. In particular, some amplitudes can develop a branch cut whose branch point lies below the physical threshold of the corresponding process. The segment of this cut below the physical threshold is called the pseudo-physical region. So-called anomalous thresholds can also occur. A bootstrap for non-identical particles therefore requires a generalisation of maximal analyticity that accounts for these additional singularities.

The goal of this paper is to initiate the study of non-identical particles in $d=4$ spacetime dimensions in the simplest possible setting. We consider two real scalar particles, denoted by $A$ and $B$, and impose no continuous global symmetry. We denote their masses by $m_A$ and $m_B$, take $m_A\leq m_B$ without loss of generality and use the notation
\begin{equation}
	m\equiv m_A,\qquad
	M\equiv m_B,\qquad
	\mu\equiv\frac{M}{m}\geq 1\,.
\end{equation}

Without further assumptions, one would in principle have to consider the full set of two-to-two processes
\begin{equation}
	\label{eq:full_set}
	\{ij\to kl\},
\end{equation}
where each of $i,j,k,l$ can be either $A$ or $B$. We simplify this setup by imposing a $\Z_2\times\Z_2$ global symmetry under which $A$ and $B$ carry charges $(-,+)$ and $(+,-)$, respectively. This symmetry has four useful consequences. First, taking $A$ and $B$ to be the lightest states in their charged sectors makes both particles absolutely stable for any value of $\mu$. Second, it forbids all cubic couplings involving $A$ and $B$ and therefore removes one-particle exchange poles due to $A$ or $B$. Third, up to relabelling it reduces the full set~\eqref{eq:full_set} to
\begin{equation}\label{eq:set_amplitudes}
	AA\to AA,\qquad
	BB\to BB,\qquad
	AB\to AB,\qquad
	AA\to BB.
\end{equation}
Fourth, it excludes anomalous thresholds generated by triangular and effective-triangular Feynman graphs. We further assume that the amplitudes contain no one-particle poles due to additional states and no anomalous thresholds of any other origin.

Rather than studying all the amplitudes in~\eqref{eq:set_amplitudes}, in this work we restrict our attention to the following subset of amplitudes:
\begin{equation}
	\label{eq:amplitudes_main}
	AB\to AB,\qquad
	AA\to BB.
\end{equation}
These two processes are related by crossing symmetry. The reduced setup~\eqref{eq:amplitudes_main} is necessarily less powerful than the full system~\eqref{eq:set_amplitudes} and is expected to give weaker bounds on observables. Its advantage is that it provides the minimal system closed under crossing symmetry on which non-trivial unitarity constraints can be imposed. Since the $\pi K\to\pi K$ and $\pi N\to\pi N$ systems are more complicated, it is useful to understand first whether this minimal setup already leads to non-trivial bounds. We show that it does.

\paragraph{Structure of the paper.}
We start in section~\ref{sec:scatteringBasics} by summarising the kinematics of the processes~\eqref{eq:amplitudes_main}, together with the corresponding crossing relations and unitarity constraints. In section~\ref{sec:analytic-structure} we discuss their analytic structure. We discuss pseudo-physical thresholds and anomalous thresholds, and show that triangular anomalous thresholds are absent in the presence of the $\Z_2\times\Z_2$ discrete symmetry.
In section~\ref{sec:observables} we define the observables that we bound in this paper.
We focus on the three dimensionless observables $\lambda_{0,0}$, $\lambda_{2,0}$ and $\lambda_{2,1}$, defined in~\eqref{eq:def-couplings}, and on the scalar scattering length $a_0$, defined through~\eqref{eq:scattering_lengths}.
In section~\ref{sec:numerics} we present our model-independent numerical bounds on these observables. In section~\ref{sec:Omnes} we show how additional assumptions on the pseudo-physical region can improve our bounds. We conclude in section~\ref{sec:discussion}.

Several technical details are collected in the appendices. Appendix~\ref{app:PWE} discusses further analytic properties of the amplitude, such
as Lehmann ellipses and double discontinuities. In appendix~\ref{app:positivity} we derive the dispersion relations and bounds from positivity and linearised unitarity used in the main text. In appendix~\ref{app:omnes} we discuss the Muskhelishvili-Omnès function relevant for section~\ref{sec:Omnes}. Finally, in appendix~\ref{app:numerics-details} we provide the details of the numerical setup needed to reproduce our results.

\paragraph{Summary of the results.}
In section~\ref{sec:numerics} we construct bounds on the couplings $\lambda_{0,0}$, $\lambda_{2,0}$ and $\lambda_{2,1}$, and on the scalar scattering length $a_0$.
We focus on three values of the mass ratio:
\begin{equation}\label{eq:mass-ratios-used}
	\mu=1 \,, \qquad \mu=1.5 \,,\qquad \mu=3.554 \,.
\end{equation}
The first value describes two distinct particles with equal masses. The other two are representative unequal-mass cases, with $\mu=3.554$ approximating the ratio of the physical kaon and pion masses in the isospin limit.

Let us first discuss the unequal-mass case, $\mu>1$.
We find that $\lambda_{0,0}$ is unbounded from both sides, as illustrated in figure~\ref{fig:L00-bound-nmax}.
The coupling $\lambda_{2,0}$ obeys the lower bound $\lambda_{2,0}\geq 0$, but has no upper bound.
The coupling $\lambda_{2,1}$ has a non-trivial absolute lower bound, shown in figure~\ref{fig:L21-bound-fit} and quite close to the bound from linearised unitarity~\eqref{eq:linearised-unitarity-L21}, but no upper bound is present.

At fixed $\lambda_{2,0}$, we obtain a convergent lower bound on $\lambda_{2,1}$ that is stronger than the positivity bound~\eqref{eq:positivity-L20-L21} and the bound from linearised unitarity~\eqref{eq:linearised-unitarity-L21}, see figure~\ref{fig:L21-L20-bound-tot}.
We find no evidence for convergent absolute upper or lower bounds on the scalar scattering length $a_0$.
At fixed $\lambda_{2,0}$, however, $a_0$ has a convergent lower bound shown in figure~\ref{fig:a0-L20-bound}.

The case $\mu=1$ is qualitatively different.
We find that all three couplings admit both upper and lower bounds.
We obtain compact allowed regions in the $(\lambda_{0,0},\lambda_{2,0})$ and $(\lambda_{2,0},\lambda_{2,1})$ planes, see figure~\ref{fig:L00-L20-L21-bound-mu-1}.
These regions are substantially larger than the corresponding regions for identical scalar particles, as shown in figure~\ref{fig:comparison-identical}.

In section~\ref{sec:Omnes} we investigate whether the unbounded directions found for $\mu>1$ can be removed by adding information about the pseudo-physical region.
As a proof of concept, we consider $\mu=1.5$ and restrict the discontinuity in the pseudo-physical region to a scalar Muskhelishvili-Omnès function.
With this additional assumption, we obtain two-sided bounds on all three couplings and a convergent lower bound on $a_0$, see figures~\ref{fig:L00-bound-nmax-omnes}--\ref{fig:a0-bound-nmax-omnes}.
In a realistic application, instead of imposing the Muskhelishvili-Omnès function, one would use the experimental information on the amplitudes entering the analytically continued unitarity relation.

\section{Scattering of unequal particles}
\label{sec:scatteringBasics}

In this section we summarise the main ingredients needed to describe the scattering of two non-identical particles and introduce the notation used throughout the paper.
In subsection~\ref{sec:scattering_amplitudes} we define the two-to-two scattering amplitudes.
In subsection~\ref{sec:kinematics} we determine the physical ranges of the Mandelstam invariants for the processes of interest.
Finally, in subsections~\ref{sec:partial_amplitudes} and~\ref{sec:unitarity} we define the partial wave amplitudes and formulate the corresponding unitarity constraints.

\subsection{Scattering amplitudes}
\label{sec:scattering_amplitudes}

In this paper we work in $d=4$ spacetime dimensions with the mostly-plus metric. We study the scattering processes $ij\to kl$, where each of $i,j,k,l$ can be either particle $A$ or $B$.

We assume the $\Z_2\times\Z_2$ global symmetry under which $A$ and $B$ carry charges $(-,+)$ and $(+,-)$, respectively.
Taking $A$ and $B$ to be the lightest states in their respective charged sectors makes them absolutely stable, while the symmetry reduces the set of amplitudes to the four given in~\eqref{eq:set_amplitudes} and forbids simple poles due to self-interactions.

We define the interacting part of the scattering amplitude by
\begin{equation}
	\label{eq:interacting_part_scattering_amplitudes}
	(2\pi)^4\delta^4(p_1+p_2-p_3-p_4)
	\times T_{ij\to kl}(s,t,u)
	\equiv
	\braket{kl|\mathbb T|ij} \,,
\end{equation}
where $\mathbb S=\id+i\mathbb T$ is the unitary scattering operator, $i$, $j$, $k$, $l$ are the particles with four-momenta $p_1$, $p_2$, $p_3$ and $p_4$.
All these momenta are on shell, for instance $p_1^2=-m_i^2$, where $m_i$ is the mass of particle $i$.
The Mandelstam variables $s$, $t$ and $u$ for the process $ij \to kl$ are defined as usual
\begin{equation}
	\label{eq:mandelstam_variables}
	s \equiv -(p_{1} + p_{2})^{2} \,, \qquad
	t \equiv -(p_{1} - p_{3})^{2} \,, \qquad
	u \equiv -(p_{1} - p_{4})^{2} \,,
\end{equation}
and satisfy the relation
\begin{equation}\label{eq:mandelstam_relation}
	s+t+u  = m_{i}^{2} + m_{j}^{2} + m_{k}^{2} + m_{l}^{2} \,.
\end{equation}
An important comment on the notation is in order.
In this paper we use the Mandelstam variables $(s,t,u)$ to describe the $s$-, $t$- and $u$-channels of \emph{each} scattering process.
For processes related by crossing symmetry, such as $AB\to AB$ and $AA\to BB$ related by $s\leftrightarrow t$, this implies that the variable $s$ used in $AB\to AB$ is not the same as $s$ in $AA\to BB$.
This notation has to be contrasted with the one often used in the literature for pion-kaon scattering (see e.g.~\cite{Pelaez:2020gnd}), where $s$ refers exclusively to the $s$-channel of $AB\to AB$, and hence the $s$-channel of $AA\to BB$ is denoted by $t$.

\subsection{Kinematics}\label{sec:kinematics}

We have formally defined the interacting parts of the scattering amplitudes in~\eqref{eq:interacting_part_scattering_amplitudes}. In this paper we focus only on the processes~\eqref{eq:amplitudes_main}, whose scattering amplitudes are related by crossing symmetry:
\begin{equation}\label{eq:crossing}
	T_{AB \to AB}(s,t,u) = T_{AB \to AB}(u,t,s) = T_{AA \to BB}(t,s,u) \,.
\end{equation}

In what follows we will carefully discuss the kinematics of the processes~\eqref{eq:amplitudes_main}. To write the equations below, we use the following notation
\begin{equation}
	\Sigma \equiv 2m^{2} + 2M^{2},
\end{equation}
together with
\begin{equation}
	m_\pm \equiv M\pm m\,.
\end{equation}
According to~\eqref{eq:mandelstam_relation} for both sets of Mandelstam variables we have
\begin{equation}\label{eq:mandelstam-constraint}
	s+t+u=\Sigma.
\end{equation}

\paragraph{Amplitude \texorpdfstring{\(AB\to AB\)}{AB to AB}.}
The Mandelstam variables in this channel are related to the particle kinematics as
\begin{equation}\label{eq:ABtoAB}
    \begin{cases}
      s = \left(\sqrt{m^{2}+\mom^{2}}+\sqrt{M^{2}+\mom^{2}}\right)^{2} \\
      t = - 2 \mom^{2}(1-z) \\
      u = \Sigma -s-t
	\end{cases}
\end{equation}
In the centre of mass (COM) frame the Mandelstam invariants can be parametrised in terms of the magnitude of the COM 3-momentum \( \mom \equiv |\vec p\,|\) and a scattering angle \( z = \cos \theta \).
This can be expressed in terms of the Mandelstam variables as
\begin{equation}\label{eq:PosABAB}
	z=\cos\theta
	=
	1+
	\frac{2st}
	{(s-m^2-M^2)^2-4m^2M^2}
	=
	1+
	\frac{2st}
	{(s-m_+^2)(s-m_-^2)} \,.
\end{equation}
Imposing \( \mom \geq 0 \) and \( z \in [-1, 1] \) in the previous formulas, we find that the physical domain of the Mandelstam variables is contained within the region
\begin{equation}\label{eq:physical}
	s \geq m_{+}^{2},\qquad t\leq 0, \qquad u \leq m_{-}^{2} \,.
\end{equation}
It is sometimes useful to express the variables $t$ and $u$ as functions of $s$ and $z$:
\begin{equation}\label{eq:t-of-s-z-ABtoAB}
    t(s,z) = - \frac{(s-m_{+}^{2})(s-m_{-}^{2})}{2s}(1-z) \,, \quad
    u(s,z) = \Sigma - s + \frac{(s-m_{+}^{2})(s-m_{-}^{2})}{2s}(1-z) \,.
\end{equation}
Applying~\eqref{eq:physical} to these expressions we immediately obtain the physical ranges of \( t \) and \( u \) at fixed \( s \),
\begin{equation}
    -\frac{(s-m_+^2)(s-m_-^2)}{s} \leq t \leq 0 \,, \qquad
    \Sigma-s \leq u	\leq \frac{{(m_{+}m_{-})}^{2}}{s} \,.
\end{equation}

\paragraph{Amplitude \texorpdfstring{\(AA\to BB\)}{AA to BB}.}
The relation between particle kinematics and Mandelstam variables is now given by
\begin{equation}\label{eq:AAtoBB}
    \begin{cases}
      s = 4(M^{2} + \momPrime^{2}) \\
      t = m^{2}-M^{2}-2 \momPrime^{2}+2\momPrime \sqrt{M^{2}-m^{2}+\momPrime^{2}} z \\
      u = \Sigma -s-t
	\end{cases}
\end{equation}
where \( \momPrime\) is the outgoing COM 3-momentum.
We recall that these Mandelstam variables are different from the ones in~\eqref{eq:ABtoAB}.
The scattering angle can be expressed in terms of the Mandelstam variables as
\begin{equation}
	\label{eq:PosAABB}
	z=\cos\theta
	=
	\frac{2t+s-\Sigma}
	{\sqrt{(s-4m^2)(s-4M^2)}}.
\end{equation}
For the process to be physical, both the incoming \(AA\) momentum and the outgoing \(BB\) momentum must be real.
Since \( M \geq m \), the physical domain for Mandelstam variables is contained within the region
\begin{equation}\label{eq:physical2}
	s \geq 4M^{2},\qquad t\leq 0, \qquad u \leq 0 \,.
\end{equation}
It is sometimes useful to express the variables $t$ and $u$ as functions of $s$ and $z$:
\begin{equation}\label{eq:t-of-s-z-AAtoBB}
    \begin{aligned}
      t(s,z)  &= \frac{\Sigma - s}{2} + \frac{1}{2}\sqrt{(s-4m^2)(s-4M^2)} \, z  \,, \\
      u(s,z) &= \frac{\Sigma - s}{2} - \frac{1}{2}\sqrt{(s-4m^2)(s-4M^2)} \, z \,.
    \end{aligned}
\end{equation}
Using~\eqref{eq:physical2} we obtain the physical ranges of \( t \) and \( u \) at fixed \( s \)
\begin{equation}
	B_-(s)\leq t\leq B_+(s) \,, \qquad
    B_{-}(s) \leq u \leq B_{+}(s) \,,
\end{equation}
where the boundary functions $B_\pm$ are
\begin{equation}\label{eq:region_t_AA_BB}
	B_\pm(s)
	\equiv
	\frac{\Sigma-s}{2}
	\pm
	\frac{1}{2}
	\sqrt{(s-4m^2)(s-4M^2)}.
\end{equation}
In particular, the physical domain in \( t \) is a convex band contained between \( t = -s \) and \( t = 0 \).

\subsection{Partial amplitudes}\label{sec:partial_amplitudes}

Consider the interacting part of the scattering amplitude~\eqref{eq:interacting_part_scattering_amplitudes} describing the process $ij \to kl$.
In this section we define the associated partial amplitudes.

Let us define the angular momentum projections as
\begin{equation}\label{eq:partial-amplitude-fl}
	f_{ij \to kl}^{\ell}(s) \equiv \int_{-1}^{1} \dif{z} \, P_{\ell}(z) T_{ij \to kl}(s,t(s,z),u(s,z)) \,.
\end{equation}
Here $\ell=0,1,2,3,\ldots$ is the angular momentum, \( P_{\ell}(z) \) are the standard Legendre polynomials and $z$ is the cosine of the scattering angle of the process $ij \to kl$.
The functions $t(s,z)$ and $u(s,z)$ depend on the kinematics of the specific process.
For $AB\to AB$ and $AA\to BB$ these are respectively given by~\eqref{eq:t-of-s-z-ABtoAB} and~\eqref{eq:t-of-s-z-AAtoBB}.
One can invert the relation~\eqref{eq:partial-amplitude-fl}, getting
\begin{equation}\label{eq:partial-wave-decomposition}
	T_{ij \to kl}(s,t,u) = \sum_{\ell=0}^{\infty} \frac{2\ell+1}{2} P_{\ell}(z(s,t)) \, f_{ij \to kl}^{\ell}(s) \,,
\end{equation}
which is valid a priori only in the physical domain of the Mandelstam variables.

The interacting part of the partial amplitude is defined as
\begin{equation}\label{eq:partial-amplitude-Tl}
	T_{ij \to kl}^{\ell}(s) = \mathcal{N}_{ij \to kl}(s) f_{ij \to kl}^{\ell}(s) \,.
\end{equation}
Here $\mathcal{N}_{ij \to kl}(s)$ is a simple kinematic factor.
In this paper we use the convention of~\cite{Hebbar:2020ukp} for the normalisation of the two-particle states.
With this convention we obtain
\begin{equation}\label{eq:partial-amplitude-normalization}
	\begin{aligned}
      s&\geq 4m^2 & \ \colon \quad
                    \mathcal{N}_{A A \to A A}(s) &= \frac{1}{32\pi}\sqrt{1-\frac{4m^{2}}{s}} \,, \\
      s&\geq 4M^2& \ \colon \quad
                   \mathcal{N}_{B B \to B B}(s) &= \frac{1}{32\pi}\sqrt{1-\frac{4M^{2}}{s}} \,, \\
      s&\geq m_+^2 & \ \colon \quad
                     \mathcal{N}_{AB \to AB}(s) &= \frac{1}{16\pi}\frac{\sqrt{(s-m_{+}^{2})(s-m_{-}^{2})}}{s} \,, \\
      s&\geq 4M^2 & \ \colon \quad
                    \mathcal{N}_{A A \to B B}(s) &= \frac{1}{32 \pi} \frac{(s-4m^{2})^{1/4}(s-4M^{2})^{1/4}}{\sqrt{s}} \,.
	\end{aligned}
\end{equation}
Using~\eqref{eq:partial-amplitude-Tl} we can also write the full partial amplitude as
\begin{equation}\label{eq:partial-amplitude-Sl}
	S_{ij \to kl}^{\ell}(s) = \delta_{(ij),(kl)} + i T_{ij \to kl}^{\ell}(s) \,.
\end{equation}
Here we have decided to present our results also for the processes $AA \to AA$ and $BB \to BB$.
This will be useful for the next section.

The angular momentum \( \ell \) in~\eqref{eq:partial-amplitude-fl} and~\eqref{eq:partial-amplitude-Tl} satisfies some selection rules.
By Bose symmetry, \( \ell \) is an even integer for the \( AA \to AA \), \( BB \to BB \) and \( AA \to BB \) processes, while it can be both even and odd for the \( AB \to AB \) process.

\subsection{Unitarity constraints}
\label{sec:unitarity}

Let us now write explicitly the unitarity constraints for our two-to-two scattering processes. For completeness, we present the constraints for all four processes in~\eqref{eq:set_amplitudes}. Unitarity couples the amplitudes $AA\to AA$, $BB\to BB$ and $AA\to BB$ in the neutral sector, while $AB\to AB$ belongs to a separate charged sector. It will then be straightforward to reduce these constraints to the two processes in~\eqref{eq:amplitudes_main}.
The unitarity constraints are most easily formulated in terms of the partial wave amplitudes defined in the previous subsection. In what follows, we use the definitions~\eqref{eq:partial-amplitude-Sl} and~\eqref{eq:partial-amplitude-Tl}.

Positivity of the Hilbert-space norm implies that the Gram matrix formed by the asymptotic in- and out-states, projected to a definite angular momentum, is positive semi-definite, as explained in~\cite{Hebbar:2020ukp}.

We now summarise the resulting unitarity constraints in the two charged sectors.

\begin{itemize}
    \item For \( 4m^{2} \leq s \leq 4M^{2} \) and \( \ell = 0, 2, 4, \ldots \) we have
    \begin{equation}\label{eq:unitarity-AAtoAA}
        \begin{pmatrix}
          1 & 1 -  i T_{AA \to AA}^{\ell}(s)^{*} \\
          1 + i T_{AA \to AA}^{\ell}(s) & 1
        \end{pmatrix} \succeq 0 \,.
    \end{equation}
    \item For \( s \geq 4M^{2} \) and \( \ell = 0, 2, 4, \ldots \) we have
    \begin{equation}\label{eq:unitarity-AB-mixed}
        \begin{pmatrix}
          1 & 0 & 1 -  i T_{AA \to AA}^{\ell}(s)^{*} & - i T_{AA \to BB}^{\ell}(s)^{*}\\
          0 & 1 & -i T_{AA \to BB}^{\ell}(s)^{*} &  1 -  i T_{BB \to BB}^{\ell}(s)^{*} \\
          1 + i T_{AA \to AA}^{\ell}(s) & i T_{AA \to BB}^{\ell}(s) & 1 & 0 \\
          i T_{AA \to BB}^{\ell}(s) & 1 + i T_{BB \to BB}^{\ell}(s) & 0 & 1
        \end{pmatrix} \succeq 0 \,.
    \end{equation}
    \item For $s \geq m_{+}^{2}$ and $\ell = 0, 1, 2,\ldots$ we have
    \begin{equation}\label{eq:unitarity-ABtoAB}
        \begin{pmatrix}
          1 & 1 -  i T_{AB \to AB}^{\ell}(s)^{*} \\
          1 + i T_{AB \to AB}^{\ell}(s) & 1
        \end{pmatrix}
        \succeq 0.
    \end{equation}
\end{itemize}

Neglecting the processes $AA\to AA$ and $BB\to BB$, we see from~\eqref{eq:unitarity-AAtoAA} and~\eqref{eq:unitarity-AB-mixed} that the following subset of unitarity constraints holds:
\begin{align}\label{eq:unitarity-AAtoBB}
  \begin{pmatrix}
    1 & - i T_{AA \to BB}^{\ell}(s)^{*} \\
    i T_{AA \to BB}^{\ell}(s) & 1
  \end{pmatrix} \succeq 0 \,, \qquad
  && s \geq 4M^{2} \,, \quad \text{and} \quad  \ell = 0, 2, 4, \ldots
\end{align}
The relations~\eqref{eq:unitarity-ABtoAB} and~\eqref{eq:unitarity-AAtoBB} are the unitarity constraints that we will use in practice in our numerical studies in section~\ref{sec:numerics}.

\section{Analytic structure of the amplitudes}\label{sec:analytic-structure}

To describe the analytic structure of the amplitudes we use a diagrammatic approach and study the singularities in various Feynman graph topologies~\cite{Landau:1959fi} compatible with the assumed spectrum and symmetries.
This approach is advocated and discussed in detail by Gribov~\cite{Gribov:2009cfk}.
For this purpose, one keeps only the scalar form of the propagators in the denominator of the expression for the diagrams and replaces numerator factors, spin structures and regular vertex functions by constants.
We call the resulting objects the reduced Feynman graphs.\footnote{%
    The same graph topologies can alternatively be generated by iterating analytically continued unitarity, see, for example,~\cite{Correia:2020xtr,Correia:2021etg}.
    In such ``unitarity graphs'', vertices represent exact scattering sub-amplitudes, while their internal lines represent on-shell intermediate states integrated over phase space.
    For the purpose of locating candidate singularities, the sub-amplitudes can be replaced by constants, reducing the unitarity graphs to the reduced Feynman graphs used in this paper.\label{foot:Naming_Landau_diagrams}
}
Their Landau equations~\cite{Landau:1959fi} determine candidate singularities in the external Mandelstam invariants.\footnote{%
    For recent works using Landau equations see~\cite{Collins:2020euz, Bourjaily:2020wvq, Hannesdottir:2021kpd, Mizera:2021icv, Hannesdottir:2022bmo, Mizera:2023tfe,Dlapa:2023cvx, Fevola:2023kaw, Fevola:2023fzn, Helmer:2024wax, Caron-Huot:2024brh}. For non-perturbative applications see~\cite{Correia:2021etg, Correia:2022dcu}.
}

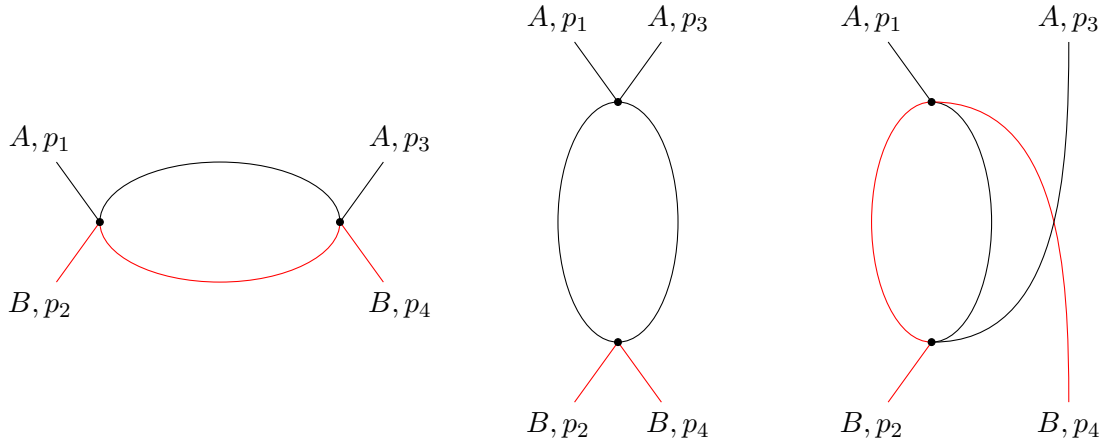
\begin{figure}[t!]
    \def\major{0.7}
    \def\minor{{\major/2}}
    \def\yshift{\minor}
    \def\xshift{\minor}
    \def\xshiftmore{0.8}

    \centering
    \begin{subfigure}[c]{0.3\linewidth}
        \centering
        \begin{tikzpicture}[x={0.5\linewidth}, y={0.5\linewidth}]
            \node[inner sep=0, outer sep=0] (A) at (-\major,0) {};
            \node[inner sep=0, outer sep=0] (B) at (\major,0) {};
            \node[above] (P1) at ($ (A) + (-\xshift,\yshift) $) {\( A, p_{1} \)};
            \node[below] (P2) at ($ (A) + (-\xshift,-\yshift) $) {\( B, p_{2} \)};
            \node[above] (P3) at ($ (B) + (\xshift,\yshift) $) {\( A, p_{3} \)};
            \node[below] (P4) at ($ (B) + (\xshift,-\yshift) $) {\( B, p_{4} \)};

            \draw[black] (B) arc[start angle=0, end angle=180, y radius=\minor, x radius=\major] -- (A);
            \draw[red] (B) arc[start angle=0, end angle=-180, y radius=\minor, x radius=\major] -- (A);
            \draw[black] (A) -- (P1);
            \draw[black] (B) -- (P3);
            \draw[red] (A) -- (P2);
            \draw[red] (B) -- (P4);
            \filldraw (A) circle (0.02);
            \filldraw (B) circle (0.02);
        \end{tikzpicture}
    \end{subfigure}
    \hspace{3em}
    \begin{subfigure}[c]{0.3\linewidth}
        \centering
        \begin{tikzpicture}[x={0.5\linewidth}, y={0.5\linewidth}]
            \node[inner sep=0, outer sep=0] (A) at (0,\major) {};
            \node[inner sep=0, outer sep=0] (B) at (0,-\major) {};
            \node[above] (P1) at ($ (A) + (-\xshift,\yshift) $) {\( A, p_{1} \)};
            \node[below] (P2) at ($ (B) + (-\xshift,-\yshift) $) {\( B, p_{2} \)};
            \node[above] (P3) at ($ (A) + (\xshift,\yshift) $) {\( A, p_{3} \)};
            \node[below] (P4) at ($ (B) + (\xshift,-\yshift) $) {\( B, p_{4} \)};

            \draw[black] (B) arc[start angle=-90, end angle=-270, y radius=\major, x radius=\minor] -- (A);
            \draw[black] (B) arc[start angle=-90, end angle=90, y radius=\major, x radius=\minor] -- (A);
            \draw[black] (A) -- (P1);
            \draw[black] (A) -- (P3);
            \draw[red] (B) -- (P2);
            \draw[red] (B) -- (P4);
            \filldraw (A) circle (0.02);
            \filldraw (B) circle (0.02);
        \end{tikzpicture}
    \end{subfigure}
    \begin{subfigure}[c]{0.3\linewidth}
        \centering
        \begin{tikzpicture}[x={0.5\linewidth}, y={0.5\linewidth}]
            \node[inner sep=0, outer sep=0] (A) at (0,\major) {};
            \node[inner sep=0, outer sep=0] (B) at (0,-\major) {};
            \node[above] (P1) at ($ (A) + (-\xshift,\yshift) $) {\( A, p_{1} \)};
            \node[below] (P2) at ($ (B) + (-\xshift,-\yshift) $) {\( B, p_{2} \)};
            \node[above] (P3) at ($ (A) + (\xshiftmore,\yshift) $) {\( A, p_{3} \)};
            \node[below] (P4) at ($ (B) + (\xshiftmore,-\yshift) $) {\( B, p_{4} \)};

            \draw[red] (B) arc[start angle=-90, end angle=-270, y radius=\major, x radius=\minor] -- (A);
            \draw[black] (B) arc[start angle=-90, end angle=90, y radius=\major, x radius=\minor] -- (A);
            \draw[black] (A) -- (P1);
            \draw[red] (B) -- (P2);
            \draw[red] (A) to[out=0,in=90] (P4);
            \draw[black] (B) to[out=0,in=270] (P3);
            \filldraw (A) circle (0.02);
            \filldraw (B) circle (0.02);
        \end{tikzpicture}
    \end{subfigure}
	\caption{Bubble diagrams associated with the leading two-particle singularities of the $AB\to AB$ amplitude. From left to right, the diagrams contain an $AB$ intermediate state in the $s$-channel, an $AA$ intermediate state in the $t$-channel and an $AB$ intermediate state in the $u$-channel. Black and red lines represent particles $A$ and $B$, respectively.\label{fig:thresholds}}
\end{figure}

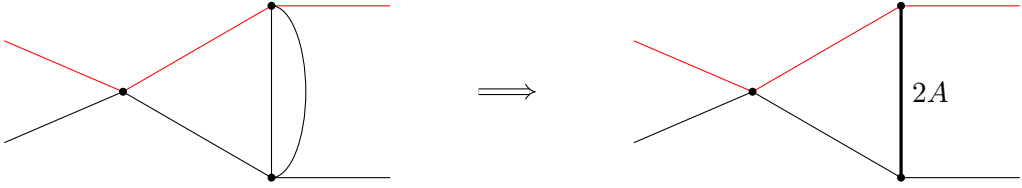
\begin{figure}[t!]
    \centering
    \hspace{-2em}
    \begin{subfigure}[c]{0.3\linewidth}
        \centering
        \begin{tikzpicture}[x={0.5\linewidth}, y={0.5\linewidth}]
            \node[inner sep=0, outer sep=0] (A) at (0,0) {};
            \node[inner sep=0, outer sep=0] (B) at (0.866,0.5) {};
            \node[inner sep=0, outer sep=0] (C) at (0.866,-0.5) {};
            \node[inner sep=0, outer sep=0] (P1) at ($ (A) + (-0.7,0.3) $) {};
            \node[inner sep=0, outer sep=0] (P2) at ($ (A) + (-0.7,-0.3) $) {};
            \node[inner sep=0, outer sep=0] (P4) at ($ (B) + (0.7,0) $) {};
            \node[inner sep=0, outer sep=0] (P3) at ($ (C) + (0.7,0) $) {};
            \draw[red] (P1) -- (A);
            \draw[red] (A) -- (B);
            \draw[red] (B) -- (P4);
            \draw (P2) -- (A);
            \draw (A) -- (C);
            \draw (B) -- (C);
            \draw (C) -- (P3);
            \draw (C) arc[start angle=-90, end angle=90, x radius = 0.2, y radius=0.5] -- (B);
            \filldraw (A) circle (0.02);
            \filldraw (B) circle (0.02);
            \filldraw (C) circle (0.02);
        \end{tikzpicture}
    \end{subfigure}
    \hspace{1em}
    \begin{subfigure}[c]{0.2\linewidth}
        \centering
        \begin{tikzpicture}[x={0.5\linewidth}, y={0.5\linewidth}]
            \draw[-implies,double equal sign distance] (-0.25,0) -- (0.25,0);
        \end{tikzpicture}
    \end{subfigure}
    \begin{subfigure}[c]{0.3\linewidth}
        \centering
        \begin{tikzpicture}[x={0.5\linewidth}, y={0.5\linewidth}]
            \node[inner sep=0, outer sep=0] (A) at (0,0) {};
            \node[inner sep=0, outer sep=0] (B) at (0.866,0.5) {};
            \node[inner sep=0, outer sep=0] (C) at (0.866,-0.5) {};
            \node[inner sep=0, outer sep=0] (P1) at ($ (A) + (-0.7,0.3) $) {};
            \node[inner sep=0, outer sep=0] (P2) at ($ (A) + (-0.7,-0.3) $) {};
            \node[inner sep=0, outer sep=0] (P4) at ($ (B) + (0.7,0) $) {};
            \node[inner sep=0, outer sep=0] (P3) at ($ (C) + (0.7,0) $) {};
            \draw[red] (P1) -- (A);
            \draw[red] (A) -- (B);
            \draw[red] (B) -- (P4);
            \draw (P2) -- (A);
            \draw (A) -- (C);
            \draw[very thick] (B) -- (C) node[midway, right] {\( 2 A \)};
            \draw (C) -- (P3);
            \filldraw (A) circle (0.02);
            \filldraw (B) circle (0.02);
            \filldraw (C) circle (0.02);
        \end{tikzpicture}
    \end{subfigure}
	\caption{A triangular diagram containing four internal lines (left) and its reduced triangular form (right).\label{fig:TriangularL2}}
\end{figure}

We apply this machinery to the amplitude $AB\to AB$. We begin with the simplest reduced Feynman graphs: the two-particle bubble graphs shown in figure~\ref{fig:thresholds}. They describe the leading thresholds in the $s$-, $t$- and $u$-channels.
The existence of possible anomalous thresholds is then addressed by analysing the triangular diagrams in figure~\ref{fig:TriangularL2}.
Further discussion of the double discontinuities arising from box diagrams is left to appendix~\ref{app:PWE}.

\paragraph{Physical and pseudo-physical thresholds.}

For the first diagram in figure~\ref{fig:thresholds}, the internal particles $A$ and $B$ can simultaneously go on shell at $s=m_+^2$.
This is the leading physical threshold in the $s$-channel.
The corresponding branch cut extends along $s\geq m_+^2$.

The second diagram contains two internal particles $A$ in the $t$-channel.
Its Landau equations give a branch point at $t=4m^2$, and the corresponding cut begins at $t\geq4m^2$.
There is also another $t$-channel bubble diagram, not shown in figure~\ref{fig:thresholds}, whose two internal lines are particles $B$.
It gives an additional branch point at $t=4M^2$. We refer to the interval
\begin{equation}\label{eq:pseudo_thresholds_AB_to_AB}
	4m^2\leq t<4M^2
\end{equation}
as the \emph{pseudo-physical region} and to the point \( 4m^{2} \) as the \emph{pseudo-physical threshold}.
In the special case $M=m$, the pseudo-physical region collapses to a point and effectively disappears.

By crossing symmetry, the analytic structure of the $AB\to AB$ amplitude in the $t$-channel coincides with that of the $AA\to BB$ amplitude in the $s$-channel.
The pseudo-physical region of the latter process is therefore $4m^2\leq s<4M^2$.
The most important feature of the corresponding cut is that its discontinuity is not directly constrained by unitarity.
Indeed, the unitarity condition~\eqref{eq:unitarity-AAtoBB} for the $AA\to BB$ process applies only in the physical region $s\geq4M^2$.

An analogous discussion applies to the amplitudes $AA\to AA$ and $BB\to BB$.
The former has only physical thresholds, while the latter has both a physical and a pseudo-physical threshold.
We summarise the relevant two-particle thresholds in table~\ref{tab:thresholds-summary}.

\begin{table}[t]
	\centering
	\begin{tabular}{c|c|c|c}
      \toprule
      amplitude
      &
        physical $s$-region
      &
        two-particle branch points
      &
        pseudo-physical region
      \\
      \midrule
      $AA\to AA$
      &
        $s\geq4m^2$
      &
        $4m^2,\ 4M^2$
      &
        none
      \\
      $BB\to BB$
      &
        $s\geq4M^2$
      &
        $4m^2,\ 4M^2$
      &
        $4m^2\leq s<4M^2$
      \\
      $AA\to BB$
      &
        $s\geq4M^2$
      &
        $4m^2,\ 4M^2$
      &
        $4m^2\leq s<4M^2$
      \\
      $AB\to AB$
      &
        $s\geq(m+M)^2$
      &
        $(m+M)^2$
      &
        none
      \\
      \bottomrule
    \end{tabular}
	\caption{Summary of the leading two-particle branch points and pseudo-physical regions for the two-to-two scattering amplitudes of particles $A$ and $B$.}
	\label{tab:thresholds-summary}
\end{table}

\paragraph{Anomalous thresholds.}
An anomalous threshold is a Landau singularity that does not coincide with an ordinary physical or pseudo-physical threshold.
The simplest reduced Feynman graph topology that can generate such a singularity is a triangle~\cite{Karplus:1958zz,Landau:1959fi}.

In our setup, the $\mathbb Z_2\times\mathbb Z_2$ symmetry forbids all cubic vertices involving the single-particle states $A$ and $B$.
Triangular diagrams whose internal lines are all single-particle propagators are therefore absent.
An effective triangle can nevertheless arise from a diagram with a multi-particle sub-graph.
A simple example is shown in the left panel of figure~\ref{fig:TriangularL2}.

The two parallel black lines in this diagram form an $AA$ bubble.
At the $AA$ threshold, the two particles have the same four-momentum.
At the level of the Landau equations, the bubble is then equivalent to a single effective internal line of mass $2m$.
Replacing the $AA$ bubble by this effective line gives the reduced triangular diagram shown in the right panel of figure~\ref{fig:TriangularL2}.

To analyse this possibility, consider a general effective triangle:
\begin{center}
    \vspace{0.5em}
    \begin{tikzpicture}[x={0.2\linewidth}, y={0.2\linewidth}]
        \node[inner sep=0, outer sep=0] (A) at (0,0) {};
        \node[inner sep=0, outer sep=0] (B) at (0.866,0.5) {};
        \node[inner sep=0, outer sep=0] (C) at (0.866,-0.5) {};
        \node[inner sep=0, outer sep=0] (P1) at ($ (A) + (-0.7,0.3) $) {};
        \node[inner sep=0, outer sep=0] (P2) at ($ (A) + (-0.7,-0.3) $) {};
        \node[inner sep=0, outer sep=0] (P4) at ($ (B) + (0.7,0) $) {};
        \node[inner sep=0, outer sep=0] (P3) at ($ (C) + (0.7,0) $) {};
        \draw (P1) -- (A);
        \draw (P2) -- (A);
        \draw (P3) -- (C) node[midway, above] {\( \mathrm{E}_{2} \)};
        \draw (P4) -- (B) node[midway, below] {\( \mathrm{E}_{1} \)};
        \draw[very thick] (A) -- (B) node[midway, above] {\( \mathrm{I}_{2} \)};
        \draw[very thick] (A) -- (C) node[midway, below] {\( \mathrm{I}_{1} \)};
        \draw[very thick] (B) -- (C) node[midway, right] {\( \mathrm{I}_{3} \)};
        \filldraw (A) circle (0.02);
        \filldraw (B) circle (0.02);
        \filldraw (C) circle (0.02);
    \end{tikzpicture}
\end{center}
The external lines $\mathrm E_1$ and $\mathrm E_2$ represent particles $A$ or $B$.
The internal lines $\mathrm I_1$, $\mathrm I_2$ and $\mathrm I_3$ may represent either single-particle or multi-particle states with the appropriate $\mathbb Z_2\times\mathbb Z_2$ charges.
We denote by $m_{\mathrm I_a}$ the corresponding threshold mass.

A necessary, but not sufficient, condition for this diagram to generate an anomalous threshold on the principal sheet is
\begin{equation}\label{eq:mEmIConditions}
	m_{\mathrm E_1}^2
	>
	m_{\mathrm I_2}^2+m_{\mathrm I_3}^2
	\qquad\text{or}\qquad
	m_{\mathrm E_2}^2
	>
	m_{\mathrm I_1}^2+m_{\mathrm I_3}^2.
\end{equation}
Neither condition can be satisfied in our setup.
Indeed, charge conservation at each vertex fixes the possible charged sectors of the adjacent internal states.
For an external particle $A$, the lightest allowed pairs are $A$ together with an $AA$ state, or $B$ together with an $AB$ state.
For an external particle $B$, the lightest allowed pairs are $B$ together with an $AA$ state, or $A$ together with an $AB$ state.
In every case, the sum of the squared internal masses is strictly larger than the squared mass of the corresponding external particle.
Thus, no anomalous threshold can arise from an elementary or effective triangular Feynman graph of the form considered above.

More complicated reduced Feynman graphs could in principle generate other anomalous thresholds.
A complete classification of all such topologies is beyond the scope of this paper.
We therefore assume that no anomalous thresholds arise from more complicated reduced Feynman graphs.

\begin{figure}[t!]
	\centering
	\tikzset{cross/.style={cross out, draw=black, minimum size=6, inner sep=0pt,
			outer sep=0pt}, cross/.default={1pt}}
	\tikzset{snake it/.style={decorate,
			decoration={snake, segment length=2mm, amplitude=0.5mm}},
		line width=0.5pt}
	\tikzset{fontscale/.style = {font=\relsize{#1}}}

	% Parameters
	\def\mA{0.3}
	\def\mB{0.4}
	\def\mAsq{\mA*\mA}
	\def\mBsq{\mB*\mB}
	\def\mApmBsq{(\mA+\mB)*(\mA+\mB)}
	\def\mAmBsq{(\mA-\mB)*(\mA-\mB)}
	\def\tval{0.5}
	\def\tinyshift{-0.01}  % arrow start, just below cross
	\def\yshift{-0.15}     % arrow end / label height
	\def\radius{0.01}
	\def\scale{0.1}

	% \polecross{x position}{x displacement of label}{label text}
	\newcommand{\polecross}[3]{
		\draw ({#1}, 0) node[cross, color=red] {};
		\draw[->, color=gray] ({#1}, \tinyshift)
		to [out=-90, in=90] ({#1 + #2}, \yshift)
		node[below, font=\footnotesize, inner sep=0pt] {#3};
	}

	% Axes macro
	\newcommand{\drawaxes}{
		\draw[-{Latex}, line width=0.5pt] (-1,0) -- (1,0);
		\draw[-{Latex}, line width=0.5pt] (0,-0.3) -- (0,0.3);
	}

	\newcommand{\rightcut}[2]{
		\filldraw ({#1},0) circle ({\radius});
		\draw[draw=blue, snake it] ({#1},0) node[above, fontscale={\scale}] {#2} -- (0.97,0);
	}

	\newcommand{\leftcut}[2]{
		\filldraw ({#1},0) circle ({\radius});
		\draw[draw=blue, snake it] ({#1},0) node[above, fontscale={\scale}] {#2} -- (-1,0);
	}

	\newcommand{\pseudocut}[3]{
		\filldraw ({#1},0) circle ({\radius});
		\draw[draw=red, snake it] ({#1},0) node[above, fontscale=1] {#3} -- ({#2},0);
	}

	\begin{subfigure}[t]{0.8\linewidth}
		\centering
		\begin{tikzpicture}[x={0.45\linewidth}, y={0.45\linewidth}]
			\drawaxes

			\rightcut{\mApmBsq}{$m_{+}^{2}$}
			\leftcut{\mAmBsq-\tval}{$m_{-}^{2}{-}t$}
		\end{tikzpicture}
		\caption{\( s \)-plane\label{fig:ABtoAB-analytic-s}}
	\end{subfigure}

	\vspace{2em}

	% -------------------------------------------------------
	% AA -> BB
	% -------------------------------------------------------
	\begin{subfigure}[t]{0.8\linewidth}
		\centering
		\begin{tikzpicture}[x={0.45\linewidth}, y={0.45\linewidth}]
			\drawaxes

			\pseudocut{4*\mAsq}{4*\mBsq}{$4m^{2}$}
			\rightcut{4*\mBsq}{$4M^{2}$}
			\leftcut{\mAmBsq-\tval}{$m_{-}^{2}{-}s$}
		\end{tikzpicture}
		\caption{\( t \)-plane\label{fig:ABtoAB-analytic-t}}
	\end{subfigure}

	\caption{Analytic structure of the \( AB \to AB \) amplitude in the complex \( s \)-plane at fixed \( t \) and in the complex \( t \)-plane at fixed \( s \). Blue lines denote physical thresholds, while the red interval in (b) denotes the pseudo-physical region.}
	\label{fig:analytic-structure}
\end{figure}
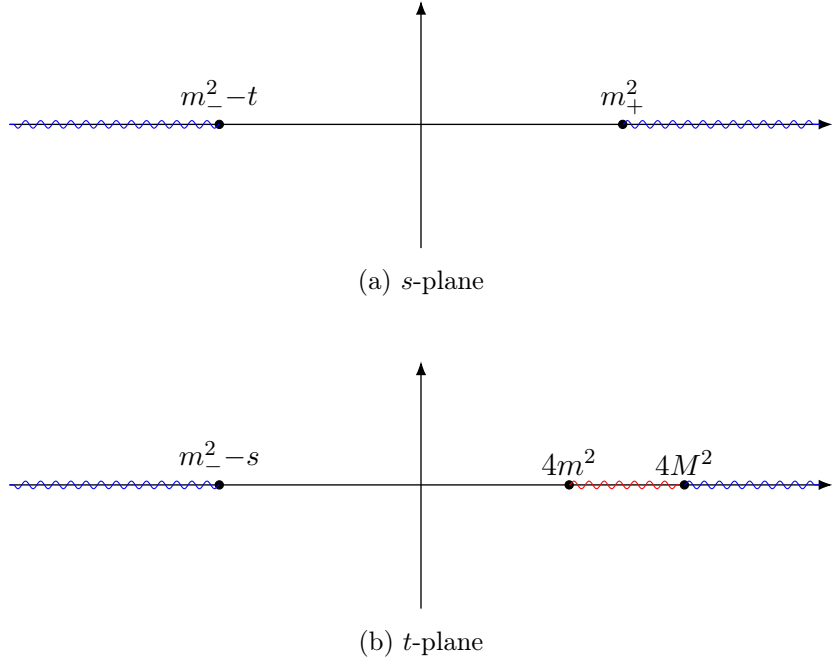

\paragraph{Assumed analytic structure.}
The resulting principal-sheet analytic structure assumed for the amplitude $AB\to AB$ is summarised in figure~\ref{fig:analytic-structure}.
This is the main conclusion of the present section.
It provides the analytic input for the numerical bootstrap of section~\ref{sec:numerics}.

\section{Observables and the space of amplitudes}\label{sec:observables}

In this short section we define the observables associated with the scattering amplitude $T_{AB\to AB}(s,t,u)$.
These observables are dimensionless quantities constructed directly from the amplitude.
A complete set of them can be viewed as coordinates on the infinite-dimensional space of scattering amplitudes: each amplitude corresponds to a point in this space, specified by the values of all its observables.
In practice, we consider only a finite subset of observables.
In the next section we construct bounds on various subsets of observables.

We consider two classes of observables.
The first class consists of couplings, defined by derivatives of the amplitude at a fixed kinematic point.
The second class consists of threshold parameters, called scattering lengths and effective ranges.

\paragraph{Couplings.}
Non-perturbative couplings are defined as derivatives of the amplitude at a fixed kinematic configuration. We define
\begin{equation}\label{eq:def-couplings}
    \lambda_{k,l} \equiv \frac{m^{2k+2l}}{k!l!} \frac{\partial^{k}}{\partial s^{k}} \frac{\partial^{l}}{\partial t^{l}} T_{AB \to AB}(s_{0},t_{0}, u_{0}) \,,
\end{equation}
with
\begin{equation}\label{eq:s0-t0-def}
    s_{0} = u_0\equiv \frac{m^{2}}{3} + M^{2} \,, \qquad
    t_{0} \equiv \Sigma - s_{0}-u_0 = \frac{4}{3} m^{2} \,.
\end{equation}
This choice is such that the couplings are real (thanks to real-analyticity), they respect the \( s-u \) crossing symmetry of the amplitude and they coincide with the couplings defined in~\cite{Chen:2022nym} when \( m = M \).
This definition is also motivated by the dispersion relations discussed in appendix~\ref{app:positivity}.
The mass-dependent prefactor is chosen so that the couplings are dimensionless.
Not all the couplings in~\eqref{eq:def-couplings} are independent or non-trivial because of \( s-u \) crossing.
For example we have
\begin{equation}\label{eq:couplings-crossing-redundancies}
    \lambda_{2k+1,0} = 0 \,, \; (k \geq 0) \,, \quad
    \lambda_{1,1} = \lambda_{2,0} \,, \quad
    \lambda_{1,2} = \lambda_{2,1} \,, \quad
    \lambda_{3,1} = 2 \lambda_{4,0} \,, \quad
    \lambda_{1,3} = \lambda_{2,2} - \lambda_{4,0} \,.
\end{equation}

We will be in particular interested in the first three couplings, namely
\begin{equation}
    \lambda_{0,0} \,, \qquad \lambda_{2,0} \,, \qquad \lambda_{2,1} \,. \label{threelambda}
\end{equation}
The first one, \( \lambda_{0,0} \), is just the value of the amplitude at the point \( (s_{0}, t_{0}, u_{0}) \).
The remaining two admit a dispersive representation\footnote{%
    If a stronger bound than the Froissart bound is assumed, then \( \lambda_{0,0} \) also admits a dispersive representation.
}
\begin{equation}\label{eq:L20-L21-dispersive-rep}
    \begin{aligned}
      \lambda_{2,0} &= \frac{m^{4}}{\pi}\int_{m_{+}^{2}}^{\infty} \dif{s} \, \frac{2 \Im T(s, t_{0})}{(s-s_{0})^{3}} \,, \\
      \lambda_{2,1} &= \frac{m^{6}}{\pi}\int_{m_{+}^{2}}^{\infty} \dif{s} \left[\frac{2 \, \partial_{t}\Im T(s, t_{0})}{(s-s_{0})^{3}}-\frac{3 \Im T(s, t_{0})}{(s-s_{0})^{4}}\right] \,, \\
    \end{aligned}
\end{equation}
where we have defined
\begin{equation}\label{eq:Tst-def}
    T(s, t) \equiv T_{AB \to AB}(s, t, \Sigma-s-t) \,,
\end{equation}
to lighten the notation.
Finally, the dispersion relations and positivity of the \( AB \to AB \) amplitude imply the following bounds
\begin{equation}\label{eq:positivity-L20-L21}
    \lambda_{2,0} \geq 0 \,, \qquad \lambda_{2,1} \geq -\frac{9}{4(1+3 \mu)} \lambda_{2,0} \,.
\end{equation}
Linearised unitarity, \( 0 \leq \Im T_{AB \to AB}^{\ell}(s) \leq 2 \), can be used to derive an analytic lower bound
\begin{equation}\label{eq:linearised-unitarity-L21}
    \lambda_{2,1} \geq h(\mu) =
    \begin{cases}
      -3.226 & \mu = 1 \\
      -1.339 & \mu = 1.5 \\
      -0.202 & \mu = 3.554
    \end{cases}
    \,,
\end{equation}
with \( h(\mu) \) given in~\eqref{eq:L21-lower-bound-integral}; its values at the mass ratios~\eqref{eq:mass-ratios-used} are reported here for future reference.
Combining linearised unitarity with numerical optimisation, we obtain a bound slightly stronger than the combination of the two analytic bounds~\eqref{eq:positivity-L20-L21} and~\eqref{eq:linearised-unitarity-L21}.
This is given by~\eqref{eq:linearised-unitarity-bound} and is shown in figure~\ref{fig:linearised-unitarity-bound} for the \( \mu = 1.5 \) case.
This bound  will be compared with the bounds we obtain from non-linear unitarity with the primal \( S \)-matrix bootstrap approach.

We refer to appendix~\ref{app:positivity} for the derivation of these results and further details.

\paragraph{Threshold parameters.}
Scattering lengths and effective ranges are natural observables that can be extracted from the expansion of the partial amplitudes at the physical threshold.
For the \( AB \to AB \) process we define the threshold parameters \( a_{\ell} \) and \( b_{\ell} \) as
\begin{equation}\label{eq:threshold-expansion}
    \Re T_{AB \to AB}^{\ell}(s) = 2 (\mom/m)^{2\ell+1}\Big(\!a_{\ell} + b_{\ell} (\mom/m)^{2} + O\big((\mom/m)^{4}\big)\!\Big) \,,
\end{equation}
where \( \mom \) is the modulus of the COM 3-momentum, related to \( s \) as in~\eqref{eq:ABtoAB}.
The observable \( a_{\ell} \) is sometimes referred to directly as the ``scattering length'' expressed in units of $m$.

We recall that in our conventions the unitary \( S \)-matrix is \( S_{AB \to AB}^{\ell}(s) = 1 + i T_{AB \to AB}^{\ell}(s) \), i.e.\ \( \left|S_{AB \to AB}^{\ell}(s)\right|^{2} \leq 1 \).
Defining phase shifts as
\begin{equation}
	\label{eq:phase_shift_AB_to_AB}
    S_{AB \to AB}^{\ell}(s) = e^{2 i \delta^{\ell}(s)} \,,
\end{equation}
we have the following commonly written threshold expansion for the scalar phase shift
\begin{equation}
	\label{eq:scattering_lengths}
    (\mom/m) \cot \delta^{0}(s) = \frac{1}{a_{0}} + O((\mom/m)^{2}) \,.
\end{equation}

For scattering of identical particles, the scattering lengths \( a_{\ell} \) with \( \ell \geq 2 \) admit a dispersive representation which can be obtained by expanding at threshold the Froissart-Gribov formula.
This implies that \( a_{\ell} \geq 0 \) for \( \ell \geq 2 \), see for example section 2.5 in~\cite{Correia:2020xtr}.
For unequal particles, dispersive representations exist for \( \ell \geq 2 \) using again the Froissart-Gribov formula, but no simple bounds can be derived because of the absence of sign-definiteness.

Still, in the case of \( \pi K \to \pi K \) and \( \pi N \to \pi N \) scattering, relative analytic bounds between scattering lengths on fixed isospin channels have been obtained~\cite{Yndurain:1972ix}.
Scalar scattering lengths are in general unbounded from above because of the possible presence of resonances close to the physical region, see~\cite{Paulos:2017fhb} for a discussion.
Several analytic lower bounds on the scalar scattering length have been obtained for equal particles~\cite{Yndurain:1972ix},\footnote{%
    Numerical lower bounds using the primal \( S \)-matrix bootstrap have been obtained in~\cite{Paulos:2017fhb}. See also~\cite{Guerrieri:2018uew} for bounds on scattering lengths in pion scattering.
} but we are not aware of generalisations to unequal particles.

\section{Numerical results}\label{sec:numerics}
In this section we investigate non-perturbative bounds on quantities that can be extracted from the \( AB \to AB \) scattering amplitude.
The bounds apply to any massive quantum field theory in \( 4d \) that contains the particles \( A \) and \( B \) in its spectrum and satisfies the assumptions stated above.

\subsection{Primal \texorpdfstring{\( S \)}{S}-matrix bootstrap}\label{sec:numerics-setup}
We obtain numerical bounds using the primal \( S \)-matrix bootstrap introduced in~\cite{Paulos:2017fhb}.
As a first step we parametrise the most general scattering amplitude for the \( AB \to AB \) process that is compatible with the analytic structure described in section~\ref{sec:analytic-structure}.
We define the \( \rho \)-variable
\begin{equation}\label{eq:rho-variable-generic}
    \rho(z, z\ped{th}, z_{0}) = \frac{\sqrt{z\ped{th}-z_{0}} - \sqrt{z\ped{th} - z}}{\sqrt{z\ped{th}-z_{0}} + \sqrt{z\ped{th} - z}} \,.
\end{equation}
This function maps the cut complex plane \( \mathbb{C} \setminus [z\ped{th}, +\infty) \) to the unit disk, sending the cut to the boundary of the disk, and has the further property that \( \rho(z_{0}, z\ped{th}, z_{0}) = 0 \).
In order to accommodate the branch cut structure in figure~\ref{fig:analytic-structure} we define
\begin{equation}\label{eq:rho-variables}
    \rho_{+}(z) = \rho(z, m_{+}^{2}, s_{0}) \,, \quad
    \rho_{m}(z) = \rho(z, 4 m^{2}, t_{0}) \,.
\end{equation}
For \( s_{0} \) and \( t_{0} \) we make the same choice as in~\eqref{eq:s0-t0-def}.

We can represent the interacting part of the scattering amplitude for the \( AB \to AB \) process as a power series expansion in the variables~\eqref{eq:rho-variables} as follows
\begin{equation}\label{eq:ABtoAB-ansatz}
    T_{AB \to AB}(s,t,u) = \sum_{a=0}^{\infty}\sum_{b=0}^{\infty} \sum_{c=0}^{\infty} \alpha_{a,b,c} (\rho_{+}(s))^{a}(\rho_{m}(t))^{b} (\rho_{+}(u))^{c} \,,
\end{equation}
where \( \alpha_{a,b,c} \) are real parameters.
The ansatz~\eqref{eq:ABtoAB-ansatz} contains several redundancies because of the relation \( s + t + u  = \Sigma \). In this work we reduce the redundancies by further imposing\footnote{%
    This condition is slightly stronger than what follows from~\eqref{eq:mandelstam-constraint}, see appendix C of~\cite{Paulos:2017fhb}, but the bounds are insensitive to the difference.
}
\begin{equation}
    \alpha_{a,b,c} = 0 \,  \qquad  \text{if} \ a b c \neq 0 \,.
\end{equation}
The crossing relations~\eqref{eq:crossing} constrain the ansatz.
The \( s \leftrightarrow u \) relation imposes a further constraint on the coefficients of the ansatz~\eqref{eq:ABtoAB-ansatz}, which have to satisfy
\begin{equation}\label{eq:alpha-sym}
    \alpha_{a,b,c} = \alpha_{c,b,a} \,.
\end{equation}
The \( s \leftrightarrow t \) relation implies that \( T_{AA \to BB} \) is not an independent amplitude and can be written as
\begin{equation}\label{eq:AAtoBB-ansatz}
    T_{AA \to BB}(s,t,u) = \sum_{a,b,c} \alpha_{a,b,c} (\rho_{+}(t))^{a}(\rho_{m}(s))^{b} (\rho_{+}(u))^{c} \,.
\end{equation}
The ansatz~\eqref{eq:AAtoBB-ansatz} is \( t-u \) symmetric thanks to~\eqref{eq:alpha-sym}.

Note that we have chosen the variable \( \rho_{m} \) so that the amplitude can accommodate a non-analyticity in the pseudo-physical region discussed in section~\ref{sec:analytic-structure}.
One could also include a \( \rho \)-variable with a cut at \( z\ped{th} = 4M^{2} \) to capture an additional discontinuity at the physical threshold.
However, this would further increase the size of the ansatz and would probably be redundant with the current ansatz in the absence of additional inputs.

In the case where we relax the symmetries imposed on the \( A \) and \( B \) particles to allow for 3-point interactions, or if we allow for additional bound states, the ansatz has to be modified to include simple poles.
This possibility, relevant for pion-nucleon scattering but not for pion-kaon scattering, will not be considered in this work, and is left for future exploration.

The ansatz~\eqref{eq:ABtoAB-ansatz} has an infinite number of terms.
In order to obtain numerical bounds, it is crucial to perform a truncation such that
\begin{equation}
    a+b+c \leq N\ped{max} \,,
\end{equation}
where \( N\ped{max} \) is a cut-off parameter, and explore convergence of the bounds as \( N\ped{max} \to +\infty \).
In practice, the computational cost increases rapidly with the cut-off and in this paper we will work with \( N\ped{max} \leq 26 \).

Once we have truncated the sums we can collect the coefficients \( \alpha_{a,b,c} \) in a finite-dimensional vector \( \vec{\alpha} \), and write the ansatz~\eqref{eq:ABtoAB-ansatz} and~\eqref{eq:AAtoBB-ansatz} as
\begin{equation}\label{eq:truncated-ansatz}
    T_{AB \to AB}(s,t,u) = \vec{\alpha} \cdot \vec{X}(s,t,u) \,, \qquad
    T_{AA \to BB}(s,t,u) = \vec{\alpha} \cdot \vec{X}(t,s,u) \,,
\end{equation}
where
\begin{equation}
    X_{a,b,c}(s,t,u) = (\rho_{+}(s))^{a}(\rho_{m}(t))^{b} (\rho_{+}(u))^{c} \,.
\end{equation}

The observables introduced in section~\ref{sec:observables} can be extracted from~\eqref{eq:truncated-ansatz} and are linear in the coefficients \( \alpha_{a,b,c} \).
The numerical strategy is to maximise or minimise such observables, subject to the unitarity constraints discussed in section~\ref{sec:unitarity}, which can be cast as a semi-definite optimisation problem (SDP).
To impose the unitarity constraint on the ansatz we compute the partial wave projection
\begin{equation}
    T_{AB \to AB}^{\ell}(s) =  \vec{\alpha} \cdot  \vec{X}_{AB \to AB}^{\ell}(s) \,, \qquad
    T_{AA \to BB}^{\ell}(s) =  \vec{\alpha} \cdot  \vec{X}_{AA \to BB}^{\ell}(s)
\end{equation}
where
\begin{equation}\label{eq:ansatz-partial-wave-projection}
    \begin{aligned}
      \vec{X}_{AB \to AB}^{\ell}(s) &= \mathcal{N}_{AB \to AB}(s) \int_{-1}^{1} \dif{z} \, P_{\ell}(z) \vec{X}(s, t(s, z), u(s, z)) \,, \\
      \vec{X}_{AA \to BB}^{\ell}(s) &= \mathcal{N}_{AA \to BB}(s) \int_{-1}^{1} \dif{z} \, P_{\ell}(z) \vec{X}(t(s, z), s, u(s,z)) \,.
    \end{aligned}
\end{equation}
See equations~\eqref{eq:partial-amplitude-fl},~\eqref{eq:partial-amplitude-Tl},~\eqref{eq:partial-amplitude-normalization} for the definitions of the partial wave projection, as well as~\eqref{eq:t-of-s-z-ABtoAB} and~\eqref{eq:t-of-s-z-AAtoBB} for the expression of \( t(s,z) \) and \( u(s,z) \) in the appropriate channel.

Unitarity of the partial amplitudes, equations~\eqref{eq:unitarity-ABtoAB} and~\eqref{eq:unitarity-AAtoBB}, can then be formulated as positive semi-definite conditions on the following matrices, linear in the coefficients \( \alpha \):
\begin{equation}\label{eq:sdp-unitarity}
    \begin{aligned}
      & \begin{pmatrix}
        1 & 1 \\
        1 & 1
      \end{pmatrix}
        + \vec{\alpha} \cdot
        \begin{pmatrix}
          0 & - i \vec{X}_{AB \to AB}^{\ell}(s)^{*} \\
          i \vec{X}_{AB \to AB}^{\ell}(s) & 0
        \end{pmatrix}
        \succeq 0 \,,
      & \quad & s \geq m_{+}^{2} \,, && \ell = 0, 1, 2, \ldots \,, \\[1em]
      & \begin{pmatrix}
        1 & 0 \\
        0 & 1
      \end{pmatrix}
        + \vec{\alpha} \cdot
        \begin{pmatrix}
          0 & - i \vec{X}_{AA \to BB}^{\ell}(s)^{*} \\
          i \vec{X}_{AA \to BB}^{\ell}(s) & 0
        \end{pmatrix}
        \succeq 0 \,,
      & \quad &s \geq 4 M^{2} \,, && \ell = 0, 2, 4, \ldots \,, \\
    \end{aligned}
\end{equation}
where \( {}^{*} \) denotes complex conjugation.
In addition, two further truncations are needed.
We put a cut-off on the spin \( \ell \leq L\ped{max} \) and sample the physical regions in \( s \) at \( n\ped{pts} \) points.
See appendix~\ref{app:numerics-details} for further details.
Again, convergence of the bounds in \( L\ped{max} \) and \( n\ped{pts} \) has to be carefully determined.

Finally, a bound on an observable \( \vec{\alpha} \cdot \vec{b} \) linear in the \( \alpha \)'s, at finite \( (N\ped{max}, L\ped{max}, n\ped{pts}) \), can be obtained from the following SDP:
\begin{equation}\label{eq:standard-SDP}
    \begin{aligned}
      & \max{\, \vec{\alpha} \cdot \vec{b}} \\
      & \text{such that} \ M_{0,j} + \vec{\alpha} \cdot \vec{M}_{j} \succeq 0 \,,
    \end{aligned}
\end{equation}
where \(  \vec{M}_{j} \) and \(  M_{0,j} \) are \( 2 \times 2 \) matrices.
More concretely, \( j \) is a multi-index labelling all the positivity conditions in~\eqref{eq:sdp-unitarity}: for \( j = (AB \to AB, \ell, s) \), \( M_{0,j} \) denotes the \( \left(\begin{smallmatrix} 1 & 1 \\ 1 & 1 \end{smallmatrix}\right) \) matrix and \(  \vec{M}_{j} \) the vector of matrices constructed out of \( \vec{X}^{\ell}_{AB \to AB}(s) \); for \( j = (AA \to BB, \ell, s) \), \( M_{0,j} \) is the identity matrix and \(  \vec{M}_{j} \) the vector of matrices constructed out of \( \vec{X}^{\ell}_{AA \to BB}(s) \).
In this work we use \texttt{SDPB}~\cite{Simmons-Duffin:2015qma,Landry:2019qug} to solve these optimisation problems, see appendix~\ref{app:numerics-details} for further details.

\begin{figure}[t]
    \centering
    \includegraphics[width=0.48\linewidth]{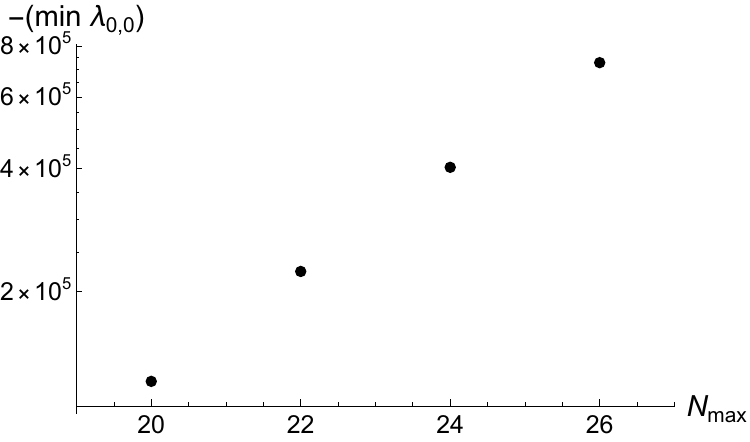}
    \includegraphics[width=0.48\linewidth]{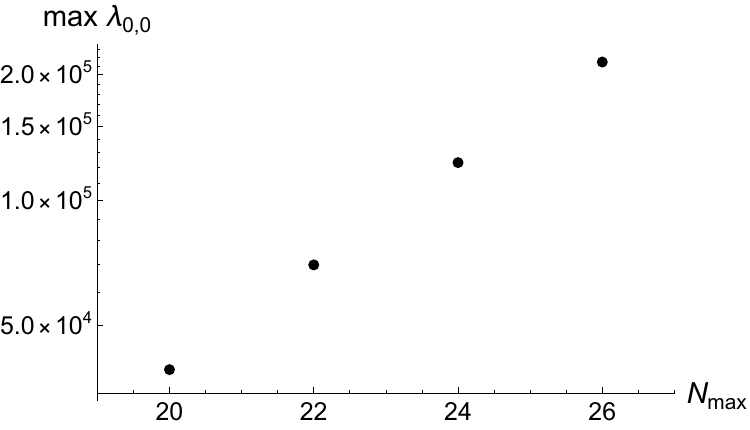}
    \caption{Minimisation and maximisation of \( \lambda_{0,0} \) as a function of \( N\ped{max} \) for the \( \mu = 1.5 \) case. The bound is obtained at \( L\ped{max} = 60 \) and \( n\ped{pts} = 200 \). The vertical axis is in logarithmic scale.\label{fig:L00-bound-nmax}}
\end{figure}
\begin{figure}[t]
    \centering
    \begin{subfigure}[c]{0.48\linewidth}
        \centering
        \includegraphics[width=\linewidth]{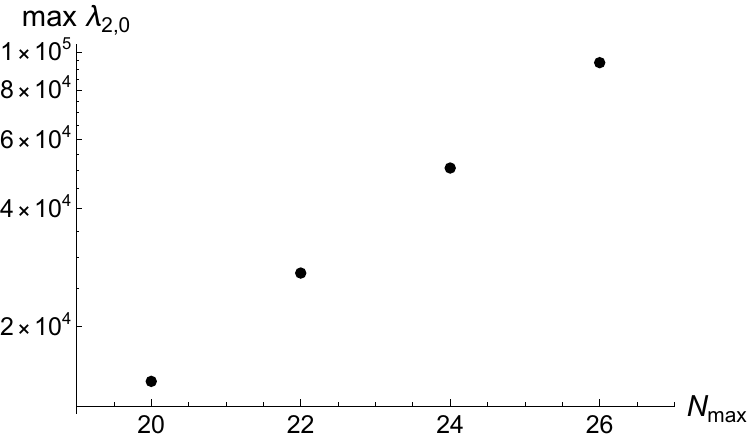}
        \caption{\( \lambda_{2,0} \) maximisation\label{fig:L20-max}}
    \end{subfigure}
    \begin{subfigure}[c]{0.48\linewidth}
        \centering
        \includegraphics[width=\linewidth]{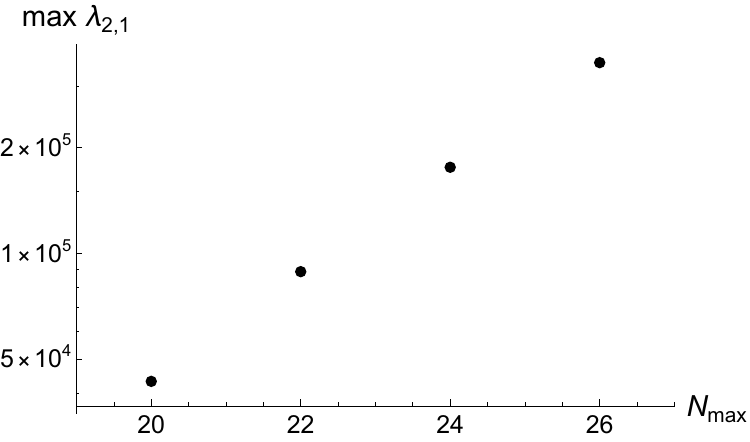}
        \caption{\( \lambda_{2,1} \) maximisation\label{fig:L21-max}}
    \end{subfigure}
    \caption{Maximisation of \( \lambda_{2,0} \) and \( \lambda_{2,1} \) as a function of \( N\ped{max} \) for the \( \mu = 1.5 \) case. The bound is obtained at \( L\ped{max} = 60 \) and \( n\ped{pts} = 200 \). The vertical axis is in logarithmic scale.\label{fig:L20-and-L21-max}}
\end{figure}

\subsection{Bounds on couplings}\label{sec:numerics-couplings}

In this section we report the bounds on the non-perturbative couplings \( \lambda_{k,l} \) defined in section~\ref{sec:observables}.
We have to make an important distinction between the special case of equal masses and the generic case of unequal masses.
The former is postponed to section~\ref{sec:numerics-couplings-equal-mass}.
Here we discuss the generic case of mass ratio \( \mu > 1 \).
In practice, due to the numerical complexity of the problem, we have decided to study two representative cases
\begin{equation}\label{eq:mass-ratios-numerics}
    \mu = 1.5, 3.554 \,.
\end{equation}
We believe that the patterns observed in these cases will hold for any \( \mu > 1 \), and indeed we have obtained partial confirmation of this also for other values of \( \mu \), but large-scale numerical optimisations have been carried out only in the aforementioned cases.
The value \( \mu = 3.554 \) might seem arbitrary at first sight, but it corresponds to the mass ratio of pions and kaons in the isospin limit\footnote{%
    More precisely, we used \( m_{\pi} = m_{\pi^{\pm}} = 139.57 \, \mathrm{MeV} \) and \( m_{K} = (m_{K^{0}} + m_{K^{+}})/2 = 496 \, \mathrm{MeV} \)~\cite{Pelaez:2020gnd}, which gives \( \mu \approx 3.554 \).
}, as we aim to investigate \( S \)-matrix bootstrap bounds on \( \pi K \to \pi K \) scattering in the future.

Using the definition of the non-perturbative couplings~\eqref{eq:def-couplings} in the ansatz~\eqref{eq:ABtoAB-ansatz}, we can express the \( \lambda_{k,l} \) as finite linear combinations of the \( \alpha_{a,b,c} \).
These are finite since we have chosen the zero of the \( \rho \)-variables to coincide with the point \( (s_{0}, t_{0}, u_{0}) \) at which the \( \lambda_{k,l} \) are defined. For example we have
\begin{equation}
    \resizebox{\linewidth}{!}{\(
        \begin{aligned}
          \lambda_{0,0} &= \alpha_{0,0,0} \,, \\
          \lambda_{2,0} &= \frac{9}{16(1+3\mu)^{2}}\left(\alpha_{0,0,1}+\frac{1}{2}\alpha_{0,0,2}-\frac{1}{4}\alpha_{1,0,1}\right) \,, \\
          \lambda_{2,1} &= \frac{27}{128(1+3\mu)^{3}}\left(\frac{1}{4} (1+3 \mu) \alpha _{0,1,1}+\frac{1}{8} (1+3 \mu) \alpha _{0,1,2}-\frac{15}{4} \alpha _{0,0,1}-3 \alpha _{0,0,2}-\frac{3}{4} \alpha _{0,0,3}+\frac{1}{2} \alpha
                          _{1,0,1}+\frac{1}{4} \alpha _{1,0,2}\right) \,.
        \end{aligned}
        \)}
\end{equation}
As anticipated in the introduction, in contrast to the case of scattering of the lightest particle, we find several unbounded directions in the space of non-perturbative couplings when imposing unitarity of the \( AB \to AB \) and \( AA \to BB \) amplitudes.
We believe this is mostly due to the absence of unitarity constraints on the pseudo-physical region where the ansatz~\eqref{eq:ABtoAB-ansatz} has an explicit discontinuity.\footnote{%
    At finite \( N\ped{max} \) the semi-definite problem is always bounded.
    As we increase the size of the ansatz, the ``freedom'' in the pseudo-physical region can make a coupling larger and larger, so that the bound as a function of \( N\ped{max} \) shows no sign of convergence.
} Perhaps the most dramatic effect is that the value of the amplitude at the symmetric point, \( \lambda_{0,0} \), is unbounded.
This is illustrated in figure~\ref{fig:L00-bound-nmax}: both the upper and lower bounds grow exponentially in \( N\ped{max} \).

As a next step, we consider the couplings \( \lambda_{2,0} \) and \( \lambda_{2,1} \).
These are likely to be on a better footing since they can be bounded analytically using dispersion relations in the \( AB \to AB \) channel and linearised unitarity.
See section~\ref{sec:observables} and equations~\eqref{eq:positivity-L20-L21} and~\eqref{eq:linearised-unitarity-L21}.

\begin{figure}[t]
    \centering
    \vspace{0.5em}
    \begin{subfigure}[c]{0.48\linewidth}
        \centering
        \includegraphics[width=\linewidth]{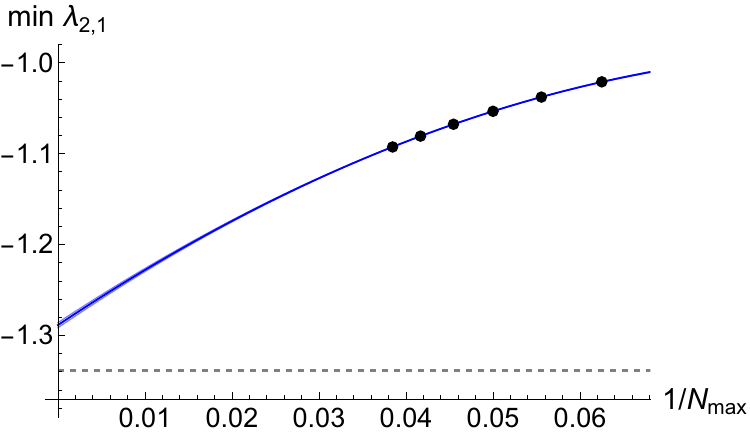}
        \caption{\( \mu = 1.5 \)}
    \end{subfigure}
    \begin{subfigure}[c]{0.48\linewidth}
        \centering
        \includegraphics[width=\linewidth]{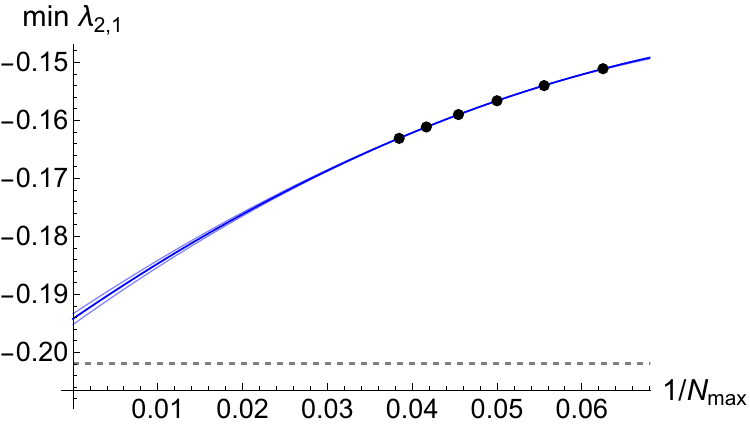}
        \caption{\( \mu = 3.554 \)}
    \end{subfigure}
    \caption{Minimisation of \( \lambda_{2,1} \) as a function of \( 1/N\ped{max} \) (black dots) for mass ratios \( \mu = 1.5 \) and \( \mu = 3.554 \). The blue line is a fit with a model \( a + b/N\ped{max} + c/N\ped{max}^{2} \). The error band denotes the estimate of uncertainty based on a leave-one-out procedure. The dashed grey line is the analytic bound~\eqref{eq:linearised-unitarity-L21} from linearised unitarity.\label{fig:L21-bound-fit}}
\end{figure}
\begin{figure}[t]
    \centering
    \begin{subfigure}[c]{\linewidth}
        \centering
        \includegraphics[width=0.63\linewidth]{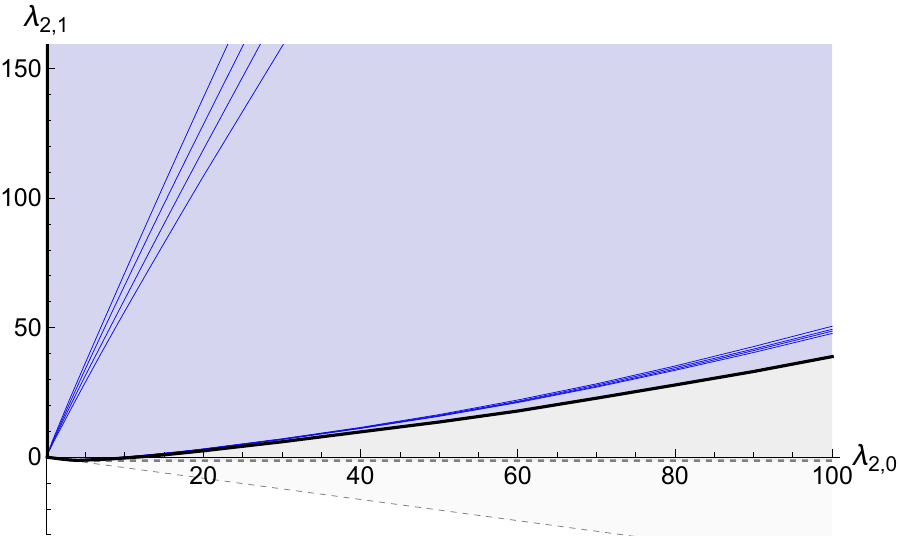}
        \caption{\( \mu = 1.5 \)\label{fig:L21-L20-bound-mu-1.5}}
    \end{subfigure}
    \begin{subfigure}[c]{\linewidth}
        \centering
        \includegraphics[width=0.63\linewidth]{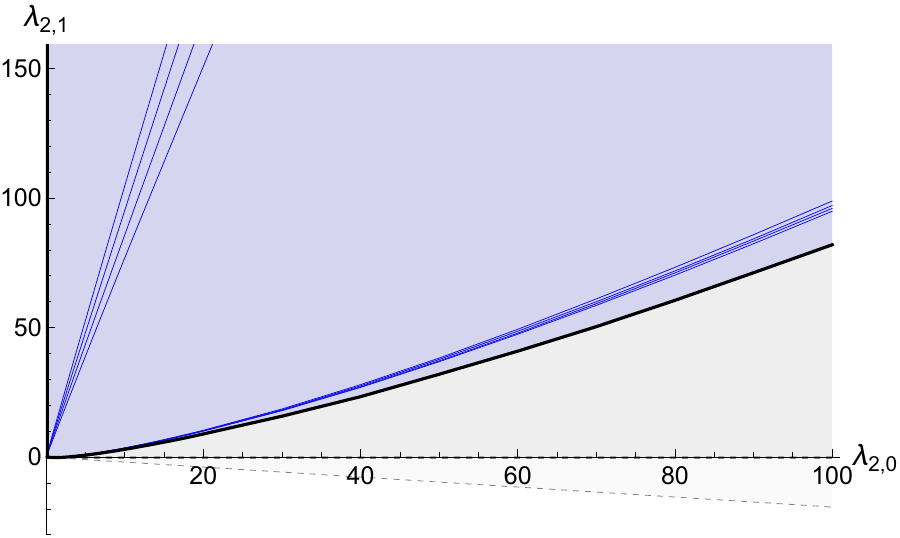}
        \caption{\( \mu = 3.554 \)\label{fig:L21-L20-bound-mu-PiK}}
    \end{subfigure}
    \caption{Minimisation and maximisation of \( \lambda_{2,1} \) at fixed \( \lambda_{2,0} \) for \( \mu = 1.5 \) and \( \mu = 3.554 \). The blue lines denote the boundary of the ``allowed'' primal region, of increasing size corresponding to increasing values of \( N\ped{max} = 20, 22, 24, 26 \). The light grey area denotes the positivity bound~\eqref{eq:positivity-L20-L21}, while the darker grey area is the linearised unitarity bound~\eqref{eq:linearised-unitarity-bound}. The black line enclosing the blue area denotes the \( N\ped{max} \to \infty \) extrapolation of the bound, which converges to a finite value only for the lower bound. The upper bound at fixed \( \lambda_{2,0} \) is observed to grow linearly in \( N\ped{max} \). All bounds use \( n\ped{pts} = 200 \) and \( L\ped{max} = 60 \), except for the \( \mu = 3.554 \), \( N\ped{max} = 26 \) case, for which \( L\ped{max} = 70 \).}
    \label{fig:L21-L20-bound-tot}
\end{figure}

For \( \lambda_{2,0} \) we obtain an absolute lower bound compatible with zero, up to the precision we used, for any \( N\ped{max} \).
This is consistent with~\eqref{eq:positivity-L20-L21} and the fact that in the free theory \( \lambda_{2,0} = 0 \).
Unfortunately, we do not find any absolute upper bound: at finite \( N\ped{max} \) it grows exponentially, as shown in figure~\ref{fig:L20-max}.

For \( \lambda_{2,1} \) we are able to obtain a non-trivial absolute lower bound.
Performing an extrapolation in \( 1/N\ped{max} \) with a quadratic model we obtain our best estimate
\begin{equation}\label{eq:min-L21-extrapolation}
    \begin{aligned}
      \mu  &= 1.5 \quad &\colon \ \quad \lambda_{2,1} &\geq -1.289(3) \,, \\
      \mu  &= 3.554 \quad &\colon \ \quad \lambda_{2,1} &\geq -0.194(1) \,.
    \end{aligned}
\end{equation}
The error is a non-rigorous estimate based on the leave-one-out uncertainty.
We report in figure~\ref{fig:L21-bound-fit} the lower bound at finite \( N\ped{max} = 16, 18, 20, 22, 24, 26 \), plotted as a function of \( 1/N\ped{max} \) for both mass ratios, together with a quadratic fit.
The lower bounds~\eqref{eq:min-L21-extrapolation} from non-linear unitarity are only a few percent away from the analytic lower bound from linearised unitarity in~\eqref{eq:linearised-unitarity-L21}, providing an important consistency check of our setup and of the \( N\ped{max} \) extrapolation.\footnote{%
    Following the derivation of~\eqref{eq:linearised-unitarity-L21} in appendix~\ref{app:positivity}, the analytic lower bound is realised by setting \( \Im T_{AB \to AB}^{\ell}(s) = 2 \delta_{\ell,0} \) for \( s \geq m_{+}^{2} \). Imposing an elastic amplitude, this implies \( S^{\ell}_{AB \to AB}(s) = 1-2 \delta_{\ell,0} \). Upon closer inspection, the extremal amplitudes for the \( \min \lambda_{2,1} \) problem at finite \( N\ped{max} \) approximate this behaviour at high energies.
}
Figure~\ref{fig:L21-max} shows instead how the absolute upper bound grows exponentially in \( N\ped{max} \).

\begin{figure}[t]
    \centering
    \begin{subfigure}[c]{0.48\linewidth}
        \includegraphics[width=\linewidth]{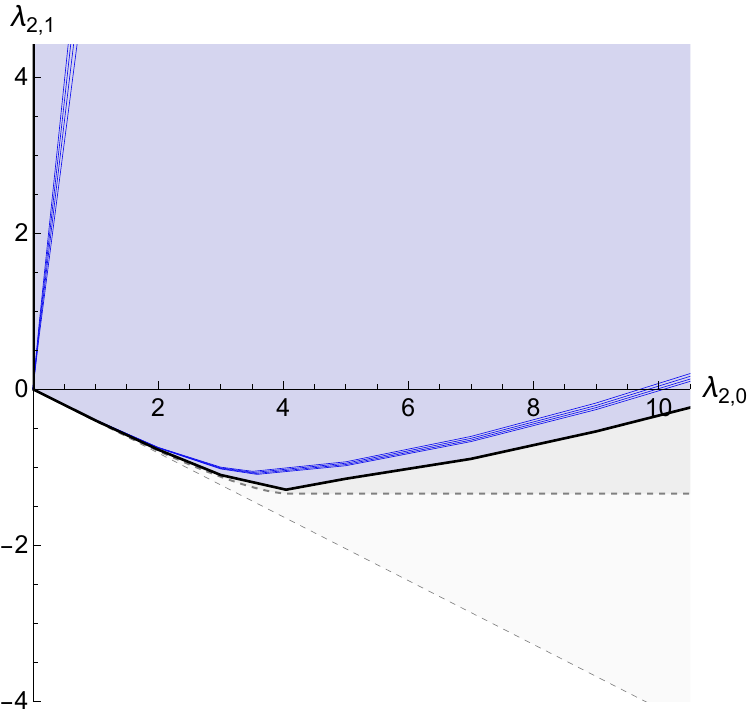}
        \caption{\( \mu = 1.5 \)}
    \end{subfigure}
    \begin{subfigure}[c]{0.48\linewidth}
        \includegraphics[width=\linewidth]{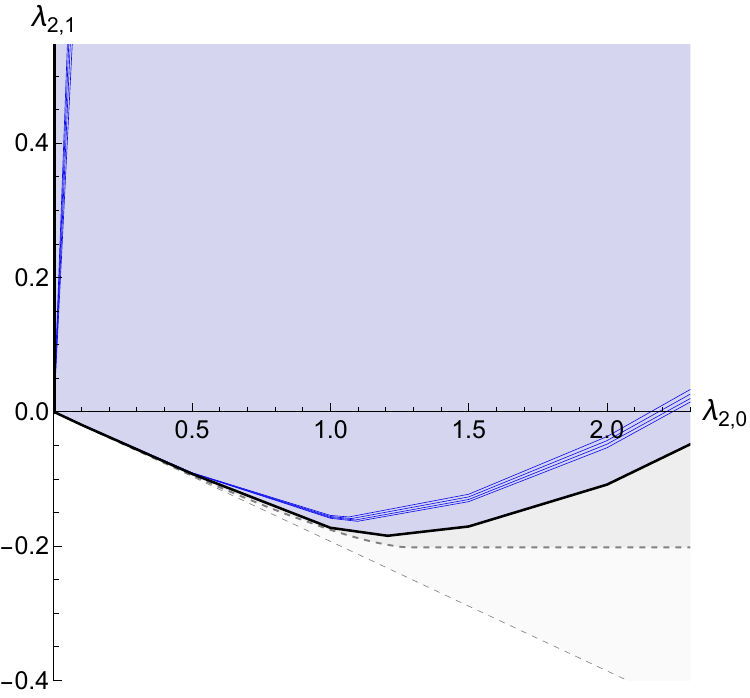}
        \caption{\( \mu = 3.554 \)}
    \end{subfigure}
    \caption{Zoom on small values of \( \lambda_{2,0} \) in figures~\ref{fig:L21-L20-bound-mu-1.5} and~\ref{fig:L21-L20-bound-mu-PiK}. The bound agrees with and asymptotes to the positivity bound~\eqref{eq:positivity-L20-L21} as \( \lambda_{2,0} \to 0 \) and almost touches the linearised unitarity bound~\eqref{eq:linearised-unitarity-bound}.\label{fig:L21-L20-bound-zoom}}
\end{figure}

We can also carve out the allowed region in the plane $(\lambda_{2,1},\lambda_{2,0})$ by fixing a value of \( \lambda_{2,0} \) and minimising/maximising \( \lambda_{2,1} \).
This again can be cast into an SDP of the form~\eqref{eq:standard-SDP}.
The result is shown in figure~\ref{fig:L21-L20-bound-mu-1.5} and in figure~\ref{fig:L21-L20-bound-mu-PiK} for \( \mu = 1.5 \) and \( \mu = 3.554 \), respectively, together with the bounds obtained by positivity~\eqref{eq:positivity-L20-L21} and linearised unitarity~\eqref{eq:linearised-unitarity-bound}.
We observe convergence for the lower bound, while the upper bound appears to be growing linearly with \( N\ped{max} \), at least for the range of values we have explored.
The bound extrapolated at \( N\ped{max} \to \infty \) shows a significant improvement compared to the positivity and linearised unitarity bounds, especially for large enough \( \lambda_{2,0} \).
For small \( \lambda_{2,0} \) our bound approaches the positivity bound,\footnote{This is expected since non-linear unitarity reduces to positivity in the perturbative limit.} and almost touches the bound from linearised unitarity, see figure~\ref{fig:L21-L20-bound-zoom}.
This is, again, an important consistency check of the numerical implementation and the extrapolation.

The bound can of course also be read as upper and lower bounds on \( \lambda_{2,0} \) at fixed \( \lambda_{2,1} \); for \( \lambda_{2,1} \geq 0 \) the lower bound reduces to \( \lambda_{2,0} \geq 0 \) but we still have a non-trivial upper bound, as we can see in figure~\ref{fig:L21-L20-bound-tot}; for \( (\min \lambda_{2,1}) \leq \lambda_{2,1} \leq 0 \), we obtain both non-trivial upper and lower bounds, as shown more clearly in figure~\ref{fig:L21-L20-bound-zoom}.

\subsection{Bounds on scalar scattering length}
In this section we explore bounds on the threshold parameters~\eqref{eq:threshold-expansion}, for small \( \ell \), focusing in particular on \( \ell = 0 \).
We consider the generic case \( \mu > 1 \) and we report bounds for the representative values of mass ratios in~\eqref{eq:mass-ratios-numerics}.

The threshold parameters can again be expressed in terms of linear combinations of the \( \alpha_{a,b,c} \) parameters.
Performing an expansion at the physical threshold of the \( \rho \)-variables~\eqref{eq:rho-variables} in the ansatz~\eqref{eq:ABtoAB-ansatz}, in the COM 3-momentum \( \mom \) at fixed scattering angle \( z = \cos \theta \), we find
\begin{equation}\label{eq:ABtoAB-ansatz-threshold-expansion}
    T_{AB \to AB}(s,t,u) = \sum_{m=0}^{+\infty}\sum_{n=0}^{\lfloor m/2 \rfloor} A_{m,n} \mom^{m} z^{n} \,,
\end{equation}
where, for example,
\begin{equation}
    A_{0,0} = \sum_{a,b,c} \alpha_{a,b,c} (2\sqrt{6}-5)^{b}\left(\frac{1+9 \mu -2 \sqrt{6} \sqrt{\mu  (3 \mu +1)}}{1-3 \mu }\right)^{c} \,.
\end{equation}
Note that, by construction, the \( A_{m,n} \) with \( m \) even are real, while the \( A_{m,n} \) with \( m \) odd are imaginary.
Performing the partial wave projection of the expansion~\eqref{eq:ABtoAB-ansatz-threshold-expansion}, we obtain the relation between the coefficients \( A_{m,n} \) and the threshold parameters. We find
\begin{equation}
    a_{\ell} = \gamma_{\ell}A_{2\ell,\ell} \,, \qquad
    b_{\ell} = \gamma_{\ell}\left(A_{2\ell+2,\ell}-\frac{1}{2\mu} A_{2\ell,\ell}\right) \,,
\end{equation}
with
\begin{equation}
    \gamma_{\ell} = \frac{\ell!}{(2\ell+1)!!} \frac{1}{8\pi(1+\mu)} \,.
\end{equation}
In particular the scalar scattering length is
\begin{equation}
    a_{0} = \frac{A_{0,0}}{8\pi(1+\mu)} \,.
\end{equation}
\begin{figure}[t!]
    \centering
    \begin{subfigure}[c]{\linewidth}
        \centering
        \includegraphics[width=0.55\linewidth]{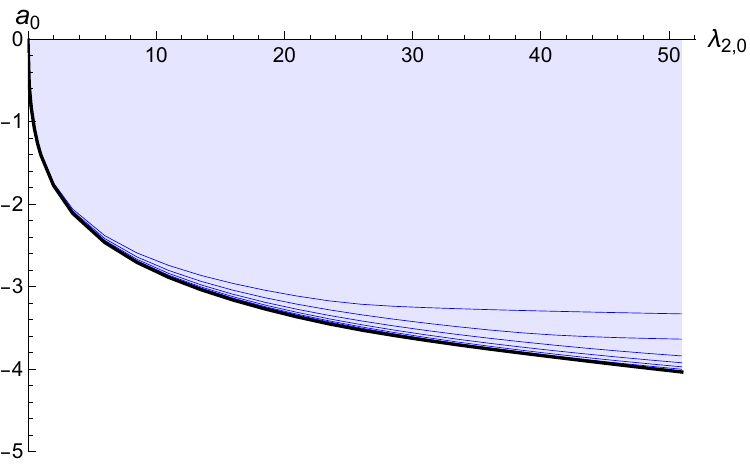}
        \caption{\( \mu = 1.5 \)\label{fig:a0-L20-bound-mu-1.5}}
    \end{subfigure}
    \begin{subfigure}[c]{\linewidth}
        \centering
        \includegraphics[width=0.55\linewidth]{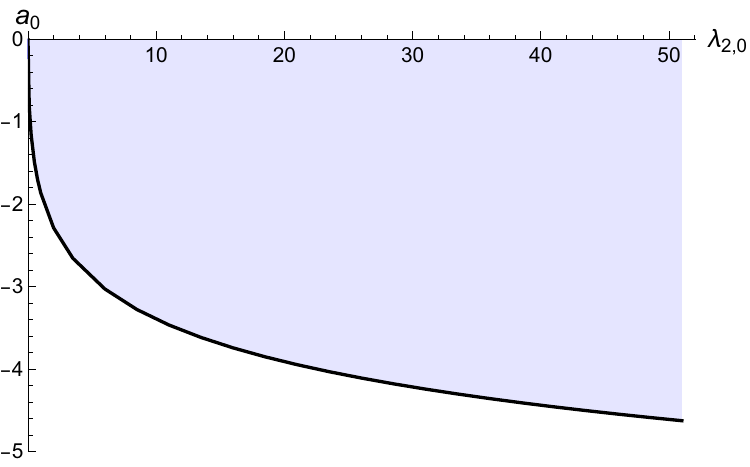}
        \caption{\( \mu = 3.554 \)\label{fig:a0-L20-bound-mu-PiK}}
    \end{subfigure}
    \caption{Minimisation of \( a_{0} \) at fixed \( \lambda_{2,0} \) for \( \mu = 1.5 \) and \( \mu = 3.554 \). The blue lines denote the boundary of the ``allowed'' primal region, of increasing size corresponding to increasing values of \( N\ped{max} = 16, 18, 20, 22, 24, 26 \) for \( \mu = 1.5 \) and \( N\ped{max} = 20, 22, 24, 26 \) for \( \mu = 3.554 \) (the bounds essentially overlap on this scale). The black line enclosing the blue area denotes the \( N\ped{max} \to \infty \) extrapolation of the lower bound. All bounds use \( n\ped{pts} = 200 \) and \( L\ped{max} = 60 \), except for the \( \mu = 3.554 \), \( N\ped{max} = 26 \) case, for which \( L\ped{max} = 70 \).\label{fig:a0-L20-bound}}
\end{figure}

In contrast to the couplings \( \lambda_{k,l} \), the threshold expansion involves all terms of the ansatz~\eqref{eq:ABtoAB-ansatz} and thus depends on the \( N\ped{max} \) truncation.

Starting with the scalar scattering length \( a_{0} \), we have not found evidence of convergent upper and lower bounds.
The absolute lower bound appears to be growing linearly in \( N\ped{max} \), while the absolute upper bound diverges exponentially in \( N\ped{max} \).
The absence of an upper bound is expected and was already observed in~\cite{Paulos:2017fhb} for the case of identical particles.
The absence of a lower bound is an unwelcome feature, similar to what happens for some couplings \( \lambda_{k,l} \).

We have also explored absolute upper and lower bounds on \( a_{1} \), \( a_{2} \) and \( b_{0} \) and found similar results.
None of the bounds appear to be convergent in \( N\ped{max} \).
Finally, we have obtained bounds on \( a_{0} \) for a fixed value of \( \lambda_{2,0} \).
In this case, perhaps surprisingly, we obtain a lower bound that converges exponentially in \( N\ped{max} \).
This is reported in figure~\ref{fig:a0-L20-bound-mu-1.5} and figure~\ref{fig:a0-L20-bound-mu-PiK} for \( \mu = 1.5 \) and \( \mu = 3.554 \), respectively, together with an \( N\ped{max} \to \infty \) extrapolation with a model \( a + b \, \exp(-c N\ped{max}) \).\footnote{%
    We find \( c \approx 0.3 \) for \( \mu = 1.5 \) and \( c \approx 0.7 \) for \( \mu = 3.554 \).
}

\subsection{Bounds on couplings for equal masses}\label{sec:numerics-couplings-equal-mass}

In this section we report bounds on the non-perturbative couplings \( \lambda_{k,l} \) for the case where the particles \( A \) and \( B \) have the same mass, namely for \( \mu = 1 \).
As discussed in section~\ref{sec:analytic-structure}, this special case is on a much better footing since the unitarity constraints bound all discontinuities of the ansatz.
Indeed we find that now all the couplings \( \lambda_{0,0} \), \( \lambda_{2,0} \) and \( \lambda_{2,1} \) admit upper and lower bounds that converge as \( N\ped{max} \to \infty \), analogously to what happens for the scattering of identical particles~\cite{Chen:2022nym}.

\begin{figure}[t!]
    \centering
    \begin{subfigure}[b]{0.47\linewidth}
        \centering
        \includegraphics[width=\linewidth]{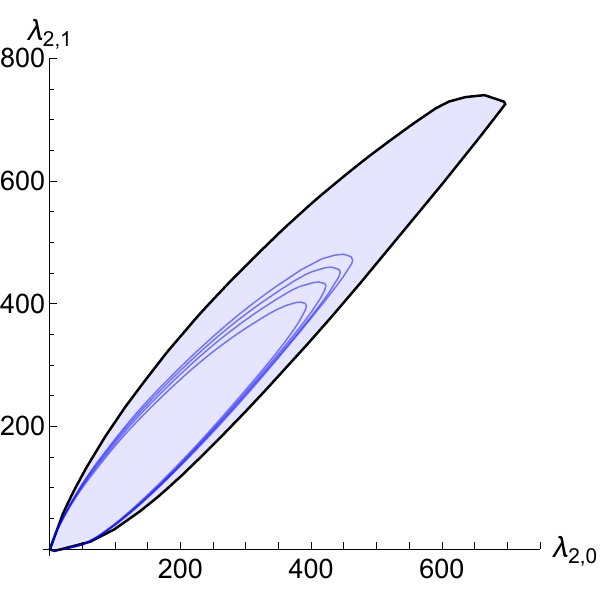}
        \caption{Bound on \( (\lambda_{2,0}, \lambda_{2,1}) \)}
    \end{subfigure}
    \hspace{0.5em}
    \begin{subfigure}[b]{0.47\linewidth}
        \centering
        \raisebox{5pt}{\includegraphics[width=\linewidth]{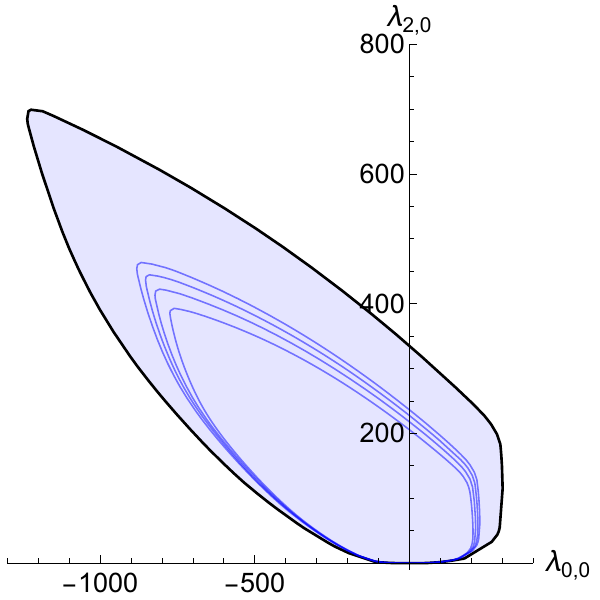}}
        \caption{Bound on \( (\lambda_{0,0}, \lambda_{2,0}) \)}
    \end{subfigure}

    \caption{Bound in the space of couplings \( (\lambda_{2,0}, \lambda_{2,1}) \) and \( (\lambda_{0,0}, \lambda_{2,0}) \) for \( \mu = 1 \). The blue lines denote the boundary of the ``allowed'' primal region for increasing values of \( N\ped{max} = 20, 22, 24, 26 \). The black line enclosing the blue region denotes the \( N\ped{max} \to \infty \) extrapolation. The bound is obtained with \( L\ped{max} = 46 \) and \( n\ped{pts} = 200 \).
        \label{fig:L00-L20-L21-bound-mu-1}}
\end{figure}

For each pair of couplings \( (\lambda_{0,0}, \lambda_{2,0}) \) and \( (\lambda_{2,0}, \lambda_{2,1}) \), we carve out the allowed region in the space of couplings at finite \( N\ped{max} \),\footnote{%
    In practice we maximise the objective \( p_{\theta} =  \cos(\theta) \, v_{1} + \sin(\theta) \, v_{2} = \vec{b} \cdot \vec{\alpha} \) for each pair \( v_{1}, v_{2} \in \{\lambda_{0,0}, \lambda_{2,0}, \lambda_{2,1}\} \)~\cite{Cordova:2019lot,EliasMiro:2026utl}. Maximising \( p_{\theta} \) selects the boundary point of the \( (v_{1}, v_{2}) \) region whose outward normal points along \( (\cos\theta, \sin\theta) \); we read its coordinates \( (v_{1}, v_{2}) \) from the optimal solution of the SDP~\eqref{eq:standard-SDP}. Scanning \( \theta \) then traces out the boundary. We sample the angle \( \theta \) by performing a dynamic bisection based on arc-length criteria.
} and extrapolate to \( N\ped{max} \to \infty \) with a linear model \( a + b/N\ped{max} \).
The result is reported in figure~\ref{fig:L00-L20-L21-bound-mu-1}.
The bound in the \( (\lambda_{2,0}, \lambda_{2,1}) \) space is again compatible with the bounds from positivity and linearised unitarity.

For completeness and comparison with~\eqref{eq:min-L21-extrapolation}, we also report our estimate for the lower bound on \( \lambda_{2,1} \) extrapolated from a linear fit in \( 1/N\ped{max} \)
\begin{equation}
    \mu = 1 \quad \colon \quad \lambda_{2,1} \geq -2.98(2) \,.
\end{equation}

\paragraph{Comparison with identical scalars.}
The scattering of identical scalars can be seen as a special case of the \( AB \to AB \) and \( AA \to BB \) system when \( \mu = 1 \).
Bounds on the non-perturbative couplings \( \lambda_{k,l} \) have been obtained for identical scalars in~\cite{Chen:2022nym} and it is useful to compare with them.
A priori, one could wonder if the bounds presented in this section are saturated by scattering amplitudes of identical scalars.
We will see that this is not the case, generically.

In order to make a fair comparison between the bounds, we need to spell out the optimisation problems that we are solving in the two cases.
For the case of non-identical particles we have two amplitudes \( T_{AB \to AB} \) and \( T_{AA \to BB} \) which are written in terms of a single scalar function \( F \)
\begin{equation}
    F(s,t,u) = T_{AB \to AB}(s,t,u) = T_{AA \to BB}(t,s,u) \,,
\end{equation}
which is \( s-u \) symmetric, and for \( m = M \) we solve the following SDP
\begin{equation}\label{eq:sdp-non-identical-mu-1}
    \begin{aligned}
      & \max \ \partial_{s}^{k}\partial_{t}^{l} F(s_{0},t_{0}, u_{0}) \,, \\[0.5em]
      & \ \text{such that} \quad
        \begin{pmatrix}
          1 & 1 - i F_{\ell}(s)^{*} \\
          1 + i F_{\ell}(s) & 1
        \end{pmatrix}
        \succeq 0 \,, \quad
      && s \geq 4m^{2}\,, \ \ell = 0, 1, 2, \ldots \,, \\[0.5em]
      & \ \phantom{\text{such that}} \quad
        \begin{pmatrix}
          1 & -i \widetilde{F}_{\ell}(s)^{*} \\
          i \widetilde{F}_{\ell}(s) & 1
        \end{pmatrix}
        \succeq 0 \,, \quad
      && s \geq 4m^{2}\,, \ \ell = 0, 2, 4, \ldots \,. \\[0.5em]
    \end{aligned}
\end{equation}
The partial amplitudes are
\begin{equation}
    \begin{aligned}
      F_{\ell}(s) &= \mathcal{N}(s) \int_{-1}^{1} \dif{z} \, P_{\ell}(z) F(s, t(s, z), u(s, z)) \,, \\
      \widetilde{F}_{\ell}(s)& = \frac{1}{2} \mathcal{N}(s) \int_{-1}^{1} \dif{z} \, P_{\ell}(z) F(t(s, z), s, u(s, z)) \,,
    \end{aligned}
\end{equation}
and
\begin{equation}\label{eq:normalization-mu-1}
    \mathcal{N}(s) = \frac{1}{16 \pi} \sqrt{1-\frac{4m^{2}}{s}} \,,
\end{equation}
as can be read from~\eqref{eq:partial-amplitude-normalization} setting \( M = m \).

For the case of identical particles we have a single amplitude \( T_{AA \to AA}(s,t,u) \equiv G(s,t,u) \) which is fully crossing symmetric and we solve the following SDP
\begin{equation}\label{eq:sdp-identical}
    \begin{aligned}
      & \max \ \partial_{s}^{k}\partial_{t}^{l} G(s_{0},t_{0}, u_{0}) \,, \\[0.5em]
      & \ \text{s.t.} \quad
        \begin{pmatrix}
          1 & 1 - i G_{\ell}(s)^{*} \\
          1 + i G_{\ell}(s) & 1
        \end{pmatrix}
        \succeq 0 \,, \qquad
      && s \geq 4m^{2}\,, \quad & \ell &= 0, 2, 4, \ldots \,, \\[0.5em]
    \end{aligned}
\end{equation}
where
\begin{equation}
    G_{\ell}(s) = \frac{1}{2}\mathcal{N}(s) \int_{-1}^{1} \dif{z} \, P_{\ell}(z) G(s, t(s, z), u(s, z)) \,,
\end{equation}
with \( \mathcal{N}(s) \) as in~\eqref{eq:normalization-mu-1}.

Let us now take \( F(s,t,u) = \frac{1}{2} G(s,t,u) \) in the SDP~\eqref{eq:sdp-non-identical-mu-1}.
This restricts the SDP to the space of amplitudes that are also \( s-t \) symmetric. The partial wave projections become in this case
\begin{equation}
    F_{\ell}(s) = G_{\ell}(s) \,, \qquad \widetilde{F}_{\ell}(s) = \frac{1}{2} G_{\ell}(s) \,,
\end{equation}
thanks to \( s-t \) symmetry of \( G \).
The semi-definite constraint for \( \widetilde{F}_{\ell}(s) \)
\begin{equation}\label{eq:identical-matching-tildeF}
    \begin{pmatrix}
      1 & - \frac{i}{2} G_{\ell}(s)^{*} \\
      \frac{i}{2} G_{\ell}(s) & 1
    \end{pmatrix}
    \succeq 0 \,, \qquad s \geq 4m^{2} \,, \quad \ell = 0, 2, 4, \ldots \,,
\end{equation}
is implied by the semi-definite constraint for \( F_{\ell}(s) \)\footnote{%
    Setting \( G_{\ell}(s) = x + i y \), the constraint~\eqref{eq:identical-matching-F}, which reads \( x^{2} + (y-1)^{2} \leq 1 \), implies~\eqref{eq:identical-matching-tildeF}, which reads \( x^{2} + y^{2} \leq 4 \) in these variables.
}
\begin{equation}\label{eq:identical-matching-F}
    \begin{pmatrix}
      1 & 1 - i G_{\ell}(s)^{*} \\
      1 + i G_{\ell}(s) & 1
    \end{pmatrix}
    \succeq 0 \,, \qquad
    s \geq 4m^{2}\,, \quad \ell = 0,1,2, \ldots
\end{equation}
Moreover,~\eqref{eq:identical-matching-F} is trivial for odd \( \ell \) since the partial wave projection is non-trivial only for even \( \ell \) thanks to \( t-u \) symmetry and the fact that \( t(s, -z) = u(s, z)  \).

\begin{figure}[t!]
    \centering
    \begin{subfigure}[b]{0.485\linewidth}
        \centering
        \includegraphics[width=\linewidth]{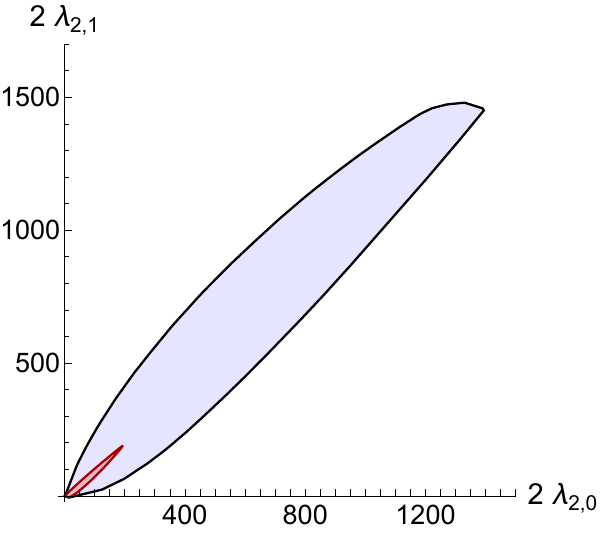}
        \caption{Bound on \( (\lambda_{2,0}, \lambda_{2,1}) \)}
    \end{subfigure}
    \hspace{0.5em}
    \begin{subfigure}[b]{0.485\linewidth}
        \centering
        \includegraphics[width=\linewidth]{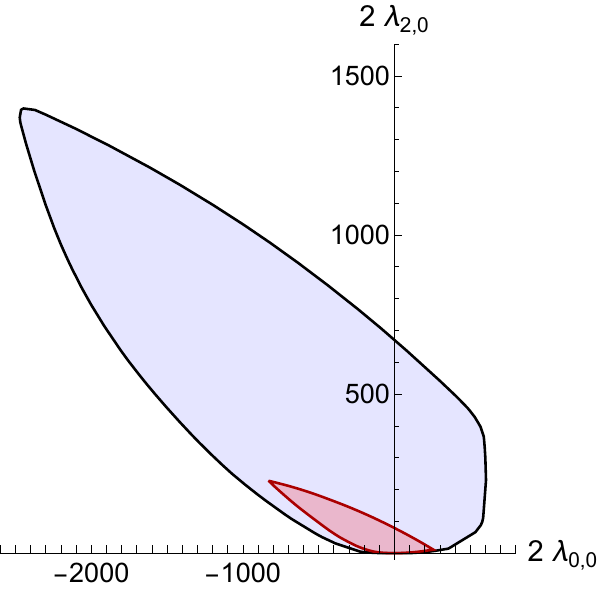}
        \caption{Bound on \( (\lambda_{0,0}, \lambda_{2,0}) \)}
    \end{subfigure}
    \caption{Comparison between the bounds on the couplings \( \lambda_{0,0} , \lambda_{2,0}, \lambda_{2,1} \) for non-identical particles (blue region) at \( \mu = 1 \) and the bounds on the same couplings for identical particles (red region), up to a factor of 2, see~\eqref{eq:identical-couplings-compare}. We report the bounds extrapolated at \( N\ped{max} \to \infty \). The bounds for identical particles are taken from~\cite{Chen:2022nym}. \label{fig:comparison-identical}}
\end{figure}

We conclude that the SDP~\eqref{eq:sdp-non-identical-mu-1} restricted to \( s-t-u \) symmetric functions \( G(s,t,u) \) is equivalent to the following SDP
\begin{equation}\label{eq:sdp-non-identical-mu-1-restricted-crossing}
    \begin{aligned}
      & \max \ \frac{1}{2}\partial_{s}^{k}\partial_{t}^{l} G(s_{0},t_{0}, u_{0}) \,, \\[0.5em]
      & \ \text{s.t.} \quad
        \begin{pmatrix}
          1 & 1 - i G_{\ell}(s)^{*} \\
          1 + i G_{\ell}(s) & 1
        \end{pmatrix}
        \succeq 0 \,, \qquad
      && s \geq 4m^{2}\,, \quad & \ell &= 0, 2, 4, \ldots \,. \\[0.5em]
    \end{aligned}
\end{equation}

To summarise, if we restrict the SDP for non-identical particles at \( \mu = 1 \) in~\eqref{eq:sdp-non-identical-mu-1} to the space of fully crossing symmetric amplitudes, we obtain the SDP in~\eqref{eq:sdp-non-identical-mu-1-restricted-crossing} which differs from~\eqref{eq:sdp-identical} only by a factor of \( 1/2 \) in the objective.
If \( \overline{\lambda}_{k,l} \) denotes the upper bound on the non-perturbative coupling, we then conclude that
\begin{equation}\label{eq:identical-couplings-compare}
    \overline{\lambda}_{k,l}^{\, (\text{id})} \leq 2 \overline{\lambda}_{k,l}^{\, (\text{non-id})} \qquad (\mu = 1) \,.
\end{equation}
In figure~\ref{fig:comparison-identical} we show the comparison between our bounds on \( (\lambda_{0,0}, \lambda_{2,0}) \) and \( (\lambda_{2,0}, \lambda_{2,1}) \) for the \( \mu = 1 \) case and the bounds on the same observables taken from~\cite{Chen:2022nym}.
This comparison shows that this slice of the space of scattering amplitudes of non-identical scalars with equal mass strictly contains, by a large amount, the space of amplitudes of identical particles.

\begin{figure}[t!]
    \centering
    \begin{subfigure}[c]{0.46\linewidth}
        \centering
        \includegraphics[width=\linewidth]{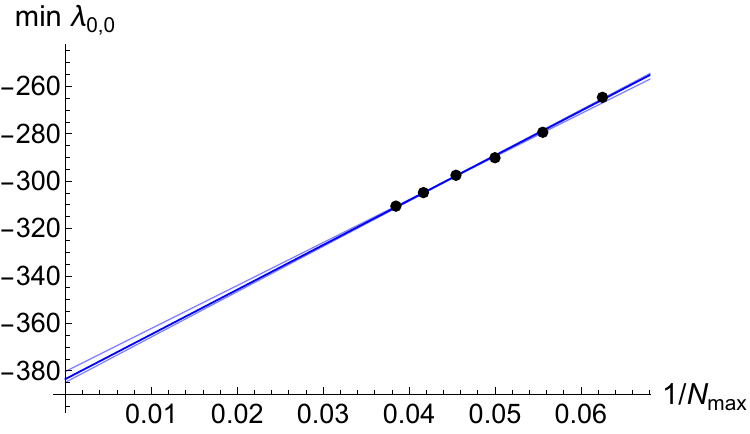}
        \caption{\( \lambda_{0,0} \) minimisation}
    \end{subfigure}
    \hspace{0.5em}
    \begin{subfigure}[c]{0.46\linewidth}
        \centering
        \includegraphics[width=\linewidth]{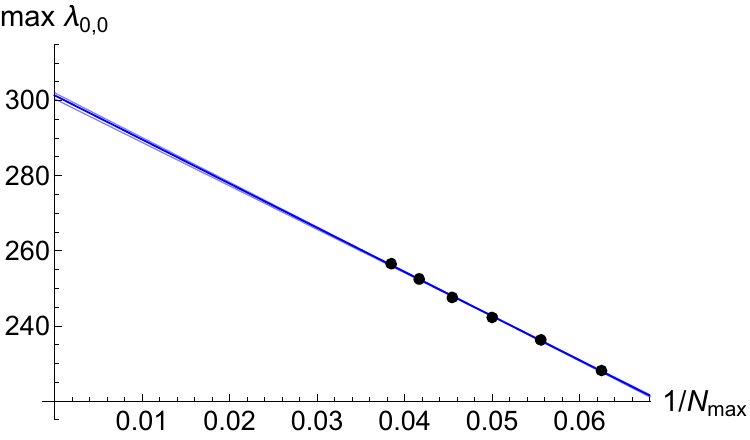}
        \caption{\( \lambda_{0,0} \) maximisation}
    \end{subfigure}
    \caption{Minimisation and maximisation of \( \lambda_{0,0} \) as a function of \( 1/N\ped{max} \) using the Omnès function ansatz. The bound is obtained for \( \mu = 1.5 \), \( m_{0}^{2} = 13/2 \) and \( g = 1/2 \), and using \( L\ped{max} = 60 \) and \( n\ped{pts} = 200 \). The blue line is a fit with a model \( a + b /N\ped{max} \). The error band denotes the estimate of the uncertainty based on a leave-one-out procedure.\label{fig:L00-bound-nmax-omnes}}
\end{figure}

\section{Taming the pseudo-physical region}\label{sec:Omnes}

In this section we present a very simple toy model that shows how additional assumptions on the pseudo-physical region can cure unbounded directions in the primal \( S \)-matrix bootstrap.
The idea is twofold: first we separate the pseudo-physical region \( 4m^{2} \leq s \leq 4M^{2} \) from the physical region \( s \geq 4M^{2} \); second we assume that the physics in the pseudo-physical region is completely dominated by a scalar resonance.
Of course the latter is a very crude assumption (and unrealistic for most scattering processes), but the aim is to show how far physical assumptions on the pseudo-physical behaviour can get us.

The parametrisation of the scalar resonance is done using the so-called Muskhelishvili-Omnès function~\cite{Omnes:1958hv,Muskhelishvili1958}.
Analytically continued unitarity for the \( AA \to BB \) process\footnote{%
    Note that \( f_{AA \to BB}^{\ell}(s)^{*} \) has to be understood as \( f_{AA \to BB}^{\ell}(s-i \epsilon) \) to perform analytic continuation.
}
\begin{equation}
    2\Im f_{AA \to BB}^{\ell}(s) = \mathcal{N}_{AA \to AA}(s) f_{AA \to AA}^{\ell}(s) f_{AA \to BB}^{\ell}(s)^{*} \,, \qquad 4m^{2} \leq s \leq  4M^{2} \,,
\end{equation}
implies Watson's theorem~\cite{Watson:1952ji}, which states that the phase of the \( AA \to BB \) amplitude coincides with the phase shift of the \( AA \to AA \) amplitude in the pseudo-physical region.
This applies for \( M < 2 m \), while for $M>2m$ it holds only if inelasticities are negligible.
In formulas, if \( \delta^{\ell}(s) \) is the phase shift of the \( AA \to AA \) scattering
\begin{equation}
    S_{AA \to AA}^{\ell}(s) = 1 + i \mathcal{N}_{AA \to AA}(s) f_{AA \to AA}^{\ell}(s) = e^{2i \delta^{\ell}(s)} \,, \qquad 4m^{2} \leq s \leq 4M^{2} \,,
\end{equation}
then
\begin{equation}\label{eq:watson}
    f_{AA \to BB}^{\ell}(s) = \left|f_{AA \to BB}^{\ell}(s)\right| e^{i \delta^{\ell}(s)} \,, \qquad 4m^{2} \leq s \leq 4M^{2} \,.
\end{equation}
The modulus is naively unconstrained, but it can be partially reconstructed from its phase using a dispersion relation.
Neglecting the left-cut for $s\leq 0$, we have
\begin{equation}
    f_{AA \to BB}^{\ell}(s) = P^\ell(s)  \Omega^{\ell}(s) \,,
\end{equation}
where
\begin{equation}\label{eq:OmegaDef}
    \Omega^{\ell}(s) = \exp\left(\frac{s}{\pi} \int_{4m^{2}}^{+\infty}\dif{s'}\, \frac{\delta^{\ell}(s')}{s'(s'-s)}\right)
\end{equation}
is called the Muskhelishvili-Omnès function, and $P^\ell(s)$ can be taken to be a polynomial in $s$, whose degree is associated to the behaviour of the amplitude and the phase shift at infinity.
We have taken the upper limit of integration in~\eqref{eq:OmegaDef} to be $\infty$ rather than $4M^2$, for simplicity.\footnote{%
    The distinction is not important since, for $s>4M^2$, the terms coming from the $\rho$-ansatz also contribute to the phase of the amplitude, see below. Such terms also account for the left-hand cut for $s\leq 0$.
}

Assuming that the \( AA \to AA \) process is dominated by the scalar partial wave, and this is dominated by a resonance, we show in appendix~\ref{app:omnes} that the
corresponding Omnès function for \( \ell = 0 \) can be written as
\begin{equation}\label{eq:omnes-ansatz}
    \Omega^{0}(s) = \frac{m_{0}^{2}}{m_{0}^{2}-s  - g^{2} C(s)} \,,
\end{equation}
where \( m_{0} \) and \( g \) are free parameters governing the mass and the width of the resonance, while $C(s)$ is the Chew-Mandelstam function~\cite{Chew:1960iv}
\begin{equation}
    C(s) = \frac{s}{\pi} \int_{4m^{2}}^{\infty}\dif{s'} \frac{r(s')}{s'(s'-s)} = \frac{1}{\pi}\left(2+r(s)\log{\frac{r(s)-1}{r(s)+1}}\right) \,, \qquad
    r(s) = \sqrt{1-\frac{4m^{2}}{s}} \,.
    \label{eq:CsDef}
\end{equation}
We refer the reader to appendix~\ref{app:omnes} for a derivation of~\eqref{eq:omnes-ansatz} and its relation to~\eqref{eq:OmegaDef} when $\delta^{0}$ is dominated by a resonance, and for the analytic properties of the function~\eqref{eq:omnes-ansatz}.

\begin{figure}[t!]
    \centering
    \begin{subfigure}[c]{0.48\linewidth}
        \centering
        \includegraphics[width=\linewidth]{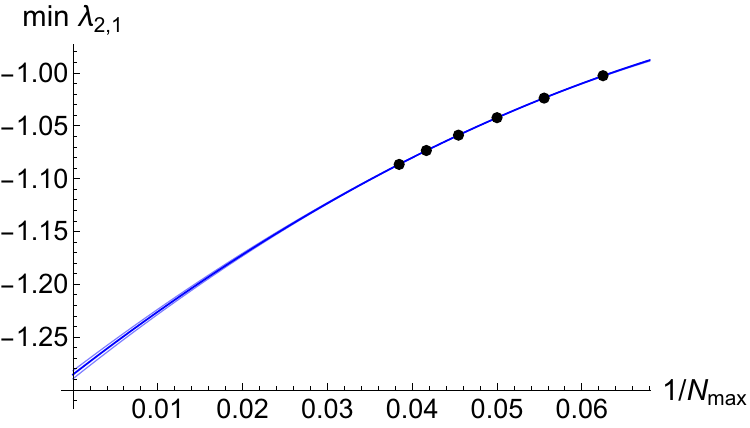}
        \caption{\( \lambda_{2,1} \) minimisation}
    \end{subfigure}
    \hspace{0.5em}
    \begin{subfigure}[c]{0.48\linewidth}
        \centering
        \includegraphics[width=\linewidth]{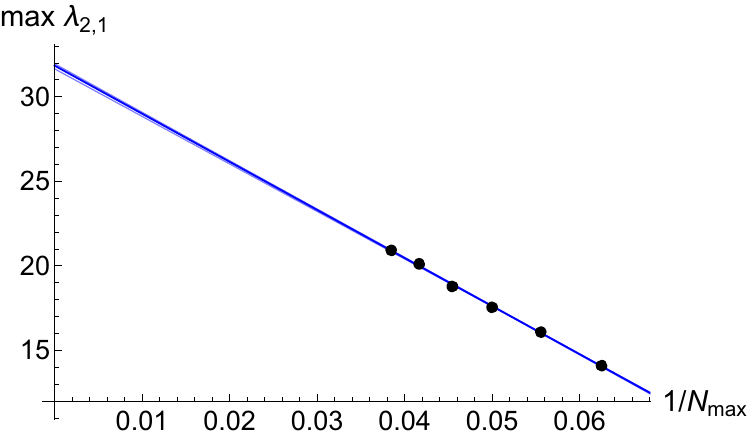}
        \caption{\( \lambda_{2,1} \) maximisation}
    \end{subfigure}
    \caption{Minimisation and maximisation of \( \lambda_{2,1} \) as a function of \( 1/N\ped{max} \) using the Omnès function ansatz. The bound is obtained for \( \mu = 1.5 \), \( m_{0}^{2} = 13/2 \) and \( g = 1/2 \), and using \( L\ped{max} = 60 \) and \( n\ped{pts} = 200 \). The blue line is a fit with a model \( a + b /N\ped{max} \) for the maximisation and a quadratic model \( a + b /N\ped{max} + c /N\ped{max}^{2} \) for the minimisation. The error band denotes the estimate of the uncertainty based on a leave-one-out procedure.\label{fig:L21-bound-nmax-omnes}}
\end{figure}

For the primal \( S \)-matrix bootstrap, we then propose to modify the ansatz in~\eqref{eq:ABtoAB-ansatz} and~\eqref{eq:AAtoBB-ansatz} as follows
\begin{align}
  T_{AB \to AB}(s,t,u) &= \alpha_{0} \, \Omega^{0}(t) + \sum_{a,b,c} \alpha_{a,b,c} (\rho_{+}(s))^{a}(\rho_{M}(t))^{b} (\rho_{+}(u))^{c} \,, \label{eq:ABtoAB-ansatz-omnes} \\
  T_{AA \to BB}(s,t,u) &= \alpha_{0} \, \Omega^{0}(s) + \sum_{a,b,c} \alpha_{a,b,c} (\rho_{+}(t))^{a}(\rho_{M}(s))^{b} (\rho_{+}(u))^{c} \,, \label{eq:AAtoBB-ansatz-omnes}
\end{align}
where, for simplicity, we assume that $P^{\ell=0}(s)$ is just a constant, the parameter \( \alpha_{0} \), and \( \rho_{M}(z) \) is a \( \rho \)-variable with a branch-cut starting at the \emph{physical} threshold
\begin{equation}
    \rho_{M}(z) = \rho\left(z, 4M^{2}, t_{0}\right) \,.
\end{equation}

The ansatz~\eqref{eq:AAtoBB-ansatz-omnes} is constructed in such a way that the discontinuity of the partial wave projection \( f_{AA \to BB}^{\ell}(s) \) in the pseudo-physical region is given by the Omnès function \( \Omega^{0}(s) \) only for \( \ell = 0 \) and is trivial for \( \ell > 0 \)
\begin{equation}
    \Im f_{AA \to BB}^{\ell}(s) = \delta_{\ell,0}  \, \alpha_{0}\Im \Omega^{0}(s) \,, \qquad 4m^{2}  \leq s < 4M^{2} \,.
\end{equation}
This is not exactly what we would like from Watson's theorem~\eqref{eq:watson}, since that would require the phase of \( f_{AA\to BB}^{0} \) to match the phase of \( \Omega^{0} \).
However, given an explicit solution to an optimisation problem, we can check a posteriori that the Omnès function term is dominant with respect to the pure \( \rho \)-ansatz part in the pseudo-physical region, and that approximately
\begin{equation}\label{eq:watson-approx}
    \left.\arg \left( f_{AA \to BB}^{\ell}(s)\right)\right\vert_{\text{optimal}} \approx \delta_{\ell,0} \arg \Omega^{0}(s) \,, \qquad 4m^{2}  \leq s < 4M^{2} \,.
\end{equation}

\begin{figure}[t!]
    \centering
    \includegraphics[width=0.48\linewidth]{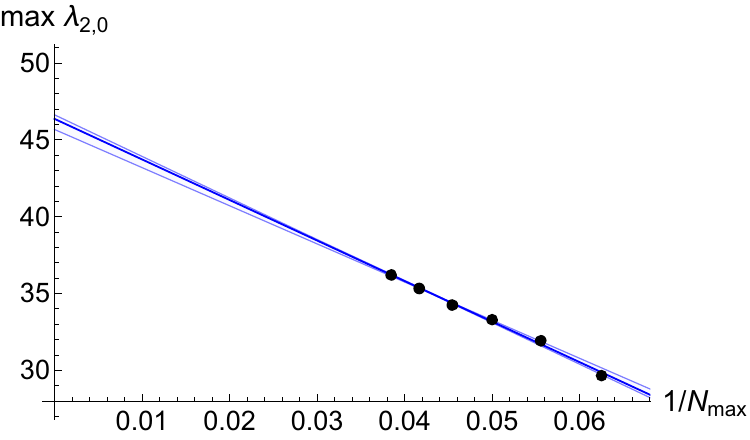}
    \caption{Maximisation of \( \lambda_{2,0} \) as a function of \( 1/N\ped{max} \) using the Omnès function ansatz. The bound is obtained for \( \mu = 1.5 \), \( m_{0}^{2} = 13/2 \) and \( g = 1/2 \), and using \( L\ped{max} = 60 \) and \( n\ped{pts} = 200 \). The blue line is a fit with a model \( a + b /N\ped{max} \). The error band denotes the estimate of the uncertainty based on a leave-one-out procedure.\label{fig:L20-bound-nmax-omnes}}
\end{figure}

\begin{figure}[t!]
    \centering
    \includegraphics[width=0.48\linewidth]{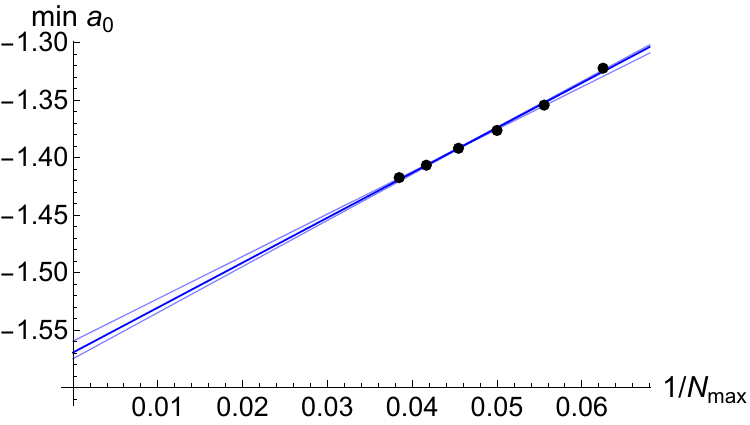}
    \caption{Minimisation  of \( a_{0} \) as a function of \( 1/N\ped{max} \) using the Omnès function ansatz. The bound is obtained for \( \mu = 1.5 \), \( m_{0}^{2} = 13/2 \) and \( g = 1/2 \), and using \( L\ped{max} = 60 \) and \( n\ped{pts} = 200 \). The blue line is a fit with a model \( a + b /N\ped{max} \). The error band denotes the estimate of the uncertainty based on a leave-one-out procedure.\label{fig:a0-bound-nmax-omnes}}
\end{figure}

\begin{figure}[t!]
    \centering
    \begin{subfigure}[c]{0.48\linewidth}
        \centering
        \includegraphics[width=\linewidth]{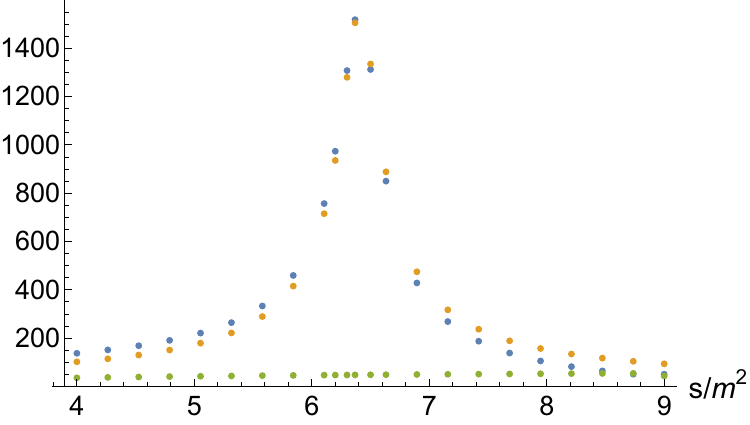}
        \caption{modulus}
    \end{subfigure}
    \hspace{0.5em}
    \begin{subfigure}[c]{0.48\linewidth}
        \centering
        \includegraphics[width=\linewidth]{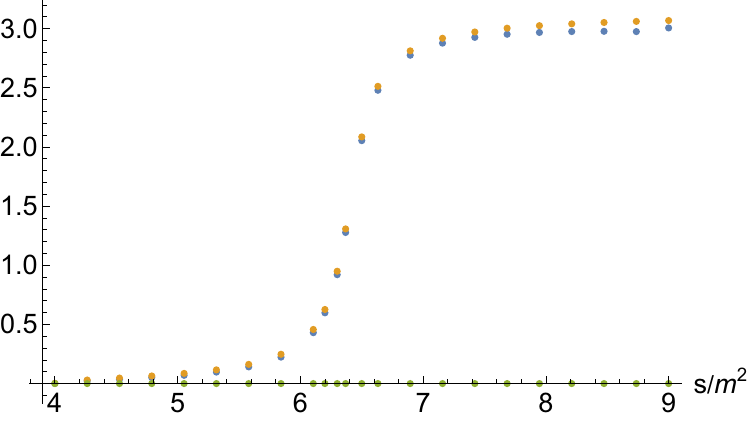}
        \caption{phase}
    \end{subfigure}
    \caption{Scalar partial amplitude in the pseudo-physical region for the \( \max \lambda_{0,0} \) problem at \( N\ped{max} = 26 \) using the Omnès function ansatz. The blue dots denote the extremal scalar partial amplitude \( f^{0}_{AA \to BB}(s) \), the orange dots denote the Omnès function contribution \( \alpha_{0} \Omega^{0}(s) \), while the green dots denote the \( \rho \)-variables part of the ansatz. This shows that the Omnès function is dominant and that~\eqref{eq:watson-approx} holds.\label{fig:omnes-check}}
\end{figure}

Adopting now this modified ansatz, we explore bounds on the non-perturbative couplings \( \lambda_{k,l} \) using the same strategy discussed in section~\ref{sec:numerics-setup}.
The problem now depends not only on the mass ratio \( \mu \), but also on the parameters \( m_{0} \) and \( g \) of the Omnès function~\eqref{eq:omnes-ansatz}.
We have not explored the full range of those parameters, but rather have chosen just a sample value since the purpose of this section is a proof of concept. We have taken
\begin{equation}
    \mu = \frac{3}{2} \,, \qquad m_{0}^{2} = \frac{13}{2} \,, \qquad g = \frac{1}{2} \,.
\end{equation}
The value of \( m_{0}^{2} \) is chosen to be the mid-point of the pseudo-physical region, while \( g \) is an arbitrary value that satisfies the constraint~\eqref{eq:omnes-g-constraint}.

In figure~\ref{fig:L00-bound-nmax-omnes} we report upper and lower bounds on \( \lambda_{0,0} \) as a function of \( N\ped{max} \).
Nicely enough, we observe now good linear convergence of the bound in \( 1/N\ped{max} \).
This is to be contrasted with figure~\ref{fig:L00-bound-nmax} which showed exponential growth of the bound with \( N\ped{max} \).
This shows that our crude assumption on the pseudo-physical region is strong enough to allow us to recover a bound on the amplitude at one point.

Similar convergent bounds are now also found for minimisation and maximisation of \( \lambda_{2,1} \), as reported in figure~\ref{fig:L21-bound-nmax-omnes}, as well as for maximisation of \( \lambda_{2,0} \), shown in figure~\ref{fig:L20-bound-nmax-omnes}.
Note how the Omnès function does not significantly affect the bounds for the minimum of $\lambda_{2,1}$. Compare
figure~\ref{fig:L21-bound-fit}(a) with figure~\ref{fig:L21-bound-nmax-omnes}(a). This is expected, since lower bounds on $\lambda_{2,0}$ and $\lambda_{2,1}$ are essentially insensitive to the pseudo-physical region.
Finally, a convergent lower bound is found also for the scalar scattering length \( a_{0} \) and is reported in figure~\ref{fig:a0-bound-nmax-omnes}.

For all these extremisation procedures, we always find the Omnès part of the ansatz~\eqref{eq:omnes-ansatz} to be dominant in the scalar partial amplitude \( f_{AA \to BB}^{\ell}(s) \) on the pseudo-physical region.
An example plot is shown in figure~\ref{fig:omnes-check} for the extremal amplitude obtained by maximising \( \lambda_{0,0} \) at \( N\ped{max} = 26 \).

\FloatBarrier

\section{Discussion}\label{sec:discussion}

We explored the constraints on the space of scattering amplitudes of two unequal scalar massive particles in \( 4d \), denoted by \( A \) and \( B \).
A \( \Z_{2} \times \Z_{2} \) global symmetry was assumed, ensuring the stability of the particles, forbidding triangular anomalous thresholds and slightly simplifying our setup.
We used the primal \( S \)-matrix bootstrap to obtain non-perturbative bounds on several observables that can be extracted from the scattering amplitude for the \( AB \to AB \) process.
This is not a trivial generalisation of the identical scalar \( S \)-matrix bootstrap because of the unavoidable presence of another source of non-analyticity, given by pseudo-physical regions.
We found that for $M>m$ there exist observables that admit one-sided bounds, or become bounded once another observable is fixed.
Upper and lower bounds on all the observables are recovered when the behaviour of the amplitude is controlled in the pseudo-physical region by some assumption, or for $M=m$, where this region collapses to a point.

There are several interesting directions for further exploration.

\paragraph{Role of pseudo-physical region.}
Singularities in the pseudo-physical region present a challenge to the bootstrap since they are not bounded by unitarity, which strictly applies to the physical region, and play a prominent role in our paper.
Lower bounds on $\lambda_{2,0}$ or $\lambda_{2,1}$ are controlled by a fixed-$t$ dispersion-relation representation and are clearly insensitive to the pseudo-physical region.
The existence of such bounds for any $\mu >1$ is then not surprising. In contrast, upper bounds require control over the $s$-integral of the sum over spins of the partial wave expansion, which in turn seems to be sensitive to the presence or absence of the pseudo-physical region.
For other observables such as $\lambda_{0,0}$ or $a_0$, the naive expectation is that maximisation or minimisation of an observable is unbounded in the \( N\ped{max} \to \infty \) limit.\footnote{%
    The basic phenomenon can be understood in a simple toy model.
    Consider the space of real analytic functions \( f(z) \) on \( \mathbb{C} \setminus [1, +\infty) \) such that \( f(z) \to 0 \) as \( z \to \infty \). Let \( x_{0} \in [1, +\infty) \) and consider the problem of maximising \( f(0) \) subject to \( |f(x)| \leq 1 \) for \( x \geq x_{0} \).
    If \( x_{0} = 1 \), we can use the $\rho$-variable $\rho(z,1,0)$ in \eqref{eq:rho-variable-generic} to map the whole cut to the boundary of the unit disk.
    The maximum modulus principle then gives \( |f(0)|\leq  1 \).
    For \( x_{0} > 1 \), the image of $[1,x_0]$ (the analogue of the pseudo-physical region) is an arc on which nothing is imposed, and $f(0)$ becomes unbounded.
    This can also be seen numerically by solving the primal semi-definite optimisation problem by taking an ansatz \( f(z) = \sum_{n=0}^{n\ped{max}} a_{n} \rho(z,1,0)^{n} \).
    One finds that a bound on \( f(0) \) does not converge in \( n\ped{max} \).
    A similar argument was presented in appendix E of~\cite{Homrich:2019cbt}.
}
It would be interesting to extend this mechanism to the actual scattering amplitude in \( 4d \) and, more generally, to better understand which observables can be bounded.

\begin{figure}[t!]
	\centering
	\includegraphics[width=0.7\linewidth]{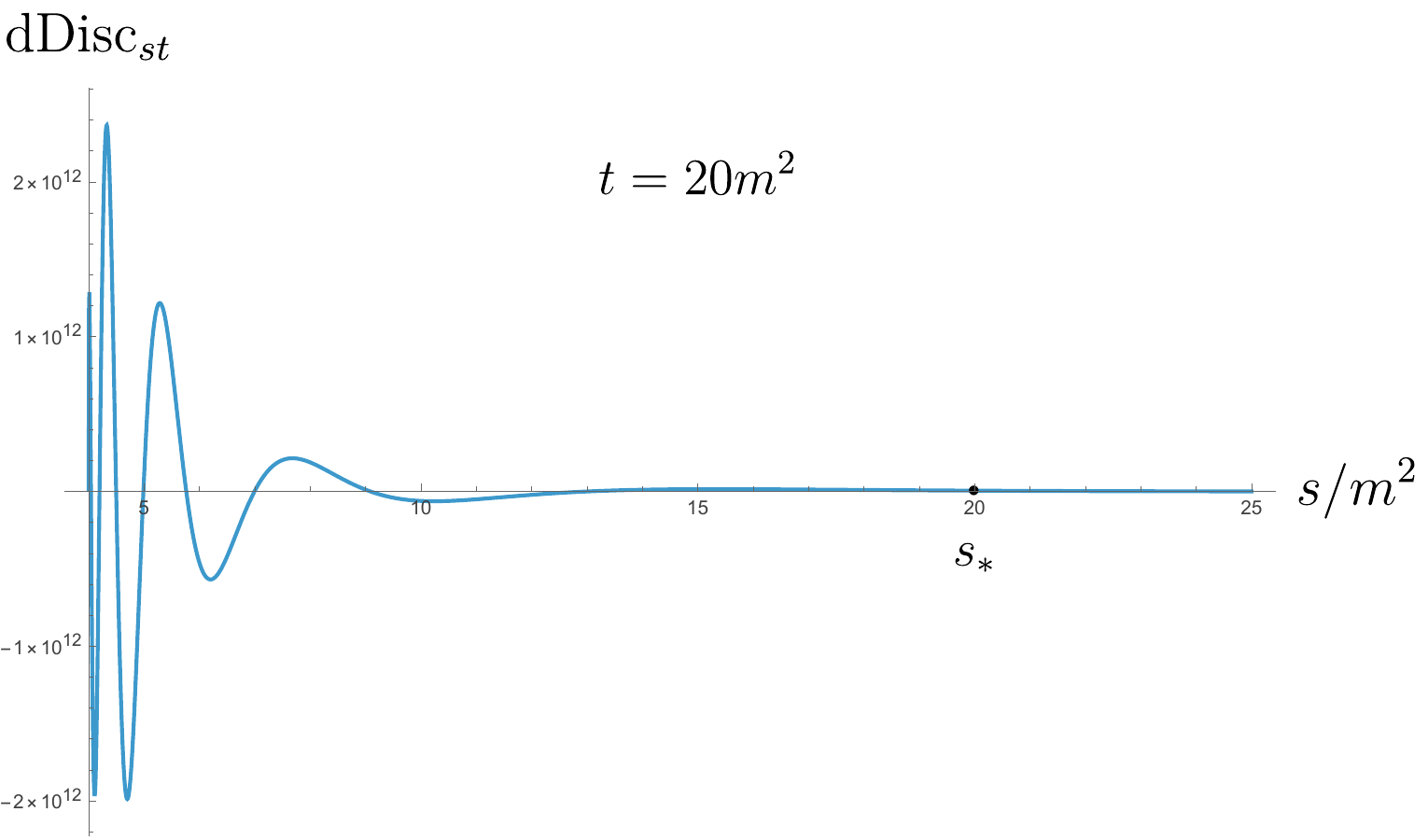}
	\caption{The value of \( \ddisc_{st}{} \) as a function of $s>s\ped{th}$ at fixed $t$ for the extremal $AB\to AB$ amplitude obtained by maximising $\lambda_{2,0}$, for the case $\mu=1$. We have $N\ped{max} = 26$, $L\ped{max}=46$.\label{fig:dDisc}}
\end{figure}

\paragraph{Double discontinuity.}
The expansion of the amplitude~\eqref{eq:ABtoAB-ansatz} was motivated by its analytic structure, but it actually fails to reproduce the correct double discontinuity (\( \ddisc{} \)) of the amplitude at finite \( N\ped{max} \).
Indeed, while the actual \( \ddisc{} \) of the $AB\to AB$ amplitude for physical values of $s$ is non-vanishing for $t> t^\text{LL}(s)>4m^2$, with $t^\text{LL}$ determined in~\eqref{eq:tKABAB}, amplitudes written as in~\eqref{eq:ABtoAB-ansatz} have a non-vanishing \( \ddisc{} \) starting directly from $t>4m^2$.
Its value is computed by noting that, for $s>m_+^2$ and $t>4m^2$, $\rho_+(s)$ and $\rho_m(t)$ are pure phases $\rho_+(s) = e^{i\theta_s(s)},\rho_m(t) =e^{i\theta_t(t)}$ and we get
\begin{equation}
    \ddisc_{st} \, T(s,t) = \sum_{a,b} \alpha_{a,b,0}  \sin \big(a\theta_s(s) \big)\sin \big(b\theta_t(t)\big)\,, \qquad (AB\to AB)\,.
\end{equation}
At first sight this might suggest that the amplitudes~\eqref{eq:ABtoAB-ansatz} are unphysical.
Very preliminary investigations indicate, however, that the observables we consider are very weakly sensitive to the double discontinuity.
A possible explanation of this welcome feature is the following.\footnote{%
    MS thanks A.\ Zhiboedov for interesting discussions on this point.
}
The coefficients $\alpha_{a,b,0}$ for the allowed amplitudes are such that \( \ddisc_{st}{} \) oscillates rapidly for $t< t^{\text{LL}}$, effectively averaging it to zero, and then changes behaviour for $t> t^{\text{LL}}$, when it is supposed to be non-vanishing.
For illustration, we show in figure~\ref{fig:dDisc} the extremal amplitude obtained by maximising $\lambda_{2,0}$ for $\mu=1$, evaluated at $t=20m^2$.
For this value of $t$, \( \ddisc_{st} \) should identically vanish in the interval $s\in[4m^2,s_*]$, with $s_*=20m^2$.
The figure shows that \( \ddisc_{st} \) oscillates in the region where it should vanish, and approaches a monotonic behaviour
when $s$ approaches $s_*$.\footnote{%
    For $s>s_*$, \( \ddisc_{st} \) is non-zero.
    This is not evident from the figure at the scales reported.
}
The same behaviour is found in other cases.
It would be interesting to undertake a systematic analysis of this kind and understand to what extent and for which observables the $\rho$-expansion
is just fine, although it does not reproduce the correct analytic behaviour of the amplitude.\footnote{%
    As far as we are aware, this question has not been settled even for the much studied case of identical scalar scattering.
}

\paragraph{Full system.}
An obvious generalisation of our work would be to consider the full system of two-to-two scattering amplitudes of the \( A \) and \( B \) particles~\eqref{eq:set_amplitudes}.
Bounds from this mixed system will be at least as strong as the ones obtained in this work, if not stronger.
It is not clear to us whether ordinary unitarity constraints in~\eqref{eq:unitarity-AB-mixed} and~\eqref{eq:unitarity-ABtoAB} would suffice
to obtain two-sided bounds on all observables.
It is possible that analytically continued unitarity makes the problem bounded.
The intuition behind this is that such a form of extended unitarity implies
\begin{equation}\label{eq:extended-unitarity-AB}
    \left|\Im f_{AA \to BB}^{\ell}(s)\right|^{2} \leq  \Im f_{AA \to AA}^{\ell}(s) \,  \Im f_{BB \to BB}^{\ell}(s) \,,
\end{equation}
also for the pseudo-physical region \( 4m^{2} \leq s < 4M^{2} \).
Indeed, in \( 2d \) extended unitarity was used to bound certain observables extracted from the \( AB \to AB \) amplitude~\cite{Homrich:2019cbt}.
Whether this persists in \( 4d \) remains to be seen.
A natural concern is that both \( AA \to BB \) and \( BB \to BB \) have a discontinuity in the pseudo-physical region, so that~\eqref{eq:extended-unitarity-AB} only imposes a relative constraint between the two, leaving the amplitudes free to behave wildly in the pseudo-physical region and rendering some observables unbounded.
Extended unitarity does, however, force the discontinuity in the pseudo-physical region of the \( BB \to BB \) process to be sign-definite, which may tame the optimisation problem.\footnote{%
    Instead, the discontinuity in the pseudo-physical region of \( AA \to BB \) can be of any sign.
    We have indeed observed that even for the optimisation problems of section~\ref{sec:numerics} that admit a convergent bound, the extremal amplitudes oscillate wildly in the pseudo-physical region.
}

\paragraph{Phenomenological applications.}
Looking ahead to phenomenological applications, it would be interesting to extend this setup to the scattering of different hadrons in low-energy QCD.
So far the applications of the primal \( S \)-matrix bootstrap in this spirit have been restricted to pion-pion scattering.
Natural targets are pion-kaon and pion-nucleon scattering.
This was, at its core, the main goal that motivated the present, more theoretical, work.
We would need to dress our setup with global symmetries, which can be approximately taken to be \( SU(2) \) isospin and \( U(1) \) strangeness, and the helicity structure of the nucleons.\footnote{%
    The analysis of \( \pi N \to \pi N \) and \( \pi \pi \to N \bar{N} \) amplitudes with dispersion relations was already carried out in the seminal work of Mandelstam~\cite{Mandelstam:1958xc}.
}
We believe non-analyticities in the pseudo-physical region will still represent the main technical complication.
In addition to that, a crucial point would be to identify both inputs from QCD and a good set of observables that make the problem bounded and, importantly, the QCD scattering amplitude close to being extremal.
In this respect, restricting to the amplitudes in~\eqref{eq:amplitudes_main} has a further potentially useful by-product. Extremal amplitudes obtained in the primal bootstrap are often elastic, i.e.\ they saturate unitarity conditions.
In studying the full system, the first particle-production threshold is at $s=16m^2$ and comes from $4A$ particles in the $AA\to AA$ process.
By contrast, the lightest multi-particle state in the $AB\to AB$ process is $3A+B$, with threshold $s=(3m+M)^2=m^2(3+\mu)^2$, parametrically larger than $16m^2$ for $\mu\gg1$.
In the absence of additional lighter channels, the $AB\to AB$ amplitude therefore remains elastic over a wider range of squared energies.
Extremal amplitudes obtained from the reduced system~\eqref{eq:amplitudes_main} may consequently be closer to physical amplitudes.

Both $\pi K\to\pi K$ and $\pi N\to\pi N$ scattering have been analysed in depth by means of Roy--Steiner equations~\cite{Roy:1971tc,Steiner:1971ms}, which exploit the same principles underlying our approach (analyticity, crossing and unitarity), supplemented by experimental input; see~\cite{Buettiker:2003pp,Pelaez:2020gnd} for $\pi K$ and~\cite{Hoferichter:2015hva} for $\pi N$.
It would be interesting to see how the \( S \)-matrix bootstrap can compare to these analyses.
The wealth of experimental information accumulated on these processes makes this a particularly exciting prospect for the bootstrap.

\section*{Acknowledgements}
We are grateful to Miguel Correia, Andrea Guerrieri and Alexander Zhiboedov for useful discussions.
GF would like to thank SISSA for the kind hospitality while this work was initiated.
DK, AP and MS would like to thank Chalmers University of Technology for the kind hospitality during the completion of this work.
DK and AP would like to thank the Aspen Center for Physics for the hospitality and the participants in the summer program ``From First Principles to Future Colliders: Amplitudes, Bootstraps and Energy Correlators'' for stimulating discussions. AP acknowledges the support from a Simons Foundation grant (1161654, Troyer).

The work of DK is funded by the Swiss State Secretariat for Education, Research and Innovation (SERI) under contract number MB25.00001. The work of DK is also supported by the SNSF Ambizione grant PZ00P2\_193411.
GF is supported by the Swedish Research Council (grant nr.\ 2024-04347), as well as by travel grants from the Carl Tryggers Foundation (CTS 24:3453) and Kungl.\ Vetenskapsakademien (PH2024-0076).
AP and MS are supported by the INFN ``Iniziativa Specifica'' ST\&FI.

The computations presented here were conducted using the facilities of the SCITAS (Scientific IT and Application Support) Center of EPFL and of the SISSA HPC cluster Ulysses. The authors used ChatGPT (OpenAI) and Claude (Anthropic) as assistive tools for discussing aspects of the physical results and for polishing the manuscript. All derivations and formulas were produced by the authors. All AI-assisted suggestions were critically assessed, independently checked and approved by the authors, who take full responsibility for the final content.

\appendix

\section{Analyticity of partial wave expansions}
\label{app:PWE}

We report in this appendix some details aimed at clarifying the analyticity properties of partial wave expansions.
We first explain why and how ellipses arise in complete generality.
We then discuss the small and large Lehmann ellipses~\cite{Lehmann:1958ita} for the cases of interest, $AB\to AB$ and $AA\to BB$ scattering amplitudes. Finally, we show how the pseudo-physical region is responsible for the appearance of a circular cut in the partial waves for $AB\to AB$ scattering.

\subsection{The origin of Lehmann ellipses}
\label{app:ellipses}

The origin of ellipses is a mathematical fact, entirely determined by the behaviour of the Legendre polynomials at large spin.
Any function $f(x)\in L^2([-1,1])$ can be expanded in partial waves as
\begin{equation}\label{app:PWE1}
    f(x) = \sum_{\ell=0}^\infty a_\ell P_\ell(x)\,,
\end{equation}
where
\begin{equation}\label{app:PWE2}
    a_\ell = \frac{2\ell+1}{2}\int_{-1}^1\dif{x}\, f(x) P_\ell(x)\,,
\end{equation}
and $P_\ell(x)$ are the Legendre polynomials. On $x\in [-1,1]$, the expansion~\eqref{app:PWE1} is convergent.
Consider now the analytic continuation of $f$ over some domain and let us determine in which region $z\in E\supset [-1,1]$
the partial wave expansion~\eqref{app:PWE1} applies.
Evidently, this requires knowing the asymptotic behaviour of $P_\ell(z)$ for $\ell\gg 1$. An easy way to determine it is to use a saddle point approximation starting from the integral representation
\begin{equation}\label{app:PWE3}
    P_\ell(z) = \frac{1}{\pi} \int_0^\pi \dif{\alpha} \, \left(z+\sqrt{z^2-1} \cos\alpha \right)^\ell \,,
\end{equation}
valid for any integer $\ell$.
For $z\notin [-1,1]$, the only saddle contributing to the integral is the one at $\alpha=0$.
Straightforward saddle-point methods give
\begin{equation}\label{app:PWE4}
    P_\ell(z) \approx \frac{\Phi(z)^{\ell+\frac 12}}{\sqrt{2\pi \ell}(z^2-1)^{\frac 14}} \left(1+O(\ell^{-1})\right)\,, \quad \ell\gg1 \,, \ z\notin [-1,1]\,,
\end{equation}
where
\begin{equation}\label{app:PWE5}
    \Phi(z) \equiv z+\sqrt{z^2-1}\,.
\end{equation}
When $z\in [-1,1]$, the relation~\eqref{app:PWE3} still applies, with $z+\sqrt{z^2-1} \cos\alpha\to z+i \sqrt{1-z^2} \cos\alpha$.
The function $\Phi$ becomes a phase:
\begin{equation}\label{app:PWE6}
    \Phi(z) = e^{i \theta}\,, \qquad z = \cos \theta\,,
\end{equation}
and now both saddle points at $\alpha=0$ and at $\alpha=\pi$ contribute.
Taking care to rotate the contour to reach the steepest-descent trajectory, eventually we get
\begin{equation}\label{app:PWE7}
    P_\ell(z=\cos\theta) \approx \sqrt{\frac{2}{\pi\ell \sin\theta}}\cos{\left[\left(\ell+\frac 12 \right) \theta-\frac{\pi}{4}\right]} \left(1+O(\ell^{-1})\right)\,,
    \quad \ell\gg1 \,, \ z\in [-1,1]\,.
\end{equation}

\begin{figure}[t!]
    \centering
    \includegraphics[scale=.33]{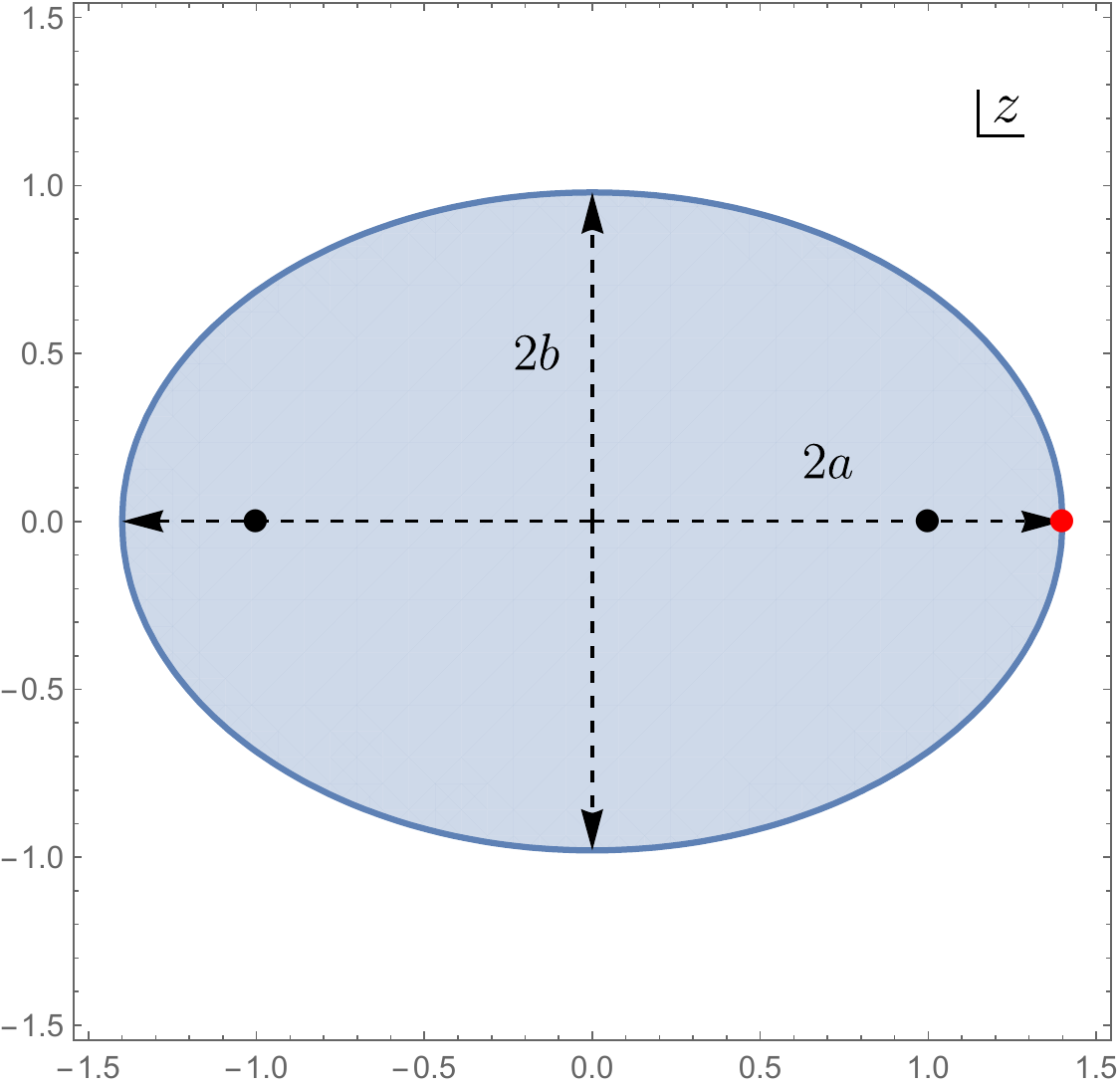}
    \caption{Ellipse of convergence of the partial wave expansion in the $z$-plane for a function with a singularity at $z_s$ (red bullet). The two black bullets denote the two foci at $(-1,0)$ and $(1,0)$. We have taken $z_s = 1.4$ in the picture.}
    \label{fig:Ellipse}
\end{figure}

Coming back to our analytic continuation, for large $\ell$, we have
\begin{equation}\label{app:PWE8}
    f(z) \sim \sum_\ell a_\ell \Phi(z)^\ell \,.
\end{equation}
In order to map this expression to an ordinary Taylor expansion, we perform the conformal map
\begin{equation}
    w = \Phi(z) \longrightarrow z = \frac 12 \Big(w+\frac{1}{w}\Big) \,.
\end{equation}
The domain of convergence of~\eqref{app:PWE8} is determined by the behaviour of $a_\ell$ or, equivalently, from the position of the singularity of $f$ closest to the origin.
Let $z_s$ be such a singularity.
Then the domain of convergence is given by a disk in the $w$-plane of radius
\begin{equation}
    R = |w(z_s)|\,.
\end{equation}

This disk is mapped to an ellipse in the $z$-plane. Indeed, if $w = R e^{i \phi}$, we have
\begin{equation}
    z = x+ i y =a \cos \phi +i b \sin \phi \longrightarrow \frac{x^2}{a^2}+\frac{y^2}{b^2} = 1\,,
\end{equation}
with
\begin{equation}
    a = \frac 12 (R+R^{-1}) \,, \qquad  b = \frac 12(R-R^{-1})\,.
\end{equation}
So, the partial wave expansion converges uniformly in the $z$-plane within an ellipse $E$ with foci at $(-1,0)$ and $(1,0)$, and major and minor axes given by $2a$ and $2b$, respectively.
See figure~\ref{fig:Ellipse} for an illustration.
We also have that
\begin{equation}
    a_\ell \sim w(z_s)^{-\ell}\,, \qquad \ell\gg 1\,.
\end{equation}
For real $z_s$, the semi-major axis of the ellipse in the $z$-plane corresponds directly to $z_s = a$.
When $z>1$, we can set $z=\cosh\alpha$, so that $\Phi(z(\alpha)) = e^{\alpha}$, and $P_{\ell}\approx e^{\ell \alpha}$ for \( \ell \gg 1 \).

\subsection{Small Lehmann ellipses}

In the context of scattering amplitudes, the function $f(z)$ introduced in section~\ref{app:ellipses} can be identified with the amplitude $T(s,t(s,z),u(s,z))$ at fixed $s$, with $t(s,z)$ and \( u(s,z) \) as given in~\eqref{eq:t-of-s-z-ABtoAB} for $AB\to AB$ and~\eqref{eq:t-of-s-z-AAtoBB} for $AA\to BB$.
The region of convergence of the partial wave expansion of $T$ determines what is called the small Lehmann ellipse (SL).
As we have seen, the size of the semi-major axis of the ellipse is determined by the singularity closest to the origin in the $z$-plane, $z_s^{\mathrm{SL}}$, at given $s>s\ped{th}$.
For scattering amplitudes, there are two natural candidates, namely the first singularities in the $t$ and $u$ channels, respectively, given by $t=t\ped{th}$ and $u=u\ped{th}$.
They respectively give $z_s(t\ped{th})>1$ and $z_s(u\ped{th})<-1$.
For identical particles, $t-u$ crossing implies $  z_s^{\mathrm{SL}}=|z_s(t\ped{th})|=|z_s(u\ped{th})|$, while for unequal particles one has
\begin{equation}
    z_s^{\mathrm{SL}}=\min\big(|z(t=t\ped{th},s)|,|z(u=u\ped{th},s)|\big) \,.
\end{equation}
For the $AB\to AB$ scattering amplitude, $t\ped{th}=4m^2$, $u\ped{th}=m_+^2$, and it is easily verified that, for any $s>s\ped{th} = m_+^2$, the singularity closest to the origin is the one given by $t=4m^2$.
Using~\eqref{eq:PosABAB}, we find
\begin{equation}\label{eq:smallLSABAB}
    z_s^{\mathrm{SL}}=  z(t=t\ped{th},s) = 1+\frac{8sm^2}{(s-m_+^2)(s-m_-^2)}\,,  \qquad  AB\to AB \,.
\end{equation}
For $AA\to BB$ scattering, $t-u$ crossing fixes $u\ped{th} = t\ped{th} = m_+^2$, and
from~\eqref{eq:PosAABB} we get
\begin{equation}\label{eq:smallLSAABB}
    z_s^{\mathrm{SL}}  =  \frac{4 m M +s}{\sqrt{(s-4m^2)(s-4M^2)}} \,,  \qquad AA\to BB\,.
\end{equation}

\subsection{Double discontinuities and large Lehmann ellipses}\label{subsec:partwavExpdD}

When the function $f$ in section~\ref{app:ellipses} is identified with the $s$-discontinuity of the amplitude,
the convergence of the partial wave expansion defines the so-called large Lehmann ellipse (LL).
Its boundary coincides with the appearance of double discontinuities in the amplitude.
We define the double $s-t$ discontinuity of an amplitude $T(s,t,u)$ as
\begin{align}
  \ddisc_{st} T
  & \equiv - \frac 14 \lim_{\epsilon\to 0} \Big( T(s+i \epsilon,t+i \epsilon) -  T(s-i \epsilon,t+i \epsilon) -  T(s+i \epsilon,t-i \epsilon)+ T(s-i \epsilon,t-i \epsilon)\Big)\nn  \\
  & = \disc_t \disc_s T(s,t) =  \disc_s \disc_t  T(s,t) =\ddisc_{ts}  T \,,
\end{align}
with \( T(s,t) \) as in~\eqref{eq:Tst-def} and where we adopt the convention~\eqref{eq:disc-def} for the discontinuity.
Similarly the double $s-u$ discontinuity is defined as
\begin{align}
  \ddisc_{su} \widetilde T
  & \equiv - \frac 14 \lim_{\epsilon\to 0} \Big(\widetilde T(s+i \epsilon,u+i \epsilon) - \widetilde T(s-i \epsilon,u+i \epsilon) - \widetilde T(s+i \epsilon,u-i \epsilon)+\widetilde T(s-i \epsilon,u-i \epsilon)\Big) \nn \\
  & = \disc_u \disc_s \widetilde T(s,u) =  \disc_s \disc_u \widetilde T(s,u) =\ddisc_{us} \widetilde T\,,
\end{align}
where $\widetilde{T}(s,u) = T(s,\Sigma-s-u,u)$.
We also have a double $t-u$ discontinuity $\ddisc_{tu}$, but this vanishes for $s>s\ped{th}$ and will not be considered in what follows.

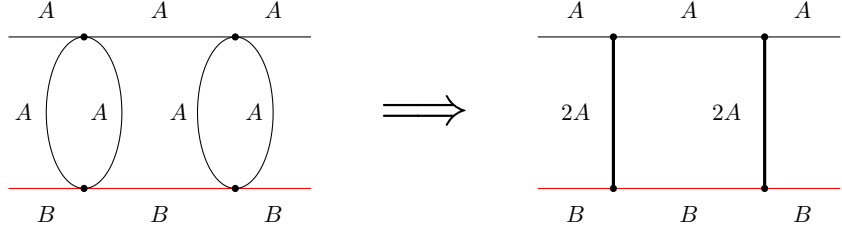
\begin{figure}[t!]
	\centering
	% \raisebox{-3.em}
	\begin{tikzpicture}
		\draw (-2, 1) to (2, 1);
		\draw  [red]  (-2, -1) to (2, -1);
		\filldraw[fill=black, draw=black]  (-1.,1.) circle (0.04 cm);
		\filldraw[fill=black, draw=black]  (-1.,-1.) circle (0.04 cm);
		\filldraw[fill=black, draw=black]  (1.,1.) circle (0.04 cm);
		\filldraw[fill=black, draw=black]  (1.,-1.) circle (0.04 cm);
		\draw (-1,0) ellipse (0.5cm and 1cm);
		\draw (1,0) ellipse (0.5cm and 1cm);
		% \draw (-1,1) arc [start angle=60, end angle=0, x radius=1, y radius=2];
		% \draw (8.5, -1) to (9., -1.);
		\node [below] at (1.5,1.6) {\scalebox{0.8}{$A$}};
		\node [below] at (0,1.6) {\scalebox{0.8}{$A$}};
		\node [below] at (-1.5,1.6) {\scalebox{0.8}{$A$}};
		\node [below] at (-1.8,0.25) {\scalebox{0.8}{$A$}};
		\node [below] at (-0.8,0.25) {\scalebox{0.8}{$A$}};
		\node [below] at (1.25,0.25) {\scalebox{0.8}{$A$}};
		\node [below] at (0.25,0.25) {\scalebox{0.8}{$A$}};
		\node [below] at (1.5,-1.1) {\scalebox{0.8}{$B$}};
		\node [below] at (-1.5,-1.1) {\scalebox{0.8}{$B$}};
		\node [below] at (0,-1.1) {\scalebox{0.8}{$B$}};

		\node [below] at (5.5,0.25) {\scalebox{0.8}{$2A$}};
		\node [below] at (7.5,0.25) {\scalebox{0.8}{$2A$}};
		\node [below] at (3.5,0.25) {\scalebox{1.9}{$\Longrightarrow$}};

		\draw (5, 1) to (9, 1);
		\draw  [red]  (5, -1) to (9, -1);
		\filldraw[fill=black, draw=black]  (6.,1.) circle (0.04 cm);
		\filldraw[fill=black, draw=black]  (6.,-1.) circle (0.04 cm);
		\filldraw[fill=black, draw=black]  (8.,1.) circle (0.04 cm);
		\filldraw[fill=black, draw=black]  (8.,-1.) circle (0.04 cm);
		\draw[very thick] (8,-1) -- (8,1);
		\draw[very thick] (6,-1) -- (6,1);
		% \draw (-1,1) arc [start angle=60, end angle=0, x radius=1, y radius=2];
		% \draw (8.5, -1) to (9., -1.);
		\node [below] at (7,1.6) {\scalebox{0.8}{$A$}};
		\node [below] at (7.,-1.1) {\scalebox{0.8}{$B$}};
		\node [below] at (8.5,1.6) {\scalebox{0.8}{$A$}};
		\node [below] at (5.5,1.6) {\scalebox{0.8}{$A$}};
		\node [below] at (8.5,-1.1) {\scalebox{0.8}{$B$}};
		\node [below] at (5.5,-1.1) {\scalebox{0.8}{$B$}};
	\end{tikzpicture}
	\caption{Example of leading box diagram contributing to the double discontinuity of the $AB\to AB$ amplitude (left) and its reduced box diagram (right). As in figure~\ref{fig:TriangularL2}, the thick vertical lines represent effective propagators for a state of $2A$ particles at threshold.}
	\label{fig:L3}
\end{figure}

The onset of the double discontinuity can be detected by considering reduced Feynman graphs, see e.g.~\cite{Pelaez:2020gnd} for the specific case of pion-kaon scattering.
Similarly to the analysis of triangular diagrams made in section~\ref{sec:analytic-structure}, the leading topologies contributing to \( \ddisc{} \) can be reduced to effective box diagrams, as illustrated in figure~\ref{fig:L3}.
As for the small Lehmann ellipse, in the absence of $t-u$ crossing symmetry ($AB\to AB$), we should check if the singularity closest to the origin in the $z$-plane, $z_s^\text{LL}$, is given by \( \ddisc_{st} \) or \( \ddisc_{su} \).
For any $s>m_+^2$, one finds that ${\ddisc}_{st}$ dominates and eventually
\begin{equation}\label{eq:largeLSABAB}
    z_s^\text{LL} = 1+ \frac{2s \, t^{\text{LL}}(s)}{\big(s-m_+^2\big)\big(s-m_-^2\big)}\,, \qquad AB\to AB\,,
\end{equation}
with
\begin{equation}\label{eq:tKABAB}
    t^{\text{LL}}=  \min (t_1, t_2)  =
    \begin{cases}
      t_1 & s\leq s_*  \\
      t_2 & s\geq s_* \\
    \end{cases}
    \,.
\end{equation}
In~\eqref{eq:tKABAB},
\begin{equation}\label{eq:t1t2s0ABAB}
    \begin{aligned}
      t_1(s)  & = 16 m^2 + \frac{64m^4 s}{\big(s-m_+^2\big)\big(s-m_-^2\big)}\,, \qquad s>m_+^2\,, \\
      t_2(s) & = 4m^2 + \frac{32m^3 (M+m)}{s-(M+3m)^2}\,,  \qquad \qquad s> s_{4p}=(3m+M)^2\,,
    \end{aligned}
\end{equation}
are the leading Karplus curves~\cite{Karplus:1959zz}, and
\begin{equation}\label{eq:t1t2s0ABABv2}
    s_*  = M^2+4 Mm + 5 m^2+2m\sqrt{5M^2+12 Mm+8m^2} > s_{4p}\,.
\end{equation}

\begin{figure}[t!]
    \centering
    \includegraphics[width=0.48\linewidth]{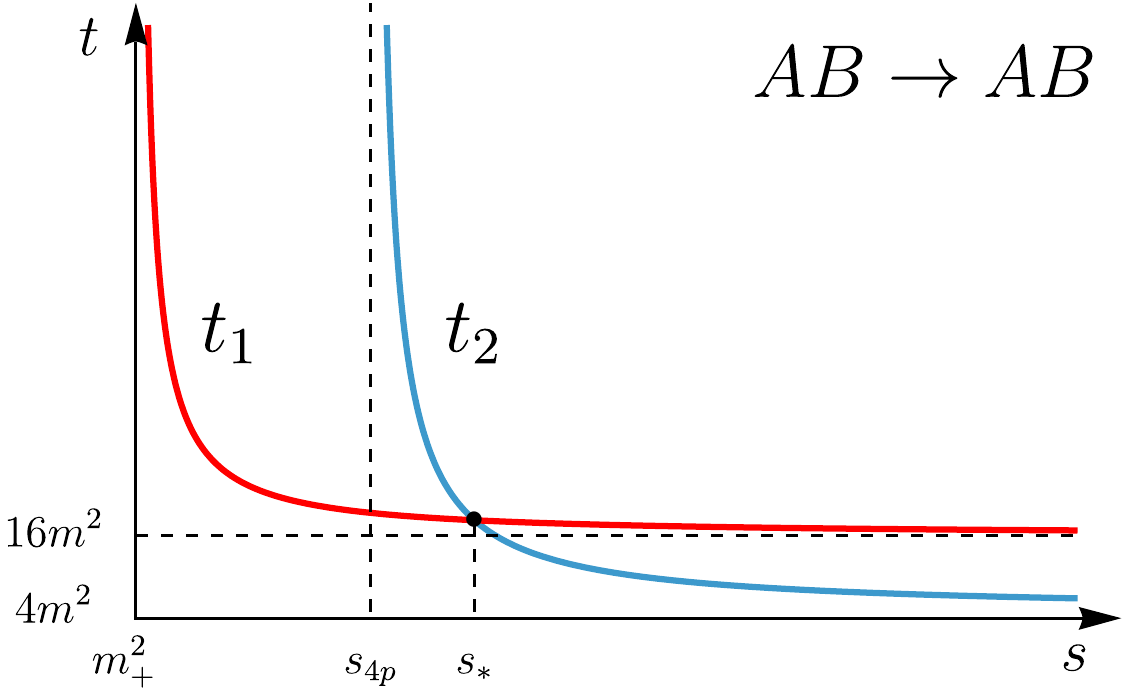}
    \hspace{0.5em}
    \includegraphics[width=0.48\linewidth]{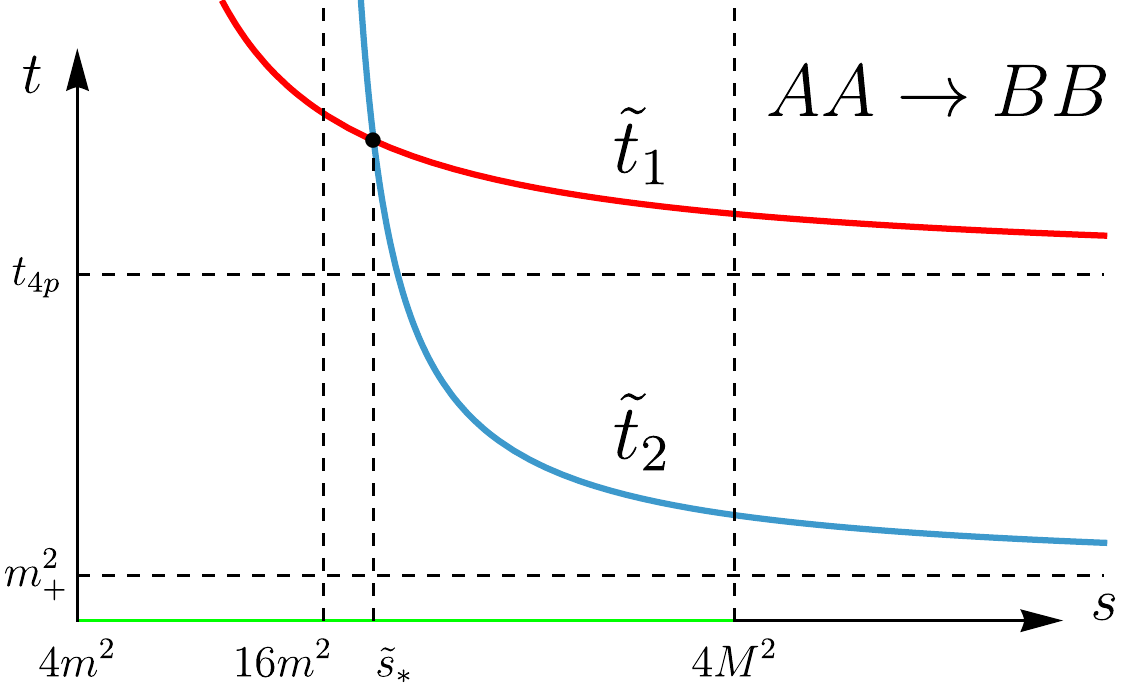}
    \caption{Onset of the double discontinuity in the real $(s,t)$ plane. (left) The two curves intersect at $s_*$, reported in~\eqref{eq:t1t2s0ABABv2}. The point $s_{4p}=
        (3m+M)^2$ is the threshold for 4-particle production. (right) The two curves intersect at $\tilde s_*$, reported in~\eqref{eq:t1t2s0AABB}, while $t_{4p}=s_{4p}$. The green line represents the pseudo-physical region. The plots shown correspond to $M=3m$.\label{fig:dDiscBoth}}
\end{figure}

As expected, for $M=m$,~\eqref{eq:t1t2s0ABAB} reduces to the leading Karplus curves for identical particles, see e.g.~\cite{Correia:2020xtr}.

For $AA\to BB$ scattering, $\ddisc_{st} =\ddisc_{su}$.
The leading singularity $z_s^\text{LL}$ is detected by considering the same Feynman graph relevant for $AB\to AB$ scattering, but with $s\leftrightarrow t$.
The final result is that, for $s>4M^2$,
\begin{equation}\label{eq:largeLSAABB}
    z_s^\text{LL} = \frac{2t^{\text{LL}}(s) +s-2m^2-2M^2}{\sqrt{(s-4m^2)(s-4M^2)}} \,,\qquad AA\to BB\,,
\end{equation}
where
\begin{equation}\label{eq:tKAABB}
    t^{\text{LL}}=  \min (\tilde{t}_1, \tilde{t}_2)  =
    \begin{cases}
      \tilde{t}_1 & s\leq \tilde{s}_*  \\
      \tilde{t}_2 & s\geq \tilde{s}_* \\
    \end{cases}
    \,,
\end{equation}
with
\begin{align}
  \tilde t_1(s) & = (3m+M)^2 + \frac{32m^3 (M+m)}{s-4m^2}\,, \quad s > 4m^2\,, \nn  \\
  \tilde t_2(s) & = \frac{16m^4+M^2s +m^2(s-16M^2)+2m \sqrt{s(sM^2-16m^2 M^2+16m^4)}}{s-16m^2}\,,  \quad s > 16m^2\,,\nn   \\
  \tilde s_* & = \frac{4m^2}{m+M}\left(3m+2M+\sqrt{5M^2+12 Mm+8m^2}\right) > 16 m^2\,.  \label{eq:t1t2s0AABB}
\end{align}
The leading Karplus curves for identical particles are again reproduced for $M=m$, as expected.

We report in figure~\ref{fig:dDiscBoth} the leading Karplus curves beyond which ${\ddisc}_{st}$ is non-vanishing for both $AB\to AB$ and $AA\to BB$ scattering amplitudes.
The picture shows the case $M=3m>2m$.
While the detailed form of the curves depends on $\mu$, their qualitative behaviour applies for any $\mu$. In contrast, the higher Karplus curves depend sensitively on $\mu$.

\subsection{Circular cut of the \texorpdfstring{$AB\to AB$}{AB to AB} partial waves}

One of the consequences of pseudo-physical regions is to give rise to a circular branch cut singularity in the complex $s$-plane for the partial waves $f^\ell_{AB\to AB}(s)$.
The emergence of this cut can be seen from~\eqref{eq:PosABAB}, rewritten as
\begin{equation}\label{eq:NonAn}
    t = \frac{z-1}{2s} \big(s-m_+^2\big)\big(s-m_-^2\big)\,.
\end{equation}
Partial waves are obtained by integrating over $z\in [-1,1]$, namely from $t(-1)$ to $t(1)=0$.
As $s$ is varied, it can happen that the integration over $z$ hits a value of $t$ where the amplitude is non-analytic.
The first singularity occurs at $t\ped{th} = 4m^2$.
Since $t\ped{th}$ is real, we can first determine for which complex values of $s$, $t(-1)$ is real.
Setting $s = r \exp(i \omega)$ and demanding $\Im t(z=-1)=0$ gives
\begin{equation}
    \sin \omega \Big(r - \frac{(M^2-m^2)^2}{r}\Big) = 0\,.
\end{equation}
This is satisfied either for $\omega = 0,\pi$, corresponding to real $s$, or $r = M^2-m^2$ and any $\omega$.
The second option corresponds to the circular cut.
Along it, we have
\begin{equation}
    t(z) =  (z-1) \Big[(M^2-m^2)\cos \omega - (M^2+m^2)\Big] \,.
\end{equation}
At $z=-1$, as $\omega$ varies from $0$ to $\pi$, $t(-1)$ ranges from $4m^2$ to $4M^2$, precisely the pseudo-physical region, signalling that indeed the integration over $z\in [-1,1]$ hits the singularity along this circle.
So, partial waves $f^\ell_{AB\to AB}(s)$, in addition to the usual physical branch cuts over the real axis, exhibit non-analyticities along a circular cut encircling the origin with radius
\begin{equation}
    R\ped{cut} = M^2-m^2\,.
\end{equation}
For equal particles the circle shrinks to the point $s=0$.
In our numerical study, partial waves are evaluated for physical values of $s$, so away from this circular cut.
However, in arguments based on dispersion relations, or for extrapolating the form of extremal partial waves in the complex $s$-plane, care must be taken with it.

\section{Dispersion relations, positivity and linearised unitarity}\label{app:positivity}
In this appendix we derive dispersive representations for some of the non-perturbative couplings \( \lambda_{k,l} \) defined in section~\ref{sec:observables} and obtain bounds using ``positivity techniques''. We further improve some of these bounds using linearised unitarity, both analytically and numerically.

In this context we follow the strategy outlined in~\cite{Caron-Huot:2020cmc}, properly generalised to unequal particles. In contrast to~\cite{Caron-Huot:2020cmc}, however, we do not rely on a Lagrangian description for the definition of the couplings, but rather use the non-perturbative definition~\eqref{eq:def-couplings}, as done in~\cite{Chen:2022nym}. The bounds so obtained are then exact and not restricted to tree-level amplitudes.
Positivity bounds on scattering of non-identical scalars have also been explored in~\cite{deRham:2025htd}, but considering the full system of two-to-two scattering amplitudes and restricting to particular ranges of mass ratios. We will instead present bounds that use only the \( AB \to AB \) amplitude but are valid for any mass ratio.

\paragraph{Dispersion relation.}
We are interested in writing fixed-\( t \) dispersion relations in the complex \( s \)-plane for the amplitude \( T(s,t) \) in~\eqref{eq:Tst-def}. The starting point is the Cauchy integral with two subtractions
\begin{equation}\label{eq:ABtoAB-cauchy}
    \oint_\gamma \frac{\dif{s'}}{2\pi i} \frac{T(s',t)}{\prod_{i=1}^3(s'-z_i)} = 0 \,,
\end{equation}
where $\gamma$ is an arbitrary closed contour around a generic complex point in the $s$-plane of figure~\ref{fig:ABtoAB-analytic-s}, where $T$ is analytic, and away from the subtraction points $z_i$, which we will fix momentarily. Blowing up the contour, we can neglect the arc at infinity thanks to the Froissart bound and the two subtractions. We then pick up the residues at \( s' = z_{i} \) as well as the Hankel contours on the branch cuts. We obtain
\begin{equation}\label{eq:ABtoAB-dispersive-step}
    \sum_{i=1}^3 \res_{s' = z_{i}} \frac{T(s',t)}{\prod_{j=1}^3(s'-z_j)}
    = \int_{-\infty}^{m_{-}^{2}-t} \frac{\dif{s'}}{\pi} K(s') \disc_s T(s',t)
    + \int^{+\infty}_{m_{+}^{2}} \frac{\dif{s'}}{\pi} K(s') \disc_s T(s',t) \,,
\end{equation}
where we define
\begin{equation}\label{eq:disc-def}
    \disc_{s} T(s, t) =\frac{1}{2i} \Big( \lim_{\epsilon \to 0^{+}} \big(T(s+i \epsilon,t) - T(s-i \epsilon,t)\big)\Big) =  \Im T(s, t)
\end{equation}
by real analyticity and \( K \) is the kernel
\begin{equation}\label{eq:K-kernel}
    K(s') = \frac{1}{(s'-z_{1})(s'-z_{2})(s'-z_{3})} \,.
\end{equation}
Using \( s-u \) crossing, \( T(s,t) = T(\Sigma-s-t,t) \), we can combine the first integral with the second one after changing variable \( s' \to \Sigma - s' - t\). After simple manipulations, we get
\begin{equation}\label{eq:ABtoAB-dispersion-s}
    \sum_{i=1}^3 \res_{s' = z_{i}} \frac{T(s',t)}{\prod_{j=1}^3(s'-z_j)} = \int^{+\infty}_{m_{+}^{2}} \frac{\dif{s'}}{\pi} \, \left(K(s') - K(\Sigma - s' - t)\right) \Im T(s',t) \,.
\end{equation}

\paragraph{Dispersive formulas for the couplings.}
By appropriately choosing \( z_{i} = z_{i}(s,t) \) to be linear functions and by Taylor expanding~\eqref{eq:ABtoAB-dispersion-s} around the point \( (s_{0}, t_{0}) \), we can obtain dispersive representations for some of the couplings \( \lambda_{k,l} \) defined in~\eqref{eq:def-couplings}.
When taking the residues and Taylor expanding at \( (s,t) = (s_{0}, t_{0}) \) the left-hand side of~\eqref{eq:ABtoAB-dispersion-s}, we generate derivatives of \( T(s,t) \) at the point \( (s, t) = (z_{i}(s_{0}, t_{0}), t_{0}) \). In order to extract the couplings \( \lambda_{k,l} \) we then want to choose \( z_{i}(s_{0}, t_{0}) = s_{0} \). A convenient choice, which does not produce accidental cancellations and respects \( s-u \) symmetry, is the following
\begin{equation}\label{eq:subtraction-choice}
    z_{1} = s_{0} \,, \qquad
    z_{2} =  s_{0} + \hat{s} + \hat{t} = s + t - t_{0}\,, \qquad
    z_{3} = s_{0} + \hat{s} - \hat{t} = s - t + t_{0} \,,
\end{equation}
where we have defined shifted Mandelstam variables
\begin{equation}
    \hat{s} = s - s_{0} \,, \qquad
    \hat{t} = t - t_{0} \,.
\end{equation}
Let us motivate the choice for \( s_{0} \) in~\eqref{eq:s0-t0-def}. For the dispersion relations to be valid, the points \( z_{i} \) must lie away from the branch cuts; moreover we want the branch cuts in figure~\ref{fig:analytic-structure} not to overlap. Working locally around \( (s,t) \sim (s_{0}, t_{0}) \) the conditions are
\begin{equation}
    m_{-}^{2} - t_{0} < s_{0} < m_{+}^{2} \,, \qquad
    m_{-}^{2} - s_{0} < t_{0} < 4m^{2} \,.
\end{equation}
Fixing \( t_{0} = \Sigma - 2 s_{0} \) to preserve \( s-u \) symmetry, and using \( m \leq M \) in the previous constraint, we obtain
\begin{equation}\label{eq:s0-constraint}
    M^{2} - m^{2} < s_{0} < m_{+}^{2} \,.
\end{equation}
Looking for an expression for \( s_{0} \) linear in \( m^{2} \) and \( M^{2} \), that reduces to the crossing-symmetric point \( 4m^{2}/3 \) in the case of equal masses, fixes \( s_{0} = 4m^{2}/3 + b(M^{2} - m^{2}) \). The constraint~\eqref{eq:s0-constraint} is then solved for any value of the masses by \( b = 1 \), which reduces \( s_{0} \) to~\eqref{eq:s0-t0-def}.

Let us now go back to~\eqref{eq:ABtoAB-dispersion-s}, plug in the subtraction points~\eqref{eq:subtraction-choice} and expand in \( (\hat{s}, \hat{t}) \) the left-hand side (LHS). Up to second order in the expansion, we find
\begin{equation}
    \text{LHS}~\eqref{eq:ABtoAB-dispersion-s}
    = \lambda_{2,0} + \lambda_{2,1} \hat{t} + 3 \lambda_{4,0} \hat{s}^{2} + (\lambda_{2,2} + \lambda_{4,0}) \hat{t}^{2} + 4 \lambda_{4,0} \hat{s} \hat{t} + \ldots
\end{equation}
We have set \( m = 1 \) in the previous formulas for simplicity. In writing this expansion we have already accounted for the constraints from \( s-u \) crossing symmetry on the couplings \( \lambda_{k,l} \), e.g.\ the ones reported in~\eqref{eq:couplings-crossing-redundancies}. Note that not all couplings appear in this expansion; those that do not are thus unconstrained by this procedure.\footnote{%
    For example the couplings \( \lambda_{0,l} \) do not appear, but admit instead a dispersive representation using fixed-\( s \) dispersion relations.
}

Performing now the same expansion on the right-hand side (RHS) of~\eqref{eq:ABtoAB-dispersion-s} we obtain
\begin{equation}
    \text{RHS}~\eqref{eq:ABtoAB-dispersion-s}
    = \int_{m_{+}^{2}}^{\infty} \frac{\dif{s'}}{\pi}\left[\frac{2 \Im T(s',t_{0})}{(s'-s_{0})^{3}} +\left( \frac{2 \partial_{t} \Im T(s', t_{0})}{(s'-s_{0})^{3}} - \frac{3 \Im T(s', t_{0})}{(s'-s_{0})^{4}}\right) \hat{t}' + \ldots \right] \,.
\end{equation}
For simplicity, we have reported only the first two orders of the expansion. We remark that in these formulas it is important to use the relation \( \Sigma = 2 s_{0} + t_{0} \) to obtain simple expressions.

By matching the two sides we find dispersive representations for the couplings. For example, up to linear order in \( (\hat{s}, \hat{t}) \), we obtain
\begin{equation}\label{eq:couplings-dispersive-formulas}
    \begin{aligned}
      \lambda_{2,0} &= \int_{m_{+}^{2}}^{\infty} \frac{\dif{s}}{\pi} \frac{2 \Im T(s,t_{0})}{(s-s_{0})^{3}} \,, \\[0.5em]
      \lambda_{2,1} &= \int_{m_{+}^{2}}^{\infty} \frac{\dif{s}}{\pi}\left( \frac{2 \partial_{t} \Im T(s, t_{0})}{(s-s_{0})^{3}} - \frac{3 \Im T(s, t_{0})}{(s-s_{0})^{4}}\right) \,.
    \end{aligned}
\end{equation}
We have dropped the primes for readability. The expansion can be carried out systematically at higher order but the expressions become increasingly complicated. An exception is given by \( \lambda_{2k,0} \) for which we find
\begin{equation}
    \lambda_{2k,0} = \int_{m_{+}^{2}}^{\infty} \frac{\dif{s}}{\pi} \frac{2 \Im T(s,t_{0})}{(s-s_{0})^{2k+1}} \,, \qquad k \geq 1 \,.
\end{equation}
We remark that, when matching powers of \( \hat{s} \) and \( \hat{t} \) on each dispersion relation and solving for the couplings, we find the linear system to be of maximal rank (at least up to order \( 8 \) in the expansion). This is in contrast to the case of dispersion relations for identical scalars, where full crossing symmetry leads to an over-determined system and so-called ``null constraints'' on dispersive integrals of \( \Im T(s,t) \).

\paragraph{Positivity bounds.} As a consequence of unitarity~\eqref{eq:unitarity-ABtoAB}, the imaginary part of the \( AB \to AB \) partial amplitude is positive
\begin{equation}
    \Im f_{AB \to AB}^{\ell}(s) \geq 0 \,, \qquad s \geq m_{+}^{2} \,, \ \ell = 0, 1, 2, \ldots \,.
\end{equation}
This property can be used to obtain bounds on the couplings \( \lambda_{k,l} \) that admit dispersive representations derived from~\eqref{eq:ABtoAB-dispersion-s}. We plug in~\eqref{eq:couplings-dispersive-formulas} the partial wave decomposition~\eqref{eq:partial-wave-decomposition} of \( T(s,t) = T_{AB \to AB}(s,t,u) \) and cast the dispersive representation as an average over a positive measure
\begin{equation}\label{eq:positivity-average}
    \Braket{X_{\ell}(s)} = \sum_{\ell=0}^{\infty} \frac{2\ell+1}{2} \int_{m_{+}^{2}}^{\infty} \frac{\dif{s}}{\pi} \Im f_{AB \to AB}^{\ell}(s) \, X_{\ell}(s) \,.
\end{equation}
For example, for the \( \lambda_{2,0} \) and \( \lambda_{2,1} \) couplings in~\eqref{eq:couplings-dispersive-formulas} we get
\begin{align}
  \lambda_{2,0} &= \Braket{\frac{2 P_{\ell}(z(s, t_{0}))}{(s-s_{0})^{3}}} \,, \label{eq:L20-positivity-representation} \\[0.5em]
  \lambda_{2,1} &= \Braket{\frac{4 s P_{\ell}'(z(s,t_{0}))}{(s-s_{0})^{3}(s-m_{+}^{2})(s-m_{-}^{2})}} - \Braket{\frac{3 P_{\ell}(z(s,t_{0}))}{(s-s_{0})^{4}}} \,, \label{eq:L21-positivity-representation}
\end{align}
with \( z(s,t) \) given in~\eqref{eq:PosABAB} and \( P_{\ell}'(z) = \partial_{z} P_{\ell}(z) \). Note that \( s \geq m_{+}^{2} \geq s_{0} \) and \( z(s, t_{0}) \geq 1 \) in the integration region of the average~\eqref{eq:positivity-average}. The Legendre polynomials satisfy \( P_{\ell}(z) \geq 0 \) for \( z \geq 1 \) and therefore the averaged object in~\eqref{eq:L20-positivity-representation} is positive. By positivity of the measure we conclude that
\begin{equation}
    \lambda_{2,0} \geq 0 \,.
\end{equation}
The two terms in~\eqref{eq:L21-positivity-representation} are separately positive, but we cannot derive an absolute bound in a similar way because of the minus sign. We can notice, however, that \( 1/(s-s_{0}) \) is monotonically decreasing for \( s \geq m_{+}^{2} \) and reaches its maximum at \( s = m_{+}^{2} \), where \( 1/(m_{+}^{2}-s_{0}) = 3/(2+6 \mu) \). Using this inequality in~\eqref{eq:L21-positivity-representation} we conclude
\begin{equation}\label{eq:L21-L20-positivity-bound-analytic}
    \lambda_{2,1} \geq - \Braket{\frac{3 P_{\ell}(z(s,t_{0}))}{(s-s_{0})^{4}}} \geq  -\frac{9}{4(1+3 \mu)} \Braket{\frac{2 P_{\ell}(z(s,t_{0}))}{(s-s_{0})^{3}}} = -\frac{9}{4(1+3 \mu)} \lambda_{2,0} \,.
\end{equation}
Bounds on higher-order couplings can be systematically obtained using moment optimisation as described in~\cite{Chen:2022nym}, which shows that the analytic bound~\eqref{eq:L21-L20-positivity-bound-analytic} is optimal.
In contrast to the case of identical scalars, in the absence of null constraints, we do not find any numerical upper bound on the ratio \( \lambda_{2,1}/\lambda_{2,0} \).

\paragraph{Linearised unitarity.} Unitarity implies not only that the partial amplitudes are positive, but also that they are bounded from above. In formulas, it implies the linear constraint \( 0 \leq \Im T^{\ell}_{AB \to AB}(s) \leq 2 \), which, recalling~\eqref{eq:partial-amplitude-Tl}, reads
\begin{equation}\label{eq:linearised-unitarity}
    0 \leq  \Im f^{\ell}_{AB \to AB}(s) \leq n(s)\,,\qquad
    n(s) \equiv \frac{2}{\mathcal{N}_{AB\to AB}(s)} \,, \qquad
    s \geq m_{+}^{2} \,, \ \ell = 0,1,2,\ldots
\end{equation}
We can now observe that in~\eqref{eq:L21-positivity-representation} the combination
\begin{equation}
    V_{2,1}^{\ell}(s) \equiv \frac{4 s P_{\ell}'(z(s,t_{0}))}{(s-s_{0})^{3}(s-m_{+}^{2})(s-m_{-}^{2})} - \frac{3 P_{\ell}(z(s,t_{0}))}{(s-s_{0})^{4}}
\end{equation}
is negative for \( \ell = 0 \), where it collapses to \( -3/(s-s_{0})^{4} \), and positive for \( \ell \geq 1 \) and for \( s \geq m_{+}^{2} \). This implies that for \( \ell = 0 \)
\begin{equation}
    \Im f^{0}_{AB \to AB}(s) V_{2,1}^{0}(s) \geq n(s) V_{2,1}^{0}(s) \,, \qquad s \geq m_{+}^{2} \,.
\end{equation}
Using this inequality and positivity of the \( \ell \geq 1 \) partial amplitudes we conclude
\begin{equation}\label{eq:L21-linearised-unitarity-analytic}
    \lambda_{2,1} \geq \frac{1}{2}\int_{m_{+}^{2}}^{\infty} \frac{\dif{s}}{\pi} n(s) V_{2,1}^{0}(s) = h(\mu) \,,
\end{equation}
where \( h(\mu) \) is a definite \( \mu \)-dependent integral
\begin{equation}\label{eq:L21-lower-bound-integral}
    h(\mu) \equiv -\int_{0}^{\infty}\dif{x}\,\frac{24\left(x+(\mu +1)^2\right)}{\sqrt{x \left(\frac{x}{4}+\mu\right)} \left(x+2 \mu+\frac{2}{3}\right)^4} \,,
\end{equation}
which does not have a particularly nice analytic expression but can be quickly evaluated numerically.

We can now combine~\eqref{eq:L20-positivity-representation},~\eqref{eq:L21-positivity-representation} and~\eqref{eq:linearised-unitarity} and moment optimisation, to obtain a bound on \( \lambda_{2,1} \) as a function of \( \lambda_{2,0} \) that is stronger than just the maximum between~\eqref{eq:L21-L20-positivity-bound-analytic} and~\eqref{eq:L21-linearised-unitarity-analytic}. To this end, let us also define
\begin{equation}
    V_{2,0}^{\ell}(s) \equiv \frac{2 P_{\ell}(z(s, t_{0}))}{(s-s_{0})^{3}} \,.
\end{equation}
We want to solve a moment problem whose primal form reads
\begin{equation}
    \begin{aligned}
      &\min \ \sum_{\ell} \int \dif{s} \, w_{\ell}(s) V_{2,1}^{\ell}(s)  \\
      &\text{such that} \quad \sum_{\ell} \int \dif{s}\, w_{\ell}(s) V_{2,0}^{\ell}(s) = \Lambda \,, \\
      & \qquad \qquad \quad \ 0 \leq w_{\ell}(s) \leq \tfrac{2\ell+1}{2\pi} n(s) \,, && s \geq m_{+}^{2} \,, \ \ell = 0,\ldots,\ell_{*} \,, \\
      & \qquad \qquad \quad \ 0 \leq w_{\ell}(s) \,, && s \geq m_{+}^{2} \,, \ \ell = \ell_{*}+1,\ell_{*} + 2,\ldots \,,
    \end{aligned}
\end{equation}
where \( w_{\ell}(s) = \tfrac{2\ell+1}{2\pi}\Im f^{\ell}_{AB \to AB}(s) \) is a measure, \( \Lambda \) is a fixed value for \( \lambda_{2,0} \) and we have suppressed for simplicity the ranges of the sum and integration. For later convenience we impose the linearised unitarity upper bound only on a finite number of spins \( \ell_{*} \).
It is useful to consider the dual problem
\begin{equation}\label{eq:linearised-unitarity-dual-problem}
    \begin{aligned}
      & \max \ \Lambda y - \sum_{\ell=0}^{\ell_{*}} \frac{2\ell+1}{2} \int\frac{\dif{s}}{\pi} \, p_{\ell}(s) n(s) \,, \\
      & \text{such that} \quad V_{2,1}^{\ell}(s) - V_{2,0}^{\ell}(s) y + p_{\ell}(s) \geq 0 \,, && s \geq m_{+}^{2} \,, \ \ell = 0,\ldots,\ell_{*} \,, \\
      & \qquad \qquad \quad V_{2,1}^{\ell}(s) - V_{2,0}^{\ell}(s) y \geq 0 \,, && s \geq m_{+}^{2} \,, \ \ell = \ell_{*}+1,\ell_{*} + 2,\ldots \\
      & \qquad \qquad \quad p_{\ell}(s) \geq 0 \,, && s \geq m_{+}^{2} \,, \ \ell = 0,\ldots,\ell_{*}
    \end{aligned}
\end{equation}
Upon discretising the values of \( s \) and putting a cutoff on \( \ell \), this linear optimisation problem can be solved numerically using either Mathematica's \texttt{SemidefiniteOptimization} or \texttt{SDPB}. This becomes, however, somewhat hard to solve and ill-conditioned as we increase the number of sampling points and try to increase \( \ell_{*} \). A better strategy is to eliminate the ``slack variables'' \( p_{\ell}(s) \) by solving the inequality constraint
\begin{equation}\label{eq:slack}
    p_{\ell}(s) \geq \left[V_{2,0}^{\ell}(s) y - V_{2,1}^{\ell}(s)\right]_{+} \,, \qquad \left[x\right]_{+} \equiv \max\{0, x\} \,.
\end{equation}
Next we notice that $V^\ell_{2,0}(s) \geq 0$ and that $r_{\ell}(s) \equiv V^\ell_{2,1}(s)/V^\ell_{2,0}(s) \geq 0$ and goes to $0$ as $s \to +\infty$ for any $\ell \geq 1$. The positivity constraints for \( \ell \geq \ell_{*} + 1 \) in~\eqref{eq:linearised-unitarity-dual-problem} require \( y \leq r_{\ell}(s) \) and since the right-hand side tends to $0$ as $s \to +\infty$, we conclude that $y \leq 0$. Substituting the optimal slack variable~\eqref{eq:slack} into the objective, the dual problem~\eqref{eq:linearised-unitarity-dual-problem} is equivalent to
\begin{equation}
    \max_{y \leq 0} \left(\Lambda\, y
        - \sum_{\ell=0}^{\ell_{*}} \frac{2\ell+1}{2} \int \frac{\mathrm{d}s}{\pi}\, n(s) \left[ V^\ell_{2,0}(s)\, y - V^\ell_{2,1}(s) \right]_+ \right)\,.
\end{equation}
Finally, we observe that the problem is independent of $\ell_{*}$ since \( \left[ V^\ell_{2,0}(s)\, y - V^\ell_{2,1}(s) \right]_+ = 0 \) for \( \ell \geq 1 \) and \( y \leq 0 \), using again that \( V^\ell_{2,0}(s) \geq 0 \) and \( r_{\ell}(s) \geq 0 \) in this domain. Thus only the $\ell = 0$ term survives.

\begin{figure}[t]
    \centering
    \includegraphics[width=0.7\linewidth]{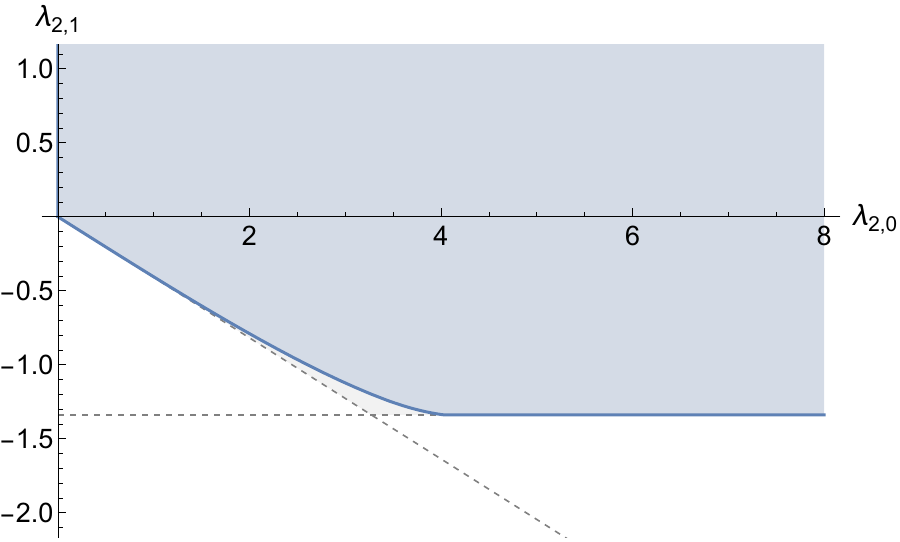}
    \caption{Bound on \( (\lambda_{2,0}, \lambda_{2,1}) \) from linearised unitarity~\eqref{eq:linearised-unitarity} at \( \mu = 1.5 \). The dashed lines are the analytic bounds~\eqref{eq:L21-L20-positivity-bound-analytic} and~\eqref{eq:L21-linearised-unitarity-analytic}, while the solid blue line is the bound~\eqref{eq:linearised-unitarity-bound} which can be computed numerically. The shaded blue area is allowed.\label{fig:linearised-unitarity-bound}}
\end{figure}

To summarise, the optimal bound from linearised unitarity on \( \lambda_{2,1} \) at fixed \( \lambda_{2,0} \) is given by
\begin{equation}\label{eq:linearised-unitarity-bound}
    \lambda_{2,1} \geq \max_{y\leq 0}\left(\lambda_{2,0} \, y - \int_{m_{+}^{2}}^{\infty}\frac{\dif{s}}{2\pi} \, n(s) \Big[V_{2,0}^{0}(s) y - V_{2,1}^{0}(s)\Big]_{+}\right) \,.
\end{equation}
This is now a simple constrained maximisation problem in one variable \( y \) that can be solved with Mathematica's \texttt{FindMaximum} combined with numerical integration. The bound is plotted in figure~\ref{fig:linearised-unitarity-bound} for \( \mu = 1.5 \), together with the analytic bounds~\eqref{eq:L21-L20-positivity-bound-analytic} and~\eqref{eq:L21-linearised-unitarity-analytic}. The bound coincides with \( \lambda_{2,1} \geq h(\mu) \) for \( \lambda_{2,0} \geq g(\mu) \equiv \int \tfrac{\mathrm{d}s}{2\pi} n(s) V_{2,0}^{0}(s) \), but for \( 0 < \lambda_{2,0} \leq g(\mu) \) it is stronger than the positivity bound~\eqref{eq:L21-L20-positivity-bound-analytic}.

\section{Details on Muskhelishvili-Omnès function}\label{app:omnes}

We explain in this appendix the origin of~\eqref{eq:omnes-ansatz}, its analytic properties and how this is related to the definition of the Muskhelishvili-Omnès function in~\eqref{eq:OmegaDef}.
Let $f\equiv f^{\ell=0}_{AA\to AA}$ be the scalar partial amplitude for $AA\to AA$ scattering.
In the elastic region, unitarity requires
\begin{equation}
    2\Im f = r(s) |f|^2 \quad \longrightarrow \quad \Im f^{-1} = -\frac{r(s)}{2}\,, \qquad 4m^2 \leq s < \min{\{4M^2, 16m^2\}}\,,
\end{equation}
where $r(s)$ is defined in~\eqref{eq:CsDef}. Consider now the Chew-Mandelstam function $C(s)$ entering~\eqref{eq:CsDef}.
Using $\lim_{\epsilon\to 0^{+}} 1/(x\pm i \epsilon) =\text{P.V.}\,(1/x) \mp i \pi \delta(x)$ in the integral form, one easily finds
\begin{equation}\label{eq:App1}
    C(s\pm i \epsilon) = \text{P.V.}\, C(s) \pm i r(s)  \longrightarrow \Im\, C(s) = r(s)\, \qquad  \text{for}\;\; s> 4m^2\,,
\end{equation}
where P.V. stands for the principal value of the integral.
The combination $K^{-1}\equiv 2/f+C$ is then an arbitrary function that
has no discontinuity across the elastic right-hand cut.
We can then express the partial wave as
\begin{equation}
    f = \frac{2}{K^{-1} - C}\,.
\end{equation}
Assuming that $f$ is dominated by a resonance, a minimal choice for $K^{-1}$ reads
\begin{equation}
    K^{-1} = \frac{m_0^2-s}{g^2}\,,
\end{equation}
where  $m_0$ and $g^2$ are related to the mass, width and coupling of the resonance.\footnote{%
    Alternative choices for $K^{-1}$ may be adopted when additional features of the amplitude need to be incorporated, such as the Adler zeros that occur, for example, in pion-pion scattering.
}
Up to an overall constant, the resulting partial wave $f$ is proportional to the Omnès function~\eqref{eq:omnes-ansatz}.
The phase shift $\delta^0$ can be determined either from $f$, or equivalently from the argument of $\Omega^0$, using~\eqref{eq:App1} and~\eqref{eq:omnes-ansatz}.
We get in both cases
\begin{equation}\label{eq:Omegadelta3}
    \delta^0(s) =   \arctan \frac{g^2 r(s)}{m_0^2-s -g^2  \text{P.V.}\, C(s)} \,.
\end{equation}
The phase shift $\delta^0$ is of the form expected for a resonance.
For a thin resonance, $\delta^0(s)$ approaches a Breit-Wigner form, where $m_0$ and $m_0 \Gamma \approx g^2 r(m_0^2)$ are approximately the mass and the width of the resonance, respectively.\footnote{%
    The formalism applies also for broad resonances. In this case, $m_0$ and $g^2 r(m_0)/m_0$ substantially deviate from the mass and width of the resonance, which are properly determined by looking at the pole of $\Omega^0$ in the second Riemann sheet of the $s$-complex plane, see below.
}
Note that $\delta^0(s)$ has the correct square-root behaviour close to the threshold $s=4m^2$:
\begin{equation}
    \delta^0(s) = \frac{g^2}{2m\Big(m_0^2-4m^2 -\frac{2g^2}{\pi}\Big)} \sqrt{s-4m^2} + {\cal O}(s-4m^2)^{\frac 32}\,,
\end{equation}
and monotonically approaches $\pi$ as $s\to \infty$.
We now show that $\Omega^0$ in~\eqref{eq:omnes-ansatz} can be written as in the defining form~\eqref{eq:OmegaDef}, with $\delta^0$ as in~\eqref{eq:Omegadelta3}.
Define
\begin{equation}
    L \equiv \log \Omega^0 = \log m_0^2 - \log D(s)\,,\qquad D(s)\equiv m_0^2 -s- g^2 C(s)\,,
\end{equation}
so that
\begin{equation}
    L(s+i \epsilon) - L(s-i \epsilon) = 2i \delta^0(s)\,.
\end{equation}
Write then a dispersion relation starting from an analytic point in the $s$-plane,
\begin{equation}
    \frac{L(s)}{s} = \frac{1}{2 i \pi}\oint_\gamma \dif{z}\frac{L(z)}{z(z-s)} \,,
\end{equation}
where $\gamma$ is a small circle around the point $s$. Blowing up the contour, we get
\begin{equation}
    L(s) = \frac{s}{\pi}\int_{4m^2}^\infty \dif{z}\,\frac{\disc L(z)}{z(z-s)} = \frac{s}{\pi}\int_{4m^2}^\infty \dif{s'}\,\frac{\delta^0(s')}{s'(s'-s)} \,.
\end{equation}
By exponentiating, we then reproduce~\eqref{eq:OmegaDef}.

We can finally verify that  $\Omega^0$ cannot have poles in the principal sheet, but it can in the second sheet, as expected for a resonance appearing in an elastic amplitude.
Away from the branch cut, the denominator in~\eqref{eq:omnes-ansatz} cannot have a zero for complex $s$, since its imaginary part never vanishes.
Indeed, simple algebra gives
\begin{equation}
    \Im \, D^{\text{I}} (s) = - \Im \,s \Big(1+ \frac{g^2}{\pi} \int_{4m^2}^\infty\dif{s'}\,\frac{r(s')}{|s'-s|^2} \Big) \,,
\end{equation}
where I indicates the principal sheet.
The term in parentheses is positive definite and hence, for Im $s\neq 0$, Im $D^\text{I}$ cannot vanish.
For real $s<4m^2$, we should check whether $D(s)$ can give rise to zeros, namely to stable bound states.
It is easy to see that $D^\text{I}(s)$ is monotonically decreasing from $\infty$ for $s\to -\infty$ to some value as $s\to 4m^2$. If $D^\text{I}(s=4m^2)>0$, no zeros can appear in the whole range $s\in(-\infty, 4m^2)$.
We have
\begin{equation}
    D^\text{I}(s=4m^2) =   m_0^2-4m^2-\frac{2g^2}{\pi} \,,
\end{equation}
so demanding the absence of poles requires
\begin{equation}\label{eq:omnes-g-constraint}
    g^2 < \frac{\pi}{2} (m_0^2 - 4 m^2)\,.
\end{equation}
In the second Riemann sheet II, we have
\begin{equation}
    D^{\text{II}} = D^{\text{I}} - 2i g^2 r(s)\,,
\end{equation}
and now a complex zero will appear at $s^*$: $D^{\text{II}} (s^*)=0$.
For simplicity, let us work out the case for a thin resonance, with $s^* = m_{\text{res}}^2- i m_{\text{res}} \Gamma$, with $\Gamma\ll m_{\text{res}}$. Expanding at linear order in $\Gamma$ and $g^2$, and recalling that $C(s) \approx \text{P.V.}\, C(m_{\text{res}}^2) + i r(m_{\text{res}}^2)$, we have
\begin{equation}
    D^{\text{II}}(s) \approx m_0^2 - m_{\text{res}}^2 + i m_\text{res} \Gamma - g^2 \text{P.V.}\,  C (m_{\text{res}}^2) - i g^2 r(m_{\text{res}}^2)= 0 \,,
\end{equation}
which gives, up to order $g^2$:
\begin{equation}
    m_{\text{res}}^2  \approx m_0^2 - g^2 \text{P.V.}\, C(m_0^2)\,, \qquad \qquad
    \Gamma  \approx \frac{g^2r(m_0^2)}{m_0}  \,.
\end{equation}

\section{Details on primal \texorpdfstring{\( S \)}{S}-matrix bootstrap}\label{app:numerics-details}
In this appendix we give further details on the numerical setup used to obtain the bounds presented in section~\ref{sec:numerics}.
We first comment on how our bounds depend on the truncations in energy and spin, controlled by the parameters \( n\ped{pts} \) and \( L\ped{max} \).
We then explain how a judicious rescaling of the SDP~\eqref{eq:standard-SDP} makes the numerical optimisation more stable and faster.
Next, we analyse the relative importance of the various positivity conditions imposed, revealing some patterns that perhaps deserve a better understanding.
Finally, we list the parameters used for the semi-definite optimisation.

\paragraph{Sampling convergence.}
The semi-definite programs that we solve depend on the parameter \( n\ped{pts} \) that controls the number of values of \( s \) where we impose the unitarity constraints~\eqref{eq:sdp-unitarity}.
As we increase \( n\ped{pts} \) the bound can only get stronger.
In practice we use a Chebyshev distribution for the angular variable \( \varphi(s) \) defined as \( e^{i \varphi(s)} = \rho_{+}(s) \) for the \( AB \to AB \) unitarity and \( e^{i \varphi(s)} = \rho(s, 4M^{2}, t_{0}) \) for the \( AA \to BB \) unitarity.
See~\eqref{eq:rho-variable-generic} and~\eqref{eq:rho-variables} for the definitions of the \( \rho \)-variables.
In both cases the physical regions \( [m_{+}^{2}, +\infty) \) and \( [4M^{2}, +\infty) \) are mapped to the interval \( [0, \pi] \) in \( \varphi \), where we pick a grid
\begin{equation}\label{eq:sampling-angles}
    \varphi_{k} = \frac{\pi}{2} + \frac{\pi}{2} \cos\left(\frac{k + 1/2}{n\ped{pts}} \pi\right) \,, \qquad k = 0, \ldots, n\ped{pts} -1 \,.
\end{equation}

For a representative set of observables, we have checked the dependence of our bounds on \( n\ped{pts} \) by solving the SDP at \( n\ped{pts} = 100, 150, 200, 250 \). We have found non-trivial dependence on this truncation up to 200, but little change between 200 and 250 (below 1\%).
We have then settled on fixing \( n\ped{pts} = 200 \) for all the numerical optimisations.

\paragraph{Spin convergence and subtracted positivity.} The semi-definite programs that we solve depend on the parameter \( L\ped{max} \) which sets the maximum spin where we impose the unitarity constraints~\eqref{eq:sdp-unitarity}.
As we increase \( L\ped{max} \) the bound can only get stronger.
Our setup turned out to be quite sensitive to this parameter, which we had to push to much higher values than those commonly used for the primal \( S \)-matrix bootstrap of identical particles.
The rule of thumb is that one needs to increase \( L\ped{max} \) as the size of the ansatz \( N\ped{max} \) increases.
In our case we also observed that we had to increase \( L\ped{max} \) as we increase the mass ratio \( \mu \).
In practice, we have adopted the strategy of fixing \( L\ped{max} \) to be a value such that the bound on a given observable at \( N\ped{max} = 26 \) is stable.
In order to accelerate the convergence of the bound in \( L\ped{max} \) we have imposed the \emph{subtracted positivity} constraints introduced in~\cite{EliasMiro:2022xaa}.\footnote{%
    We thank Andrea Guerrieri for suggesting this.
}
Subtracted positivity is the statement that
\begin{equation}
    \begin{aligned}
      &\Im T_{AB \to AB}(s,t,u) - \sum_{\ell \leq L\ped{max}} \frac{2 \ell+1}{2} P_{\ell}(z(s,t)) \Im f_{AB \to AB}^{\ell}(s) \\
      &\qquad = \sum_{\ell > L\ped{max}} \frac{2 \ell+1}{2} P_{\ell}(z(s,t)) \Im f_{AB \to AB}^{\ell}(s) \geq 0 \,,
    \end{aligned}
\end{equation}
for \( s \geq m_{+}^{2} \) and \( 0 \leq t < 4m^{2} \), with \( z(s,t) \) given in~\eqref{eq:PosABAB}.
Indeed in this kinematic regime the partial wave decomposition converges, as recalled in appendix~\ref{app:PWE}, the partial amplitudes \( f^{\ell}_{AB \to AB}(s) \) have positive imaginary part as a consequence of~\eqref{eq:unitarity-ABtoAB}, and the Legendre polynomials \( P_{\ell}(z) \) are positive since \( z(s,t) \geq 1 \).
This condition can again be formulated in the form~\eqref{eq:standard-SDP}.
We have imposed this constraint on a grid in \( (s,t) \), by sampling \( s \) as before and picking \( 10 \) values of \( t_{n} \) in the interval \( [0, 4m^{2}] \) accumulating at the endpoint
\begin{equation}\label{eq:subtracted-positivity-t-values}
    t_{n} \in \{0, 0.95, 1.9, 2.85, 3.8, 3.9, 3.99, 3.999, 3.9999, 3.99999\} \,.
\end{equation}

Here we report an example check of spin convergence for the minimisation of \( \lambda_{2,1} \) at \( N\ped{max} = 26 \) and \( \mu = 1.5 \), imposing the subtracted positivity constraints.
\begin{center}
    \begin{tabular}{r|lllll}
      \toprule
      \( L\ped{max} \) & 30 & 40 & 50 & 60 & 70 \\
      \( \min \lambda_{2,1} \) & \( -10.063271 \) & \( -1.144910 \) & \( -1.092609 \) & \( -1.092587 \) & \( -1.092583 \)\\
      \bottomrule
    \end{tabular}
\end{center}
A common pattern we observe is that in the presence of the subtracted positivity constraint the bound gets very stable after a threshold \( L\ped{max} \), while in its absence it changes quite smoothly in the above range but extrapolates to the same value. Subtracted positivity is then useful to avoid the need for a double extrapolation in \( L\ped{max} \) and \( N\ped{max} \).

\paragraph{Integrals precision.}
The partial wave projection of the \( \rho \)-ansatz~\eqref{eq:ansatz-partial-wave-projection} is computed by numerical integration with \texttt{Mathematica}.
This requires working at quite high precision because the integrals for \( s \) close to the physical threshold and large \( \ell \) are exponentially suppressed as in~\eqref{eq:threshold-expansion}.
For large \( n\ped{pts} \) and large \( L\ped{max} \), the smallest partial wave projection scales roughly as \( \sim n\ped{pts}^{-4L\ped{max} - 2} \).
This means that numerical integration for points close to the physical threshold has to be performed at very high \texttt{WorkingPrecision} and it is essentially insensitive to \texttt{PrecisionGoal}.
Away from the physical threshold, instead, the precision of the numerical integration is controlled by \texttt{PrecisionGoal}.
In practice we have used the following
\begin{align*}
  \text{\texttt{NIntegrate[\ldots,}} & \text{\texttt{"WorkingPrecision" -> 900, "MaxRecursion" -> 150,}} \\
                                     & \text{\texttt{"PrecisionGoal" -> 200, "AccuracyGoal" -> 200,}}\\
                                     & \text{\texttt{"Method" -> \{"GlobalAdaptive", Method -> "ClenshawCurtisRule"\}]}}
\end{align*}
which guarantees at least 200 correct digits for the smallest integrals that we faced.

\paragraph{SDP rescaling.} The suppression of the partial wave projections at the physical threshold~\eqref{eq:threshold-expansion} also leads to ill-conditioning of the SDP~\eqref{eq:standard-SDP} in its naive formulation.
Unitarity of the \( AB \to AB \) amplitude~\eqref{eq:unitarity-ABtoAB} is imposed in the numerics as positive semi-definiteness of the real matrix which takes the form
\begin{equation}\label{eq:sdp-real-formulation}
    \begin{pmatrix}
      2 - I_{j} & R_{j} \\
      R_{j} & I_{j}
    \end{pmatrix}
    \succeq 0 \,,
    \qquad j = (AB \to AB , \ell, s) \,,
\end{equation}
where
\begin{equation}
    R_{j} = \vec{\alpha} \cdot \Re\left[\vec{X}^{\ell}_{AB \to AB}(s)\right] \qquad
    I_{j} = \vec{\alpha} \cdot \Im\left[\vec{X}^{\ell}_{AB \to AB}(s)\right]
\end{equation}
This is equivalent to the first constraint in~\eqref{eq:sdp-unitarity} for hermitian matrices.
Close to the physical threshold, both the real and the imaginary parts of \( \vec{X}^{\ell}_{AB \to AB}(s) \) are exponentially suppressed in the COM momentum \( \mom \), implying that for generic \( \vec{\alpha} \) the matrix in~\eqref{eq:sdp-real-formulation} has an eigenvalue which is extremely close to zero.
Conditions in the SDP with eigenvalues very close to zero are problematic for the solver \texttt{SDPB} and can lead to instabilities where the algorithm never manages to achieve a \emph{dual jump}, unless the error thresholds are set extremely small, roughly below the size of the smallest eigenvalue involved, which, in turn, also requires working at extremely high precision.

A useful trick to overcome this difficulty is to perform a diagonal rescaling of the SDP conditions. Defining
\begin{equation}
    D_{j} = \begin{pmatrix}
      1 & 0 \\
      0 & 1/\sqrt{d_{j}}
    \end{pmatrix}
    \qquad d_{j} = \left\lvert\left\lvert\Im \vec{X}^{\ell}_{AB \to AB}(s)\right\rvert\right\rvert_{\infty}
\end{equation}
where \( ||\vec{x}||_{\infty}= \max_{i}|x_{i}| \) is the supremum norm,
we construct the following equivalent SDP constraint
\begin{equation}\label{eq:sdp-rescaling}
    D^{t}_{j} \begin{pmatrix}
      2 - I_{j} & R_{j} \\
      R_{j} & I_{j}
    \end{pmatrix}
    D_{j}
    =
    \begin{pmatrix}
      2 - I_{j} & R_{j}/\sqrt{d_{j}} \\
      R_{j}/\sqrt{d_{j}} & I_{j} / d_{j}
    \end{pmatrix}
    \succeq 0
\end{equation}
The matrix \( D_{j} \) is invertible, so the congruence transformation preserves the cone of positive semi-definite matrices and the rescaled constraint is equivalent to~\eqref{eq:sdp-real-formulation}.
An important observation is that, at fixed \( \ell \) and \( s \), both the real and the imaginary parts of the components of \( \vec{X}_{AB \to AB}^{\ell}(s) \) are roughly of the same size and do not span orders of magnitude.
This implies then that the new positive semi-definite condition no longer has almost-zero eigenvalues, since \( I_{j}/d_{j} \) is now made of roughly order-one numbers.

In practice we apply this rescaling to the SDP not only close to the physical threshold but for all the \( (\ell, s) \) that we sample.
This has the additional benefit that the SDP constraints are all roughly of the same size, independently of \( \ell \) and \( s \).
After performing this rescaling we observe that the \texttt{SDPB} solver always has both primal and dual jumps, that the internal precision needed can be dramatically reduced, and that the internal convergence of the SDP algorithm is much faster.

\begin{figure}[t]
    \centering
    \begin{subfigure}[c]{\linewidth}
        \centering
        \includegraphics[width=0.85\linewidth]{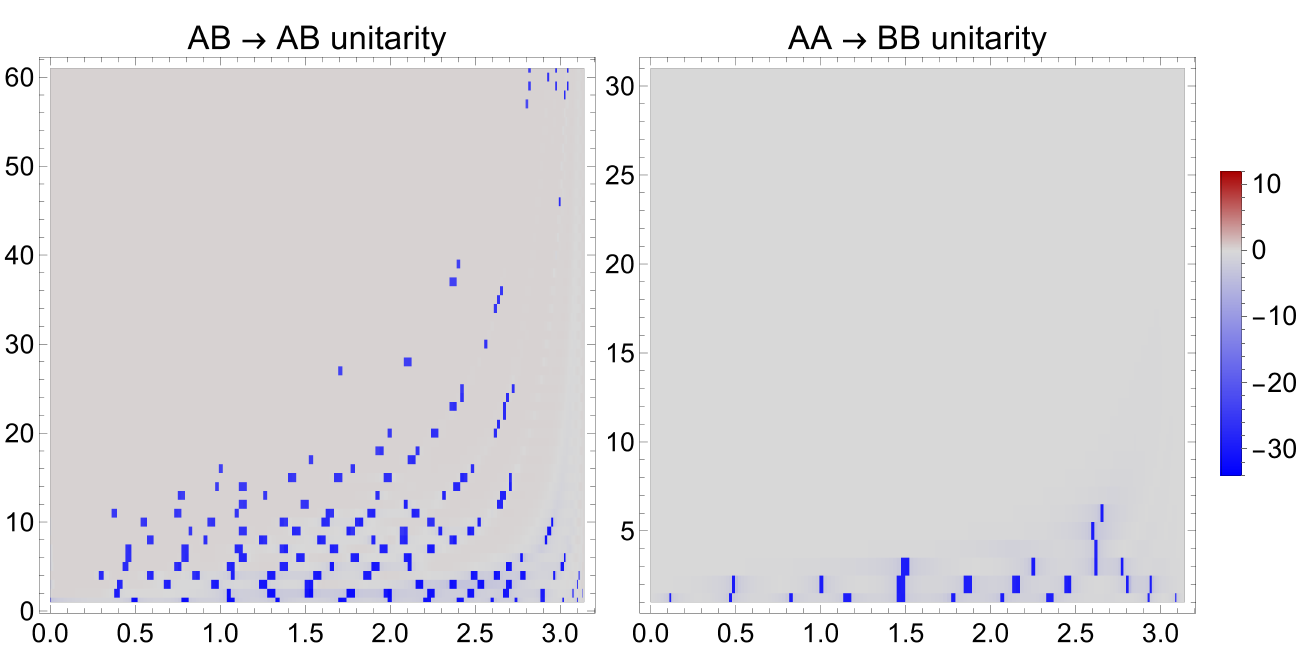}
        \caption{\( \min \lambda_{2,1} \) at fixed \( \lambda_{2,0} = 20 \)}
    \end{subfigure}

    \vspace{1em}

    \begin{subfigure}[c]{\linewidth}
        \centering
        \includegraphics[width=0.85\linewidth]{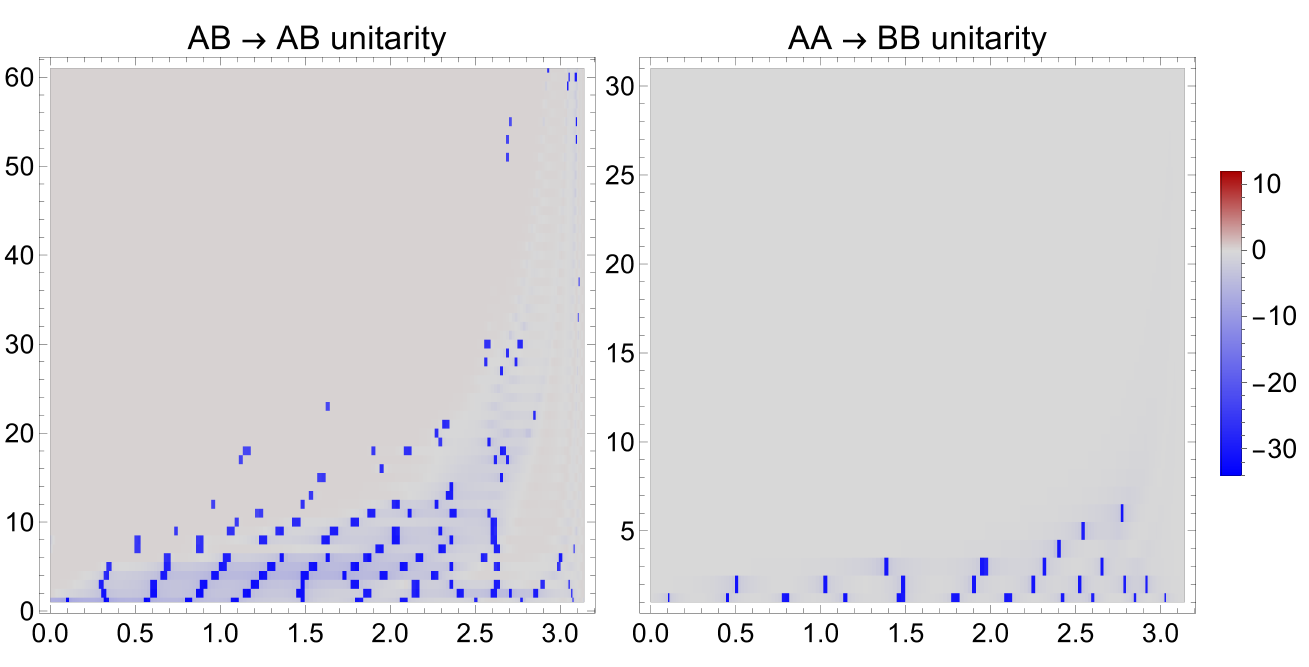}
        \caption{\( \min a_{0} \) at fixed \( \lambda_{2,0} = 11 \)}
    \end{subfigure}
    \caption{Density plot of \( \log \lambda_{j} \), the logarithm of the minimum eigenvalue of the condition matrix \( M_{j}(\alpha^{*}) \) at the optimal solution of two optimisation problems (\( \mu = 1.5, N\ped{max} = 18, L\ped{max} = 60 \)). The index \( j \) is expanded into a two-dimensional grid \( (\varphi_{k}, \ell) \) for the unitarity conditions. A blue colour denotes the conditions that have the most impact on the bound, while conditions with \( \lambda_{j} > 1 \) can be removed from the SDP with little change in the optimal objective.\label{fig:conditioning}}
\end{figure}

\paragraph{Which conditions are more important?} Let \( \vec{\alpha}^{*} \) be the optimal solution to the SDP~\eqref{eq:standard-SDP} and let
\begin{equation}
    M_{j}(\alpha^{*}) = M_{0,j} + \vec{\alpha}^{*} \cdot \vec{M}_{j} \,.
\end{equation}
We recall that \( j \) is a multi-index labelling all the positive semi-definite constraints that we compose indexed by \( j = (\text{channel}, \ell, \varphi_{k}) \) (here we trade \( s \) for \( \varphi_{k} \) in~\eqref{eq:sampling-angles}).
A useful measure of how much a given constraint contributes to the optimal bound is given by the minimal eigenvalue of the matrix, \( \lambda_{j} = \lambda\ped{min}(M_{j}(\alpha^{*})) \).
Typically, the condition \( j \) has an important impact on the bound if \( \lambda_{j} < 1 \) and the conditions with the smallest eigenvalue are the most important, in the sense that removing them can alter significantly the value of the optimal objective.\footnote{%
    The minimum value of \( \lambda_{j} \) is of the same size as the duality gap used in the SDP solver.
}
In figure~\ref{fig:conditioning} we report a plot of such quantity for a sample of optimisation problems.
The index \( j \) is expanded into a two-dimensional grid: for the \( AB \to AB \) unitarity conditions this is \( (\varphi_{k}, \ell) \), with \( k = 1,\ldots,n\ped{pts} \) and \( \ell = 0,\ldots,L\ped{max} \); for the \( AA \to BB \) unitarity conditions this is \( (\varphi_{k}, \ell) \) with \( k \) as before and \( \ell = 0, 2, \ldots, L\ped{max} \).
We observe that the bound on this specific observable is most sensitive to the \( AB \to AB \) unitarity conditions for \( \ell \lesssim 20 \) and for a wide range of energies, and also to the high-energy and high-spin region.
The bound is also sensitive to \( AA \to BB \) unitarity conditions with small \( \ell \lesssim 5 \).

It would be interesting to understand better this pattern, which is shared also by other observables.
In particular, it would be interesting to understand why it is not so sensitive to low-energy and high-spin for \( AB \to AB \) and not sensitive to high spin, for any energy, for \( AA \to BB \).
Figure~\ref{fig:conditioning} suggests that the grid used in \( (\varphi_{k}, \ell) \) is probably not the most efficient choice.
A smarter, and perhaps adaptive, sampling scheme in \( s \) and \( \ell \) could be adopted to speed up the numerics in the future.

\paragraph{SDP solver.} The semi-definite optimisation problems considered in this work were solved using the SDP solver \texttt{SDPB}~\cite{Simmons-Duffin:2015qma,Landry:2019qug}.
On certain occasions, especially before applying the SDP rescaling described above, we also used a modified version to overcome some stalling problems, available at~\url{https://gitlab.com/apiazza134/sdpb-midck-stallingrecover}.
The parameters we used for the solver are the default ones, in particular a \texttt{dualityGap} of \( 10^{-30} \), and a binary \texttt{precision} of 655.

\bibliographystyle{JHEP}
\bibliography{refs}

\end{document}